\documentclass{aa}  

\usepackage{graphicx}
\usepackage{natbib}
\usepackage{array}
\usepackage{rotating}
\usepackage{txfonts}
\usepackage{multirow}
\usepackage{longtable}
\usepackage{lscape}

\usepackage{multirow}

\begin{document} 

\title{Chemical abundances of fast-rotating massive stars  
\thanks{Based on observations obtained with the Heidelberg Extended Range Optical Spectrograph (HEROS) at the Telescopio Internacional de Guanajuato (TIGRE) with the SOPHIE \'echelle spectrograph at the Haute-Provence Observatory (OHP; Institut Pytheas; CNRS, France), and with the Magellan Inamori Kyocera Echelle (MIKE) spectrograph at the Magellan II Clay telescope. Based also on archival data from the Galactic O-Star Spectroscopic Survey (GOSSS), the Anglo-Australian Telescope (AAT) equipped with the University College London Echelle Spectrograph (UCLES), the ESO/La Silla Observatory with the Fiber-fed Extended Range Optical Spectrograph (FEROS; programmes 70.D-0110, 075.D-0061, 076.C-0431, 081.D-2008, 083.D-0589, 086.D-0997, 087.D-0946, 089.D-0189, 089.D-0975, 179.C-0197, and the High Accuracy Radial velocity Planet Searcher (HARPS; programme 60.A-9036), the Pic du Midi Observatory equipped with the NARVAL spectropolarimeter, the San Pedro M{\'a}rtir (SPM) observatory with the Echelle SPectrograph for Rocky Exoplanet and Stable Spectroscopic Observations (ESPRESSO), the OHP with the AURELIE and ELODIE \'echelle spectrographs, the Nordic Optical Telescope (NOT) with the FIbre-fed Echelle Spectrograph (FIES), the Canada-France-Hawaii Telescope (CFHT), with the Echelle SpectroPolarimetric Device for the Observation of Stars (ESPaDOnS) spectrograph, the Leonhard Euler Telescope with the CORALIE spectrograph.}
$^{\rm ,}$\thanks{Table \ref{tabResults} is available in electronic form at the CDS via anonymous ftp to cdsarc.u-strasbg.fr (130.79.128.5) or via http://cdsweb.u-strasbg.fr/cgi-bin/qcat?J/A+A/}
}
\subtitle{I. Description of the methods and individual results}

   \author{Constantin Cazorla\inst{1}
          \and
          Thierry Morel\inst{1}
          \and
          Ya\"el Naz\'e\inst{1}\thanks{Research associate FNRS.}
          \and
          Gregor Rauw\inst{1}
          \and
          Thierry Semaan\inst{2,1}
          \and
          Simone Daflon\inst{3}
          \and
          M. S. Oey\inst{4}
          }

   \institute{Space sciences, Technologies and Astrophysics Research (STAR) Institute, Universit\'e de Li\`ege, Quartier Agora, All\'ee du 6 Ao\^ut 19c, B\^at. B5C, B4000-Li\`ege, Belgium
   \and
   Observatoire de Gen\`eve, Universit\'e de Gen\`eve, Chemin des Maillettes 51, 1290 Versoix, Switzerland
   \and
   Observat\'orio Nacional, Rua General Jos\'e Cristino 77 CEP 20921-400, Rio de Janeiro, Brazil
   \and
   University of Michigan, Department of Astronomy, 311 West Hall, 1085 S. University Ave, Ann Arbor, MI 48109--1107, USA
   \newline
              \email{cazorla@astro.ulg.ac.be}
             }

   \date{Received ... ; accepted ...}

 
  \abstract
   {}
   {Recent observations have challenged our understanding of rotational mixing in massive stars by revealing a population of fast-rotating objects with apparently normal surface nitrogen abundances. However, several questions have arisen because of a number of issues, which have rendered a reinvestigation necessary; these issues include the presence of numerous upper limits for the nitrogen abundance, unknown multiplicity status, and a mix of stars with different physical properties, such as their mass and evolutionary state, which are known to control the amount of rotational mixing.}
   {We have carefully selected a large sample of bright, fast-rotating early-type stars of our Galaxy (40 objects with spectral types between B0.5 and O4). Their high-quality, high-resolution optical spectra were then analysed with the stellar atmosphere modelling codes DETAIL/SURFACE or CMFGEN, depending on the temperature of the target. Several internal and external checks were performed to validate our methods; notably, we compared our results with literature data for some well-known objects, studied the effect of gravity darkening, or confronted the results provided by the two codes for stars amenable to both analyses. Furthermore, we  studied the radial velocities of the stars to assess their binarity.}
   {This first part of our study presents our methods and provides the derived stellar parameters, He, CNO abundances, and the multiplicity status of every star of the sample. It is the first time that He and CNO abundances of such a large number of Galactic massive fast rotators are determined in a homogeneous way.}
   {}


   \maketitle
%

\section{Introduction}
\label{intr}
Massive stars are defined {as objects born with O or early B spectral types (subsequently evolving to later types during their life) and by their death as a supernova (thus having initial masses larger than $\sim$8 M$_{\odot}$)}. These OB stars are the true cosmic engines of our Universe. They emit an intense ionising radiation and eject large quantities of material throughout their life, shaping the interstellar medium, affecting star formation, and largely contributing to the chemical enrichment of their surroundings. It is therefore of utmost importance to develop a good understanding of the physical processes at play in these objects and to {properly model their evolution.}

One important feature of massive stars is their high rotational velocities, which can be up to {at least} 400 km s$^{-1}$ (\citealt{how97}{; \citealt{duf11}}). Such a fast rotation can be produced by several mechanisms: it can be acquired at birth as a result of their formation or develop subsequently during their evolution as they interact with a companion \citep[through tidal forces, mass accretion, or even merging;][]{zah75,hut81,pac81,pol91,pod92,lan03b,pet05a,pet05b,dem09,dem13,der10,tyl11,son13}.

Despite the rotational velocity of OB stars, which can amount to a significant fraction of the critical (break up) velocity\footnote{The critical velocity of a star is reached when the centrifugal acceleration is equal to the gravitational one at the equator.}, rotation had been considered for a long time as a minor ingredient of stellar evolution until some important discrepancies between model predictions and observations were brought to light \citep[e.g.][]{mae95}. The importance of rotation on the evolution of massive stars is now considered to be comparable to that of stellar winds \citep{mey00}, influencing all aspects of stellar evolution models \citep{mae15}. For example, rotation increases the main-sequence (MS) lifetime by bringing fresh combustibles to the core. It also modifies the stellar temperature, thus the radiative flux.

Rotation also triggers the transport of angular momentum and chemicals in the interior \citep{mae96}. This can notably lead to a modification of the wind properties and to changes in the chemical abundances seen at the stellar surface. In this context, it might be useful to recall that massive stars burn their central hydrogen content through the CNO cycle, which can be partial or complete depending on the temperature. For stars whose mass does not exceed 40 M$_{\odot}$, the \element[][16][]{O} abundance can be considered constant and that of \element[][12][]{C} depleted in the core. For more massive stars, the constancy applies to the \element[][12][]{C} abundance, while the core is depleted in \element[][16][]{O}. In all cases, the slow reaction rate of \element[][14][7]{N} $\longrightarrow$ \element[][15][8]{O} leads to an excess of nitrogen nuclei in the core. These elements may then be dredged up to the stellar surface, but the actual amount transported depends on the mixing efficiency, which is primarily a function of the rotation rate. Because it is the most affected, the nitrogen abundance at the stellar surface is considered the best indicator of rotational mixing (along with boron, but UV spectra are needed to study the abundance of this latter element; \citealt{pro01}). In contrast, slow rotation is expected in principle not to lead to any detectable nitrogen enrichment during the main-sequence phase, at least for stars in the mass range 5--60 M$_{\odot}$ \citep{mae14}.

However, recent observations of B stars in the Galaxy and the Magellanic Clouds (MCs) in the framework of the VLT-FLAMES Survey of Massive Stars \citep{eva08} have revealed two stellar populations that exhibit surface nitrogen abundances not predicted by single-star evolutionary models incorporating rotational mixing \citep{hun07,hun09}. For instance, in the LMC, the first population (15\% of the sample) is composed of slow rotators that unexpectedly exhibit an excess of nitrogen, while stars of the second group (also 15\% of the sample) are fast rotators with $v\sin\,i$ up to $\sim$ 330 km s$^{-1}$ showing no strong nitrogen enrichment at their surface, if any \citep{bro11b}. Additional examples of the former category have been found amongst O stars in the LMC \citep{riv12a,riv12b,grin16}. The origin of this population is a matter of speculation, but has been proposed to result from the action of magnetic fields (\citealt{mey11}, \citealt{pot12}). On the other hand, it is conceivable that stars in the second group are binaries that have undergone an episode of highly non-conservative mass transfer, with transport of angular momentum, but little transfer of CNO-processed material (see \citealt{lan08}). 

A clear interpretation of these observations is, however, hampered by the limited quality of the abundance determinations. The reported nitrogen abundance of the fast rotators frequently are upper limits and {information is unavailable or uncertain} for other key elements, such as helium or carbon {(e.g. \citealt{hun09} in the case of carbon)}. Furthermore, \citet{mae09} pointed out the different evolutionary stages (on and away from the main sequence) and the large range of masses (from 10 to 30 M$_{\odot}$) of the stars studied in the VLT-FLAMES Survey. These authors found a better agreement with model predictions after the sample was split into groups of stars with similar properties (but see \citealt{bro11b}, who addressed this issue through population synthesis). \citet{mae14} also questioned some results obtained by \citet{hun07,hun09} based on a reanalysis of their data. Finally, \citet{bou13} and \citet{mar15a} argued that the CNO abundances of most O stars in their studies are compatible with the expectations from single-star evolutionary models, although their samples only contain few fast rotators. The observed efficiency of rotational mixing thus appears unclear, and more data is required to make progress.

\section{Rationale of our study}
Up to now, only a few comprehensive investigations of the metal content of fast-rotating, Galactic OB stars have been undertaken. \object{HD 191423} (ON9\,II-IIIn, \citealt{sot11}; $v$\,sin\,$i$ $\sim$ 420 km s$^{-1}$) has been studied by \citet{vil02}, \citet{mah15}, and \citet{mar15a}. \object{HD 149757} ($\zeta$\,Oph; O9.2\,IVnn, \citealt{sot11}; $v$\,sin\,$i$ $\sim$ 378 km s$^{-1}$) has been studied by \citet{vil05}. {In addition, the CNO abundances of two O-type supergiants, two O dwarfs, five additional O giants, and four other O-stars with $v$\,sin\,$i \ge 200$ km s$^{-1}$ have been derived by \citet{bou12}, \citet{mar12b}, \citet{mar15b}, and \citet{mar15a}, respectively.} The small number of high-resolution studies combined with the heterogeneity of the analyses has motivated us to undertake an in-depth study of bright OB stars with high rotational velocities.

The stars in our sample span a limited range in rotational velocities and evolutionary status {(as they are all core-hydrogen burning stars)}. This restricts the number of parameters potentially affecting the abundances and allows us to more easily interpret our results. Enhancement of the surface nitrogen abundance (and accompanying carbon depletion) arising from rotational mixing is expected to be more subtle at Galactic metallicities than in the MCs. However, the detailed study of fast rotators in the MCs (with typically m$_{V}$ $\sim$ 13 mag) would be a major observational undertaking (see \citealt{grin16}). In contrast, focussing on nearby stars permits a detailed abundance study with only a modest investment of telescope time. As we show below, a large body of spectroscopic data is even already available in public archives.

For all stars, we have self-consistently determined the stellar properties from high-resolution spectra: effective temperature, $T_{\rm{eff}}$, surface gravity, $\log g$, projected rotational velocity, $v$\,sin\,$i$, macroturbulence, $v_{\rm{mac}}$, as well as He and CNO abundances. An interaction with a companion may dramatically affect the evolution of the rotational and chemical properties of stars in binary systems. However, little is known about the binary status of the fast rotators previously studied in the literature. Therefore, another important aspect of our analysis is the determination of the multiplicity as a result of a radial-velocity (RV) study of our targets. To reinforce the point made above, such an investigation for the faint MC targets is also too demanding in terms of observing resources.

The results of our spectroscopic study of fast rotators are presented in two parts. This first paper describes the methods that have been used and the numerous checks performed to ensure the quality of the results. It also presents the results obtained for each star, while a follow-up paper (Paper II; \citealt{caz17}) will focus on the global interpretation of these results.

This paper is organised as follows. The sample, observations, and data reduction are outlined in Sect. \ref{samObs}; the spectroscopic analysis is described in Sect. \ref{speAna}; uncertainties in the derived physical parameters and abundances are discussed in Sect. \ref{sectUnce}; several checks of our methods are presented in Sect. \ref{secMethVal}; and conclusions are given in Sect. \ref{secCon}. Finally, {Appendices \ref{jourRV} and \ref{diagLinesCNO} provide some individual information in tabular format, while} notes on the binary and runaway status of individual stars are given in Appendix \ref{res}, Appendix \ref{resB} compares our results to those in the literature and Appendix \ref{secCompaCMFGENFig} provides a comparison between the observations of the hotter stars and their best-fit CMFGEN models.

\section{Sample, observations, and data reduction}
\label{samObs}

Our sample is composed of Galactic OB stars that have a projected rotational velocity exceeding 200 km s$^{-1}$ ; {the vast majority have m$_{V}$ $\lesssim$ 10} to ensure good quality spectra. This is further separated into two subsamples.

The first subsample comprises dwarfs and (sub)giants with spectral types between B0.5 and O9. The constraints on the spectral type and luminosity class arise from the applicability domain of our first analysis tool, DETAIL/SURFACE, which is only suitable for stars with weak winds. In addition, \ion{He}{II} features must be present, which excludes cooler objects. The second subsample contains hotter stars with spectral types up to O4, which were studied with CMFGEN, as this code can treat stars with extended atmospheres. For the sake of homogeneity, it would have been relevant to analyse the whole sample with CMFGEN. However, it is intractable in practice because of the time-consuming nature of the CMFGEN analysis. To demonstrate the validity of our approach, in Sect. \ref{subsecCMF} we compare the results provided by the two codes for a few representative cases {and show that they are consistent}. 

We excluded double-lined spectroscopic binaries because a correct extraction of each spectral component through disentangling techniques is very difficult when spectral lines are heavily broadened. Besides, it requires a large number of spectra with a good phase coverage, which are often not available. We also excluded {classical} Oe and Be stars because circumstellar discs cannot be modelled with the chosen tools. {The weak H$\alpha$ emission observed in a few stars rather originates from a stellar outflow (e.g. \object{HD 184915}; \citealt{riv13}).} In addition, we also avoided confirmed $\beta$ Cephei stars \citep{sta05} for which revealing binarity can be challenging because of line-profile variations arising from pulsations. Furthermore, {this peculiarity makes} the atmospheric parameter and abundance determinations difficult. 

We ended up with 40 targets (Table \ref{tabObj}) {that fulfilled the aforementioned criteria. While this sample of massive Galactic fast rotators is certainly not complete, it does represent a very large portion of those known in the solar vicinity. For example, SIMBAD lists only 50 stars with spectral type earlier than B0.5, m$_{V} \le$ 13, and $v\sin\,i >$ 200 km s$^{-1}$, while \citet{how97} list 32 O-type stars with $v\sin\,i >$ 200 km s$^{-1}$ but it has to be noted that these catalogues include SB2 systems, Oe/Be stars, and pulsating stars that were discarded from our sample.}

Part of the high-resolution spectra were obtained through our dedicated programmes on the following {\'e}chelle spectrographs: 

\begin{itemize}
        \item   The CORALIE spectrograph mounted on the 1.2\,m EULER Swiss telescope located at the ESO La Silla Observatory (Chile). CORALIE has the same optical design as ELODIE \citep{bar96}. All the steps of the reduction were carried out with the dedicated pipeline called DRS. The spectra cover the wavelength range 3870--6890 $\AA$ with a resolving power, $R$, of 60\,000.
        \item The HEROS spectrograph mounted on the 1.2\,m TIGRE telescope at La Luz Observatory \citep[Mexico;][]{sch14}. The spectral domain covered by HEROS spans from 3500 to 5600 $\AA$ and from 5800 to 8800 $\AA$ (blue and red channels, respectively) for $R\sim20\,000$. The spectra were automatically reduced with an Interactive Data Language (IDL) pipeline based on the reduction package REDUCE written by \citet{pis02}.
        \item The MIKE spectrograph mounted on the 6.5\,m Magellan II Clay telescope located at the Las Campanas Observatory (LCO; Chile). MIKE is a double {\'e}chelle spectrograph yielding blue (3350--5000 $\AA$) and red (4900--9500 $\AA$) spectra simultaneously. In the blue part, $R\sim$ 53\,000. The spectral reduction was carried out using the Carnegie Observatories python pipeline\footnote{\url{http://obs.carnegiescience.edu/Code/mike}} \citep{bra12,gar15}.
        \item The SOPHIE spectrograph at the 1.93\,m telescope at Observatoire de Haute-Provence (OHP; France). The spectra cover the wavelength range 3872--6943 $\AA$ with $R\sim40\,000$ (high-efficiency mode). The data were processed by the SOPHIE fully automatic data reduction pipeline. As a check, we reduced the raw data using standard IRAF\footnote{\url{http://iraf.noao.edu}} routines, but found negligible differences with respect to the pipeline products.
\end{itemize}   

The rest of the data were collected from several archives (unless otherwise noted, the spectra were reduced with the instrument pipeline):

\begin{itemize}
        \item The AURELIE spectrograph mounted on the 1.52\,m telescope at OHP \citep{gil94}. The spectra have $R\sim$ 9\,000 and either cover the wavelength range 4100--4950 (see \citealt{deb04}) or 4450--4900 $\AA$ \citep[see][]{mah13}. The data reduction procedure is described in \citet{rau03} and \citet{rau04}. Other reduced AURELIE data were retrieved from the Information Bulletin on Variable Stars (IBVS; \citealt{deb08})\footnote{\url{http://ibvs.konkoly.hu/cgi-bin/IBVSetable?5841-t1.tex}}.
        \item The ELODIE {\'e}chelle spectrograph mounted on the 1.93\,m telescope at OHP, which was operational from 1993 to 2006 \citep{bar96}. This instrument\footnote{\url{http://atlas.obs-hp.fr/elodie/}} covers the spectral range from 3850 to 6800 $\AA$ and has $R\sim42\,000$.
        \item The ESPaDOnS {\'e}chelle spectrograph mounted on the Canada-France-Hawaii Telescope (CFHT) on Mauna Kea. Spectra were retrieved from the Canadian Astronomy Data Centre\footnote{\url{http://www.cadc-ccda.hia-iha.nrc-cnrc.gc.ca}} and cover the wavelength range 3700--10500 $\AA$ with $R\sim81\,000$ in ``object only'' spectroscopic mode.      
        \item The ESPRESSO {\'e}chelle spectrograph mounted on the 2.12\,m telescope at Observatory Astron{\'o}mico Nacional of San Pedro M{\'a}rtir (SPM; Mexico). The spectra cover the wavelength domain 3780--6950 $\AA$ with $R\sim$ 18\,000 \citep{mah13}. The data reduction was completed using the {\'e}chelle package included in the ESO-MIDAS software\footnote{\url{http://www.eso.org/sci/software/esomidas/}}, as carried out by \citet{mah13}.
        \item The FEROS {\'e}chelle spectrograph mounted on the 2.2\,m telescope at La Silla. The ESO archives provide already reduced data for most of the sample but, when this was not the case, we reduced the raw data with the standard dedicated ESO pipeline (except for the \object{HD 52266} data taken in 2011 for which J.\,Pritchard's personal pipeline\footnote{\url{http://www.eso.org/~jpritcha/jFEROS-DRS/index.html}} was used). The FEROS spectrograph covers the spectral domain from 3500 to 9200 $\AA$ and provides spectra with $R\sim48\,000$.
        \item The FIES {\'e}chelle spectrograph at the 2.5\,m Nordic Optical Telescope (NOT) located at the Observatorio del Roque de los Muchachos (La Palma, Spain). This spectrograph covers the spectral range 3700--7300 $\AA$ with $R\sim46\,000$ (in medium-resolution mode) or 25\,000 (in low-resolution mode). FIES data were reduced with the dedicated reduction software FIEStool\footnote{\url{http://www.not.iac.es/instruments/fies/fiestool}}.
        \item The Galactic O-Star Spectroscopic Survey (GOSSS). The normalised spectra were retrieved from the GOSSS database\footnote{\url{http://ssg.iaa.es/en/content/galactic-o-star-catalog/}} \citep{mai11}. These spectra come from two facilities: the 1.5\,m telescope at Observatorio de Sierra Nevada (OSN; Loma de Dilar, Spain) with the Albireo spectrograph (spectral range coverage: 3740--5090 $\AA$) and the 2.5\,m du Pont telescope at LCO with the Boller \& Chivens spectrograph (spectral range coverage: 3900--5510 $\AA$). Because the spectral resolution of both instruments ($R\sim3000$) is much lower than that of the other spectrographs used in this work, GOSSS spectra were only used for the RV study (see Sect. \ref{sectRadVal}). 
        \item The HARPS {\'e}chelle spectrograph mounted on the 3.6\,m telescope at La Silla. The spectrograph covers the spectral range 3780--6910 $\AA$ with $R\sim120\,000$.   
        \item The NARVAL spectropolarimeter mounted on the 2\,m Telescope Bernard Lyot (TBL). NARVAL covers the wavelength range $\sim$ 3700--10\,500 $\AA$ with $R\sim75\,000$ in ``object only'' mode. Spectra were retrieved from the PolarBase database\footnote{\url{http://polarbase.irap.omp.eu}}.
        \item The UCLES {\'e}chelle spectrograph mounted on the 3.9\,m Anglo-Australian Telescope (AAT; Siding Spring Observatory, Australia). UCLES covers the wavelength range $\sim$ 4340--6810 $\AA$ with a resolving power of at least 40\,000, depending on the slit width. The raw data\footnote{\url{http://site.aao.gov.au/arc-bin/wdb/aat_database/user/query}} were reduced in a standard way with the IRAF {\'e}chelle package. 
\end{itemize}
Some spectra extracted from the archives were already normalised and, in that case, we simply checked that the normalisation was satisfactory.
Otherwise, the spectra were normalised within IRAF using low-order polynomials in selected continuum windows. These ``clean'' windows were identified after a SOPHIE spectrum of the slow rotator 10\,Lac (O9\,V) was broadened\footnote{This broadening was performed by the \begin{tt}ROTIN3\end{tt} programme that is part of the SYNSPEC routines; \url{http://nova.astro.umd.edu/Synspec43/synspec.html}} with the $v$\,sin\,$i$ value corresponding to each target.

All spectra were considered for the RV study. However, only a limited number were used to derive the parameters and abundances. The choice was based on several criteria (spectral resolution, wavelength coverage, S/N). Further details on this point can be found in Sect. \ref{sectRadVal}.

\section{Spectroscopic analysis}
\label{speAna}

\subsection{Radial velocities and binary analysis}
\label{sectRadVal}
For each stellar spectrum, the first step of our analysis was to determine the radial velocity with a cross-correlation technique available in the IRAF package RVSAO\footnote{\url{http://tdc-www.harvard.edu/iraf/rvsao}} \citep{kur98}. The closest TLUSTY synthetic spectrum \citep[BSTAR06 and OSTAR02 grids;][]{lan03,lan07} for each star was determined by a $\chi^2$ analysis and used as template. The correlation was performed only in the wavelength range from about 4350 to 4730 $\AA$. This region was chosen because of the relatively large number of spectral features (mostly helium lines), the absence of Balmer lines (which may be affected by emissions linked to stellar winds and colliding wind effects in binaries), and the fact that it was covered by all the spectrographs used in this work. Undesirable features (e.g. diffuse interstellar bands) were masked out. Table \ref{tabJour} provides the RVs measured for each spectrum alongside the observation date. 

To get the best quality data for the determination of physical parameters, we then corrected the individual spectra for their radial velocity and, when necessary, averaged on an instrument-by-instrument basis with a weight depending on the signal-to-noise ratio. These spectra, which were subsequently used for the stellar parameters determination, are identified in boldface in Table \ref{tabJour}.

To establish whether the measured RVs are variable or not, {we adopt a criterion inspired by that of \citet{san13b}}: the maximum RV difference larger than 4\,$\sigma$ and above a given threshold (20 km s$^{-1}$ as appropriate for O stars). The multiplicity status of our targets depends on the outcome of this test. If the differences are not significant, then the star is presumably considered to be single; otherwise the star is considered a RV variable {(and thus a probable binary)}. Among the latter category, we further classify as SB1 those for which a full orbital solution can be calculated (see below). For some targets, additional information is available in the literature and the multiplicity status may then be revisited (see Appendix \ref{res} for details). 

Finally, when there were at least 15 RV measurements, including all available literature values (even if their error is unknown), we also analysed the RV datasets using the following period search algorithms: (1) the Fourier algorithm adapted to sparse/uneven datasets \citep{hmm,gos01,zec09}; (2) two different string length methods \citep{lafkin,renson}; (3) three binned analyses of variances (\citealt{whi44}; \citealt{jur71}, which is identical, with no bin overlap, to the ``pdm'' method of \citealt{ste78}; and \citealt{cuy87}, which is identical to the ``AOV'' method of \citealt{sch89}); and (4) conditional entropy \citep[see also \citealt{gra13}]{cin99,cin99b}. {Although the most trustworthy technique is the Fourier method, a reliable detection is guaranteed by the repeated recovery of the same signal with different methods.} When a potential period was identified, an orbital solution was then calculated using the Li{\`e}ge Orbital Solution Package \citep[LOSP; see][]{san13}. The results of these variability tests and period searches are presented in Appendix \ref{res} for each star.

\subsection{Rotational velocities}
\label{secRotVel}  
The second step of our analysis was to derive the projected rotational velocity through Fourier techniques \citep{gra05,sim07}. In the Fourier space, the rotational broadening indeed expresses itself through a simple multiplication with the Fourier transform of the line profiles, hence providing a direct estimate of $v\sin\,i$. We considered as many lines as possible (notably \ion{He}{I}\,4026, 4471, 4713, 4922, 5016, 5048, 5876, 6678; \ion{He}{II}\,4542, 5412; \ion{C}{IV}\,5801, 5812;  and \ion{O}{III}\,5592) in order to enhance the precision of our determinations. We also made use of the \begin{tt}{iacob-broad}\end{tt} tool \citep{sim14} to determine the macroturbulent velocities. As this tool also provides an independent estimate of $v\sin\,i$ -- albeit it is also based on Fourier techniques -- it allows us to check the robustness of our $v\sin\,i$ values. These values were consistently recovered within the error bars. We caution that the derived macroturbulent velocities are upper limits only since they cannot be determined reliably for fast rotators \citep{sim14}. No significant change in stellar parameters and abundances was found whether or not the macroturbulence was considered in the computation of the synthetic spectra; the macroturbulence broadening of our synthetic spectra was performed with the \begin{tt}macturb\end{tt} programme of the SPECTRUM suite of routines\footnote{\url{http://www.appstate.edu/~grayro/spectrum/spectrum276/node38.html}} that makes use of the formulation of \citealt{gra05}. Furthermore, our spectral fits are already satisfactory when rotational velocity is the only source of broadening considered. After some preliminary tests, we therefore chose not to consider macroturbulence in our determination of the stellar parameters.

To further validate our method, we compared the $v\sin\,i$ for nine stars with those obtained by \citet{bra12}, \citet{daf07}, and \citet{gar15} with a different method based on the full width at half-minimum [FWHM] of \ion{He}{I} lines. The results are presented in Table \ref{tabCompVsiniRio}: they show a good agreement within errors, although there is some indication of slightly larger values in our case. This might be attributed to differences in the normalisation. 

\subsection{Atmospheric parameters and abundances}
\label{subSecParaHe}
Two methods were used to determine the atmospheric parameters ($T_{\rm eff}$, $\log g$) and chemical abundances depending on the sample considered. They are both based on spectral synthesis whereby a search is made for the best match between each observed spectrum and a grid of synthetic profiles broadened with the appropriate instrumental and rotational velocity profiles. They are now presented in turn. Our full results can be found in Table \ref{tabResults}, and a comparison with literature values, when available, is given in Table \ref{tabResComp}. 

We provide $\log g_{\rm{C}}$, which is the surface gravity corrected for the effects of centrifugal forces: $g_{\rm C}$ = $g$ + ($v\sin\,i$)$^2$/$R_{*}$, where $R_{*}$ is the star radius \citep{rep04}. {The radius was always estimated, for consistency, from the gravity value ($g$ = G$M$/$R_{*}^{2}$) taking the appropriate mass $M$ for each star (see Paper II for details) into account}. {Radii can also be computed from the temperatures (our best-fit $T_{\rm{eff}}$ ) and the luminosities, which are derived from the magnitude and distance of the target under consideration. While distances are not available for all our targets, two stars are believed to be part of clusters and, therefore, have their distance $d$ estimated: \object{HD 46056} and \object{HD 46485} in \object{NGC 2244} ($d$ = 1.4 kpc). Furthermore, the Hipparcos distances of \object{HD 66811} and \object{HD 149757} are known: 335$_{-11}^{+12}$ and 112$\pm$3 pc, respectively \citep{vanl97,mai08}. In addition, we used V magnitudes taken from SIMBAD, reddenings taken from WEBDA\footnote{\url{https://www.univie.ac.at/webda/}} for the cluster members or from \citet{bas92} and \citet{mor75} for \object{HD 66811} and \object{HD 149757}, respectively, as well as typical bolometric corrections for the appropriate spectral type \citep{mar05}. The radii derived from both methods agree well; however, a full comparison must await the availability of accurate distances from {\it{Gaia}} \citep{gai16}.}

\subsubsection{Method for the cooler stars}
\label{subSecMethTM}
The synthetic spectra for the stars whose spectral types are comprised between B0.5 and O9 were computed using Kurucz LTE atmosphere models assuming a solar helium abundance and the non-LTE line-formation code DETAIL/SURFACE \citep{gid81,but85}. The choice of a solar helium abundance was motivated by the fact that no appreciable differences in stellar parameters and CNO abundances were found when considering model atmospheres with a helium abundance that is twice solar, as is the case for some of our targets (Table \ref{tabResults}). The model atoms implemented in DETAIL/SURFACE are the same as those employed in Morel et al. (2006). This combination of LTE atmospheric models and non-LTE line-formation computations has been shown to be adequate for late O- and early B-type stars for which wind effects can be neglected \citep{nie07,prz11}.

We assumed a typical microturbulence to compute the synthetic spectra \citep[$\xi$ = 10 km s$^{-1}$; e.g.][]{hun09}. However, we explore the impact of this choice on our results in Sect. \ref{sectUncCoo}.

We performed the analysis  in three steps (see \citealt{rau12} for further details).  The stellar parameters and helium abundance (by number, noted $y = \mathcal{N}(\rm{He})/[\mathcal{N}(\rm{H})\,+\,\mathcal{N}(\rm{He})]$) were first determined for each star. We only summarise the procedure briefly here. The grid of synthetic spectra used was constructed by varying $\log g$ in the range 3.5--4.5 dex with a step of 0.1 dex, $T_{\rm{eff}}$ in the domain 27--35 kK with a step of 1 kK, and $y$ in the range 0.005--0.250 with a step of 0.005. A few models with both large $T_{\rm{eff}}$ and low $\log g$ are lacking because of convergence issues. We selected four Balmer lines (\ion{H}{$\epsilon$}, \ion{H}{$\delta$}, \ion{H}{$\gamma$}, and \ion{H}{$\beta$}) to derive the surface gravity, and we chose nine prominent helium lines (\ion{He}{I}\,4026, 4388, 4471, 4713, 4922, 5016, and \ion{He}{II}\,4542, 4686, 5412) because they are sensitive to both the stellar temperature, through the ionisation balance of \ion{He}{I}\,and \ion{He}{II}\,lines, and the abundance of helium. Metallic lines falling across the Balmer and He lines, but that are not modelled by DETAIL/SURFACE, were masked out during the fitting procedure. For the other metallic features, abundances typical of early-B stars determined with the same code were assumed \citep[see Table 6 of][]{mor08}. 

\begin{figure}
\centering
\begin{turn}{0}
\includegraphics[scale=0.40]{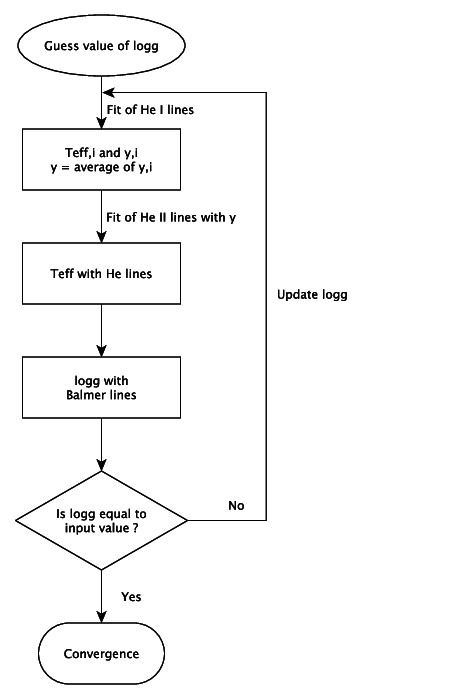}
\end{turn}
\caption{Flowchart of the method used for the cooler stars to derive the atmospheric parameters and helium abundance.}
\label{TMMethFlow}
\end{figure}

\begin{figure*}
\centering
\begin{turn}{0}
\includegraphics[scale=0.35]{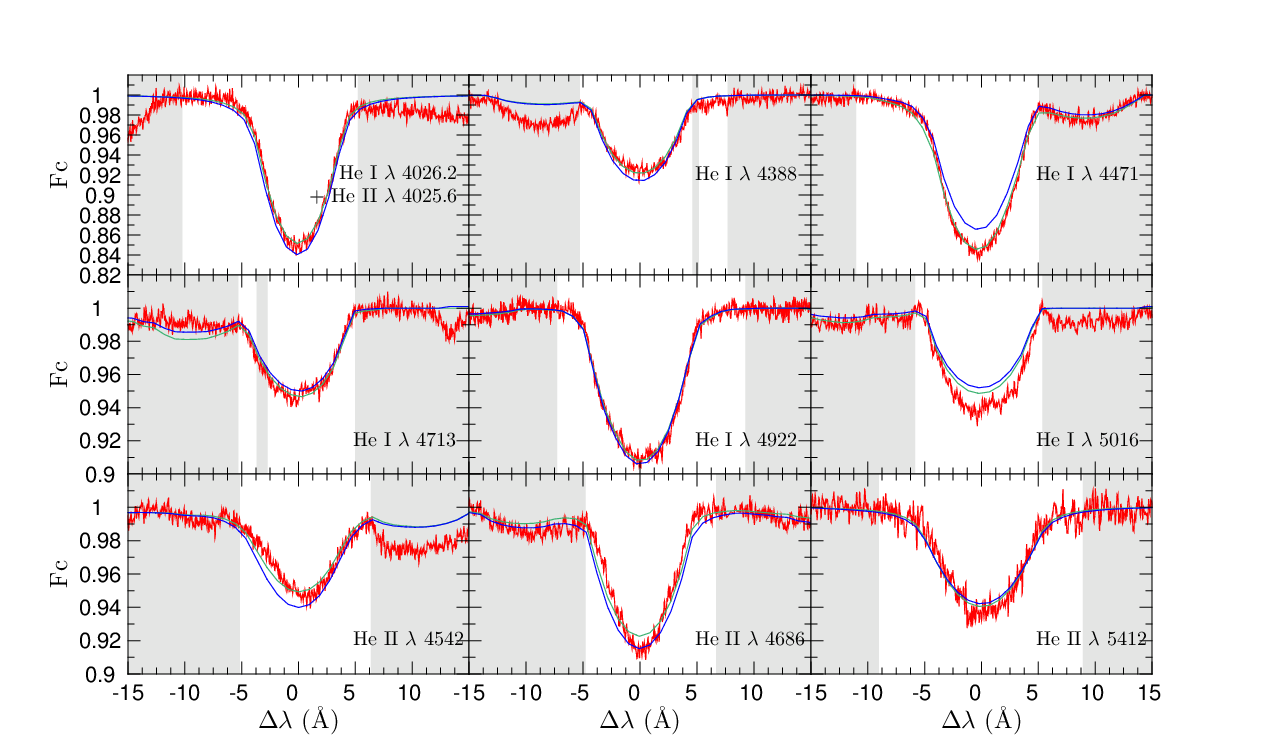}
\end{turn}
\caption{Comparison between the observed FEROS spectrum of \object{HD 90087} (red) and the best-fitting synthetic He line profiles (green). The line profiles computed for the final, mean parameters are shown in blue. The white areas delineate the regions where the fit quality was evaluated.}\label{figHeLinesParam}
\end{figure*}

\begin{figure*}
\centering
\begin{turn}{0}
\includegraphics[scale=0.35]{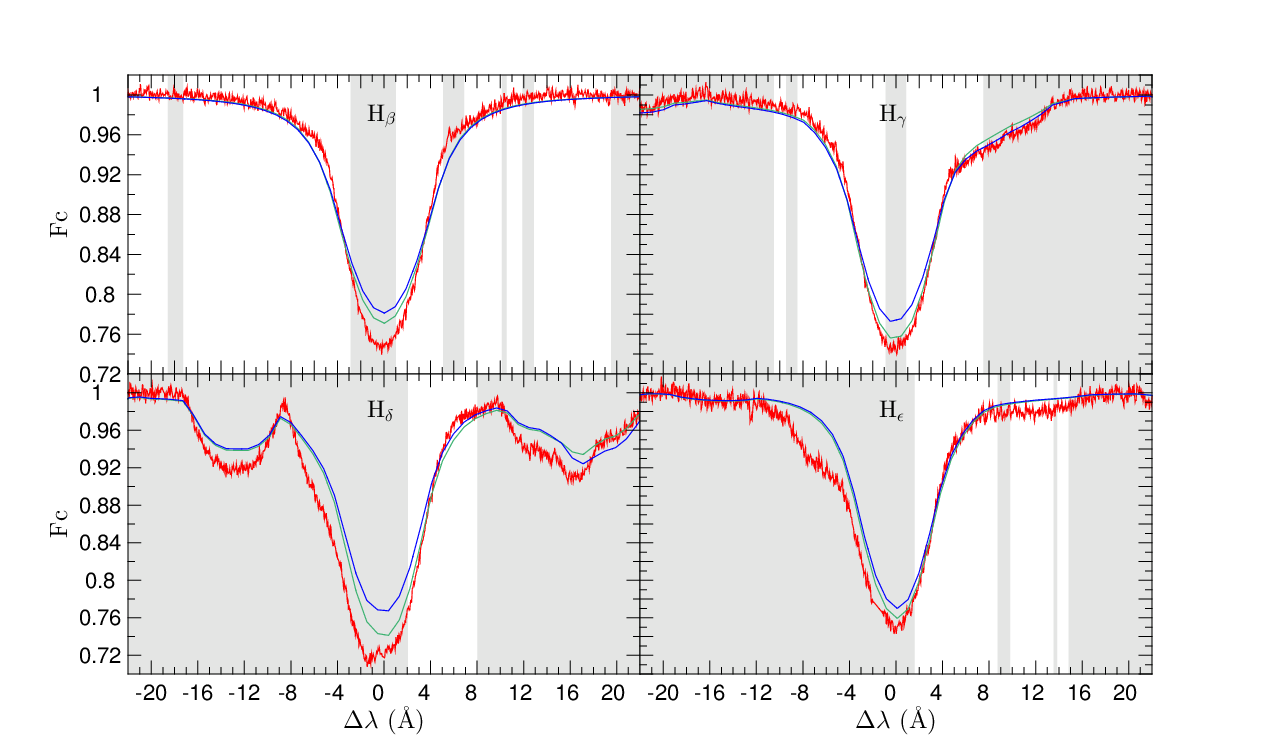}
\end{turn}
\caption{Same as Fig. \ref{figHeLinesParam}, but for the Balmer line profiles. {As in previous studies (e.g. \citealt{fir12}), achieving a good fit may be difficult, but selecting specific regions helps in this regard.}}
\label{figBalmerLinesParam}
\end{figure*}

For the initial step, we chose a value of $\log g$ (either 3.5 or 4.0) as a first guess. Both values were tried and, if results differed after convergence, those associated with the input $\log g$ yielding the smallest residuals were kept. A comparison between the observed and synthetic spectra for the aforementioned \ion{He}{I} lines provides values of $T_{\rm{eff}}$ and $y$ for each line. The helium abundances were then averaged by weighting the results according to the residuals. The $y$ value of the grid closest to this mean helium abundance was then fixed for the next step, the fit of the \ion{He}{II} lines, which was performed in a similar way. We calculated the mean temperatures for each ion separately and results from individual lines were weighted according to their residuals. We then averaged the two mean values, considering equal weights for the two ions, to derive a new $T_{\rm{eff}}$ value. The values of $T_{\rm{eff}}$ and $y$ in the grid, which are closest to the values just derived, were then fixed to determine $\log g$ by fitting the wings of the Balmer lines. If the value of $\log g$ was not equal to the input value, we performed additional iterations until convergence (see sketch on Fig. \ref{TMMethFlow}). Caution must be exercised when fitting spectral regions where orders of the {\'e}chelle spectra are connected, especially when this occurs over the broad Balmer lines. It should, however, be noted that no deterioration of the fit in these regions was apparent. {An illustration of the fits of He line profiles is given in Fig. \ref{figHeLinesParam}, demonstrating that the observed features are satisfactorily reproduced. Achieving a good fit for the Balmer lines using DETAIL/SURFACE is more challenging (Fig. \ref{figBalmerLinesParam}), as found in previous studies (e.g. \citealt{fir12} for Galactic BA supergiants), but remains possible when carefully selecting the regions that are deemed reliable.}

The next step is to determine the CNO abundances\footnote{{CNO abundances are given in the form $\log \epsilon$(X) = 12 +  $\log [\mathcal{N}\rm{(X)} / {\mathcal{N}\rm{(H)}}]$, where X $\equiv$ C, N, O.}}. To this end, we built a grid of CNO synthetic spectra for the ($T_{\rm{eff}}$, $\log g$) pair determined previously. We created these grids by varying $\log \epsilon$(C) in the range 7.24--8.94 dex, $\log \epsilon$(N) in the range 7.24--8.64 dex, and $\log \epsilon$(O) in the range 7.74--9.24 dex, with a step of 0.02 dex  in each case. We used synthetic spectra linearly interpolated to the exact $T_{\rm{eff}}$ values because the CNO abundances may be very sensitive to the temperature in certain $T_{\rm{eff}}$ regimes.

\begin{figure*}
\centering
\begin{turn}{0}
\includegraphics[scale=0.35]{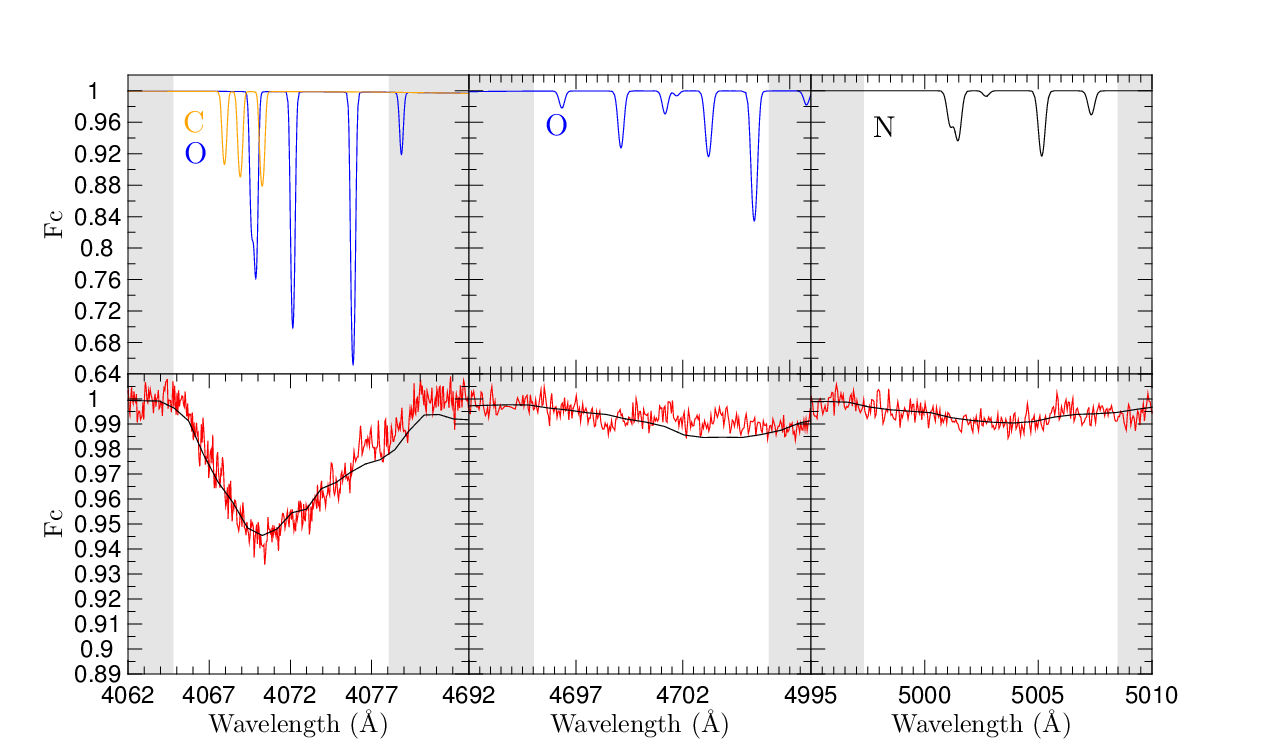}
\end{turn}
\caption{Comparison for \object{HD 90087} between the observed profiles (red; FEROS spectrum) and best-fitting synthetic metal line profiles (black). The white areas delineate the regions where the fit quality was evaluated. The top panels show the non-rotationally broadened synthetic profiles computed for the final parameters and abundances. 
}
\label{figCNO}
\end{figure*}

The choice of suitable CNO lines is complicated by the high rotation rates of our targets. We chose to consider some spectral domains that have been shown not to be significantly contaminated by lines of other species and to provide results that are consistent for a set of well-studied stars with those of more detailed and much more time consuming analyses (see \citealt{rau12} for a discussion). These regions are illustrated in Fig. \ref{figCNO}: the features in the first region (4060--4082 $\AA$) are mostly \ion{C}{III} and \ion{O}{II} lines, whereas \ion{O}{II} lines contribute predominantly to the second region (4691--4709 $\AA$) and \ion{N}{II} to the third (4995--5011 $\AA$). The associated CNO abundances were then found by minimising the residuals between the observed and synthetic spectra. 

\onecolumn
\begin{sidewaystable}
\begin{tiny}
\caption[]{Main characteristics of our targets, along with the source of data.}
\label{tabObj}
\centering
\begin{tabular}{ccccccccccccccccccccccccc}
\hline\hline
\multicolumn{2}{c}{\multirow{1}{*}{Name}} & \multirow{3}{*}{Spectral type} & \multirow{3}{*}{Ref.}& \multirow{2}{*}{m$_{V}$} & \multicolumn{14}{c}{Number of spectra} \\ \cline{1-2}\cline{6-20}
\multirow{2}{*}{ALS/BD/HD}  & \multirow{2}{*}{Other} & & & & \multirow{2}{*}{AU}&\multirow{2}{*}{C} & \multirow{2}{*}{EL} &  \multirow{2}{*}{ESPa} & \multirow{2}{*}{ESPR}&\multirow{2}{*}{FE} & \multirow{2}{*}{FI} & \multicolumn{2}{c}{G} & \multirow{2}{*}{HA} & \multirow{2}{*}{HE} & \multirow{2}{*}{M}  & \multirow{2}{*}{N} & \multirow{2}{*}{S} & \multirow{2}{*}{U} \\\cline{13-14}
  &                               &                                                      &             & [mag]                                  & & & &  & & & & Alb & B\&C  & & &  & &  &  \\
%
\hline\hline\multicolumn{20}{l}{\it Slow rotators}\\
\object{HD 214680} & 10\,Lac                      & O9V                         &1                    &4.88                                &               &                  &              &            &            &             &             &  &      & &                 1            &                  &                      &          &   \\
\object{HD 46328} & $\xi^1$\,CMa                & B0.5IV                         &2                 &4.33                        &               &                  &              &            &            &             &             &  &      & &                1              &                 &                    &            &   \\
\object{HD 57682} &                                     & O9.2IV                 &1                 &6.43                        &               &                  &              &            &            &             &           &    &         & &               5               &    &                       &          &\\
\object{HD 149438} & $\tau$\,Sco              & B0V                       &3               & 2.81                    &           &                &             &               &              &          &           &  &      & &                 2             &          &                      &          &\\
\hline\multicolumn{20}{l}{\it Fast rotators}\\
\object{ALS 864}&                               & O9V                           &4                   & 10.63                     &               &                &               &            &            &             &           &   &         & &                                   & 1       &                      &            & \\
\object{ALS 18675}&                                     & ...                           &1                 & 13.60                 &               &                &               &            &            &             &           &   &         & &                                   & 1       &                      &            &   \\
\object{BD +60$^{\circ}$594}&                   & O9V                   &3                   & 9.30              &               &                &      1       &            &            &               &           &   &         & &                                  &          &                      &1          &   \\
\object{BD +34$^{\circ}$1058}&                  & O8nn                          &2                   & 8.84              &               &                &              &            &            &               &           &   &         & &                                  &          &                      &1          &   \\
\object{HD 13268}&                              & ON8.5IIIn             &1                  & 8.18               & 62            &                & 2            &            &            &               & 3         &   &         & &                                  &          &                      &          &  \\
\object{HD 14434}&                              & O5.5Vnn((f))p         &1                  & 8.49               & 6             &                &              &  8        &            &                &           &   &         & &                                  &          &                      & 1        &  \\
\object{HD 14442}&                              & O5n(f)p               &1                  & 9.27               & 38            &                &              &            &            &               &           &   &         & &                                  &          &                      &          &   \\
\object{HD 15137}&                              & O9.5II-IIIn                   &1                  & 7.86               &               &                & 1            &            &            &               & 3         &   &         & &                                  &          &                      &          &   \\
\object{HD 15642}&                              & O9.5II-IIIn                   &1                  & 8.55               &               &                &              &            &            &               & 3         &   &         & &                                  &          &                      &          &   \\
\object{HD 28446A} & 1\,Cam\,A                  & B0IIIn                        &2                & 5.77                           &               &                &               &            &            &             &           &   &         & &                14                &           &                     &            &   \\
\object{HD 41161}&                              & O8Vn                          &1                  & 6.76               &               &                & 1            &            &            &               & 3         &   &         & &                   22           &          &                        &          &    \\
\object{HD 41997}&                              & O7.5Vn((f))           &5                 & 8.46                &               &                & 1            &            &            &               &           &   &         & &                                  &          &                      &          &   \\
\object{HD 46056}&                              &  O8Vn                         &1                    & 8.16               &               &                &               &  8        & 1         &               &           &   &         & &                                   &          &                     &            &   \\
\object{HD 46485}&                              &  O7V((f))nz           &1                  & 8.27               &               &                & 1            &            &            &       1       &           &   &         & &                                  &          &                      &          &   \\
\object{HD 52266} &                                     & O9.5IIIn                       &1                 &7.23                        &               &                &                &            &            &      6   & 2       &&1     & &                                &          &                     &         &   \\
\object{HD 52533} &                                     & O8.5IVn                       &1                  &7.68                        &               & 1      &              &            &            &      2        & 2       &&1      & &                               &  1      &                     &      &   \\
\object{HD 53755} & V569\,Mon                   & B0.5V                         &6                  &6.49                        &               &                &               &            &            &             &           &   &         & &                                &    &                       &         &1\\
\object{HD 66811}&  $\zeta$\,Pup                & O4I(n)fp              &1                  & 2.25               &               &                &              &  2        &            &                &           &   &         & &                                  &         &                        &         &  \\
\object{HD 69106}&                              & O9.7IIn               &1                  & 7.13               &               & 3              &              &  4        &            &                &           &&1      & &                                       &         &                        &         &  \\
\object{HD 74920}&                              & O7.5IVn((f))          &1                  & 7.54               &               & 3              &              &            &            &               &           &&1      & &                                       &         &                      &           &  \\
\object{HD 84567} &                                     & B0.5IIIn                       &7                 &6.45                        &               & 2        &              &            &            &             &           &  &      & &                               &                 &                      &           &   \\
\object{HD 90087} &                                     & O9.2III(n)                     &1                 &7.80                        &               &                &               &            &            &      2      &           &   &           & &                                &   &                      &            &  \\
\object{HD 92554} &                                     & O9.5III               &8                  & 9.50                       &               &2               &               &            &            &             &           &   &           & &                                &    &                     &            &  \\
\object{HD 93521} &                                     & O9.5IIInn              &1                 &7.03                        &               &                  & 20   &            &            &             &           &  &      & &                 2            &           &                    &  1  &   \\
\object{HD 102415}&                             & ON9IV:nn              &1                  &9.28                        &               & 1              &               &            &            &      4   &      &  &      & &                                &          &                     &          &  \\
\object{HD 117490}&                             & ON9.5IIInn                    &1                 &8.89                 &               &               &               &            &             &  8     &         &  &      & &                                &         &                      &         &   \\
\object{HD 124979}&                             & O7.5IV(n)((f))        &1                 & 8.51                &               &               &               &            &             &    2    &        &  &      & &                                &         &                      &         &   \\
\object{HD 149757} & $\zeta$\,Oph           & O9.2 IVnn             &1               & 2.56                     &                &                &       1      &              &              &1        &            &  &     &1 &                               &         &                       &          &\\
\object{HD 150574}&                             & ON9III(n)             &1                  & 8.50               &               &                &              &            &             &  2      &        &  &      & &                                &         &                       &        &   \\
\object{HD 163892}&                             & O9.5IV(n)             &1                  &7.44                        &               & 3              &               &            &             &    12     & 2          &   &         & &                                &     &                      &         &  \\
\object{HD 172367}&                             & B0V                           &3                  & 9.54               &               &                &              &            &             &              &           &   &         &  &                               &     &                      & 1       &   \\
\object{HD 175876}&                             & O6.5III(n)(f)         &1                 & 6.94                &               &1               &              &            &             &      11      &           &   &         &  &                               &     &                      &         &   \\
\object{HD 184915}&$\kappa$\,Aql                & B0.5III                       &3                 &4.96                 &               &                &              &            &             &          & 4     &   &         &  &            3                    &           &                      &   1   & \\
\object{HD 188439}&V819\,Cyg                    & B0.5IIIpvar           &3                  &6.28                        &               &                &               &            &             &            & 4         &   &         &  &            5                    &           &                      &  1   &  \\
\object{HD 191423}&                                     & ON9II-IIInn           &1                  &8.03                        & 7             &                & 2             &            &   2        &             & 2         &1&      &  &                              &           &                      &          &  \\
\object{HD 192281}&V2011\,Cyg           & O4.5V(n)((f))         &1                  & 7.55                       &37             &                & 4             &            &             &            &           &   &           &  &                                &           &                      &1     &   \\
\object{HD 198781}&                                     & B0.5V                          &2                  &6.45                       &               &                  &              &            &             &            &             &   &         &  &                     5          &           &                      &          &   \\
\object{HD 203064}& 68\,Cyg                     &  O7.5IIIn((f))                &1                    &5.00                                &               &                  &              & 1         &             &             & 3           &   &         &  &                                &           &    4                &           &   \\
\object{HD 210839}& $\lambda$\,Cep      &  O6.5I(n)fp           &1                  &5.05                                &               &                &               & 1         &             &             &           &   &           &  &                                &           &    6                & 1         &   \\
\object{HD 228841}&                             & O6.5Vn((f))           &1                    &9.01                                &               &                  &              &            &             &            &             &   &         &  &                                &           &                      & 1        &   \\
\hline
\end{tabular}   
\tablefoot{The slow rotators are used for validation purposes (see Sect. \ref{secMethVal}). Magnitudes in the $V$ band are from the SIMBAD database. References for spectral types: [1] \citet{sot11,sot14}; [2] \citet{les68}; [3] \citet{mor55}; [4] \citet{fit75}; [5] \citet{bla61}; [6] \citet{mur69}; [7] \citet{hil69}; [8] \citet{hum73}. For data sources, AU corresponds to AURELIE, C to CORALIE, EL to ELODIE, ESPa to ESPaDOnS, ESPR to ESPRESSO, FE to FEROS, FI to FIES, G to GOSSS (Alb to Albireo and B\&C to Boller \& Chivens), HA to HARPS, HE to HEROS, M to MIKE, N to NARVAL, S to SOPHIE, and U to UCLES.}
\end{tiny}
\end{sidewaystable}

\twocolumn 

\begin{table}
{\begin{center}
\caption[]{Comparison between our projected rotational velocities and those in the literature based on the FWHM of \ion{He}{I} lines.}
\label{tabCompVsiniRio}
\centering
\begin{tabular}{cccc}
\hline\hline
\multirow{2}{*}{Star}              & \multicolumn{2}{c}{$v$\,sin\,$i$ [km s$^{-1}$]} & \multirow{2}{*}{Reference}\\ \cline{2-3}
                                           & This work    &    Literature                                     &                     \\
\hline\hline
\object{ALS 491}                                  & 228$\pm$15          &  223$\pm$56                                         & 1 \\      
\object{ALS 535}                                  & 200$\pm$15          &  179$\pm$14                                         & 1 \\                      
\object{ALS 851}                                  & 167$\pm$15          &  165$\pm$31                                         & 1 \\      
\object{ALS 897}                                  & 180$\pm$15          &  175$\pm$10                                         & 1 \\      
\object{ALS 864}                                  & 249$\pm$15          &  232$\pm$22                                         & 1 \\      
\object{ALS 18675}                                & 236$\pm$15          & 212$\pm$11                                         & 1 \\       
\object{HD 42259}                                 & 256$\pm$15          &  249$\pm$25                                                 & 2 \\      \object{HD 52533}                                 & 305$\pm$15          &  291$\pm$29                                                 & 2 \\      
\hline
\end{tabular}
\tablefoot{References: [1] \citet{gar15}; [2] \citet{bra12}; [3] \citet{daf07}.}
\end{center}}
\end{table}

The final abundance of oxygen is the unweighted mean of the values found for the first and second regions. Since the \ion{C}{III}\,lines allowing us to probe the carbon abundance are weak for the coolest stars in our sample and become a minor contributor to the blend with the nearby \ion{O}{II}\,lines, carbon abundances cannot be reliably determined for the B0.5 stars.

\subsubsection{Method for the hotter stars}
\label{subSecMethCMFGEN}
For the hotter stars that possess strong winds, we used the non-LTE spherical atmosphere code CMFGEN to derive stellar parameters. Full details about this code (e.g. atomic data) can be found in \citet{hil98}\footnote{See also \url{http://kookaburra.phyast.pitt.edu/hillier/web/CMFGEN.htm} for upgrades since the original publication.}.

As a starting point, CMFGEN makes use of a hydrodynamical structure, characterising the velocity and density profiles, which is created from TLUSTY models \citep{lan03}. The wind is described by a mass-loss rate, $\dot{M}$, a $\beta$-like velocity law, $v = v_{\infty}\,\left(1 - {R_{*}}/{r}\right)^\beta$, where $R_{*}$ is the stellar radius, and $r$ the distance from the stellar centre, $\beta$ a parameter with typical values for massive stars close to 0.8--1, and $v_{\infty}$ the terminal velocity. {We adopted a volume filling factor at terminal velocity of 0.1 and a clumping velocity factor of 100 km s$^{-1}$; for} the clumping formalism implemented in CMFGEN, see e.g. \citet{rau16}. The following elements are included in the calculations of our models: H, He, C, N, O, Ne, Mg, Al, Si, P, S, Ca, Fe, and Ni. Computing time was reduced thanks to the use of the super-level approach, but remains much longer than for DETAIL/SURFACE.

A synthetic spectrum was created after finding the formal solution of the radiative transfer equation. A microturbulent velocity varying linearly from the photosphere to $0.1\,v_{\infty}$ at the top of the atmosphere was considered. The value at the photosphere depends on the luminosity class: 10 km s$^{-1}$ for dwarfs, 12 km s$^{-1}$ for (sub)giants, and 15 km s$^{-1}$ for supergiants \citep{bou12}. A typical X-ray flux corresponding to $L_{\rm X}$/$L_{\rm BOL}$ $\sim$ 10$^{-7}$ is considered in our models, as X-rays have an impact on the ionisation balance. After transforming vacuum wavelengths into air wavelengths, the spectrum was then broadened in order to take the appropriate instrumental resolution and object's projected rotational velocity into account. 

Given the large number of free input parameters entering the CMFGEN code and the fact that the computing time necessary to create a new model is in general very lengthy, computation of a complete grid of models is virtually impossible. We therefore adopted a procedure slightly different from that described in the previous subsection. A first guess of stellar parameters, wind parameters, and surface abundances for each star was adopted \citep[either from the literature, if available, or from typical values for the considered spectral type given by][]{mui12}. Wind parameters are not investigated in this study, hence they were {not fitted} since our main concern was to unveil surface abundances {(an approach previously used by \citealt{mar15b})}. In particular, $v_{\infty}$ is fixed, when possible, to values provided by \citet{pri90}. {We nevertheless checked that the fits of wind-sensitive lines were reasonable, and the wind parameters were slightly modified for stars with strong outflows (e.g.\ \object{HD 66811}) when these fits were not deemed satisfactory.} We calculated a small grid of CMFGEN spectra with five temperature ($\Delta\,T_{\rm eff}$ = 500\,K) and five gravity values ($\Delta\log g$ = 0.125\,dex) around the initial guesses. We then computed the residuals for each point of the grid between the observed spectrum and the synthetic spectra. This was performed over the same regions, encompassing the Balmer and He lines, as those used for the cooler objects (Sect. \ref{subSecMethTM}). A surface corresponding to a piecewise cubic interpolation was fitted to the $\chi^2$ results of this analysis. The best-fit values of $T_{\rm eff}$ and $\log g$ are at the minimum of this surface fit. The good agreement for the hotter stars between CMFGEN spectra and observations is illustrated in Appendix \ref{secCompaCMFGENFig}.

\begin{figure*}
\centering
\begin{turn}{0}
\includegraphics[scale=0.35]{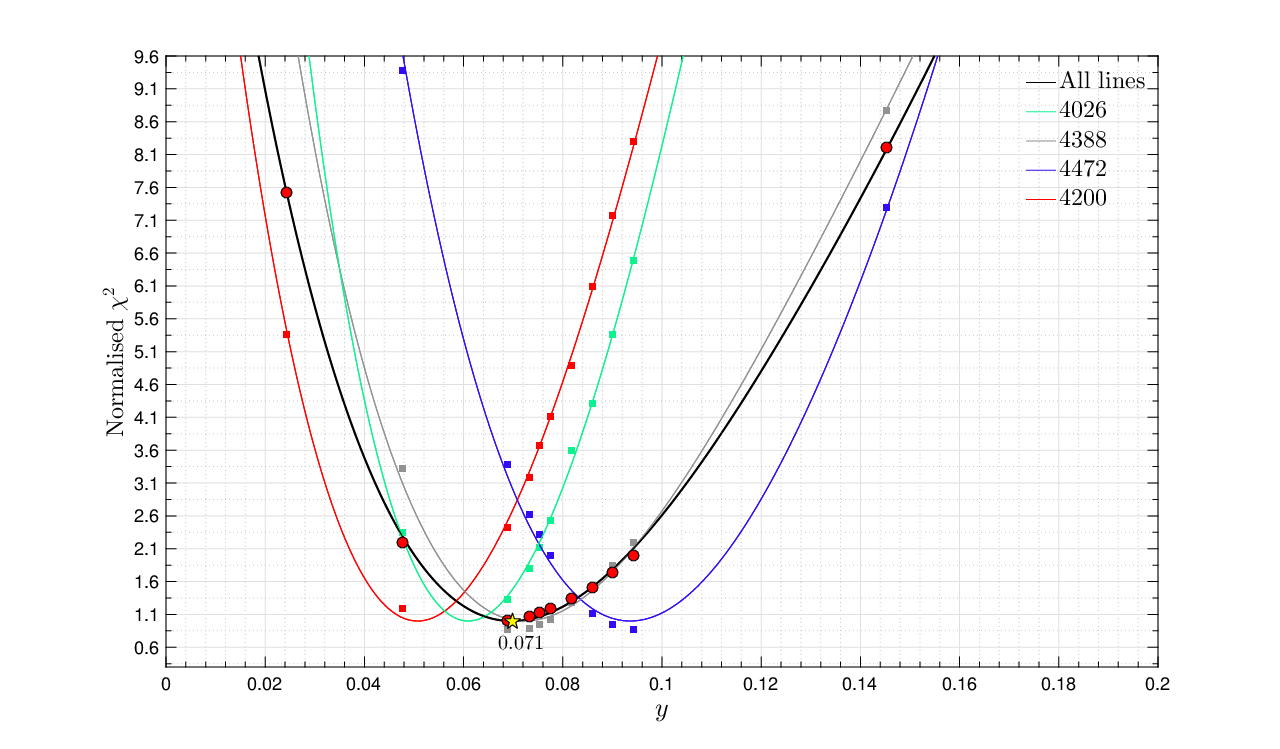}
\end{turn}
\caption{Illustration of the helium abundance determination for \object{HD 163892} with CMFGEN (see Sect. \ref{subsecCMF}). Only a few lines are shown for clarity. Results of the $\chi^{2}$ analysis are shown as solid red circles. The black curve is the global, polynomial fit for all lines. The solid yellow star indicates the abundance providing the best fit (Table \ref{tabMethValCMFGEN}).}
\label{detHe}
\end{figure*}

Next, we determined the helium abundance by performing a $\chi^{2}$ analysis similar to that of \citet{mar15b}, considering the same helium lines as in Sect. \ref{subSecMethTM}, with the addition of \ion{He}{II}\,4200. This time points in the grid were separated by $\Delta y \sim$ 0.025. A polynomial fit (of degree smaller than or equal to 4) of individual features first allowed us to identify discrepant lines (the fit quality was an additional criterion). Then, a global fit of the remaining lines enabled us to find the best value of $y$ (see illustration in Fig. \ref{detHe}).

The carbon, nitrogen, and oxygen abundances were derived following the same approach, with the grids usually having $\Delta [ \mathcal{N}\rm{(X)} / {\mathcal{N}\rm{(H)}} ]$ = 2\,$\times$\,10$^{-4}$, where X $\equiv$ C, N, O. The initial line list used to derive the CNO abundances is taken from \citet{mar12a,mar15a,mar15b}, {and shown in plots of Appendix \ref{secCompaCMFGENFig},} while the lines actually used for each star are listed in {Table \ref{tabLinesCNO}}.

\section{Uncertainties of the results}
\label{sectUnce}

\subsection{RVs}
\label{sectUnceRV}
High-resolution spectrometers usually yield low errors on RVs. For example, RV dispersions below 1 km s$^{-1}$ are commonly found for narrow interstellar features \citep{bat92}. In our case, however, the lines are very broad, generally leading to larger errors. Indeed, RVSAO calculates errors on RVs, which are of the order of 1--20 km s$^{-1}$ (and typically 7 km s$^{-1}$) for our sample stars, depending on noise level, spectral type, $v\sin\,i$, and spectral resolving power. The uncertainty arising from the wavelength calibration ($\sim 1$ km s$^{-1}${, as determined from narrow interstellar lines}) is generally negligible in comparison.

\begin{figure}[!h]
\centering
\begin{turn}{0}
\includegraphics[scale=0.5]{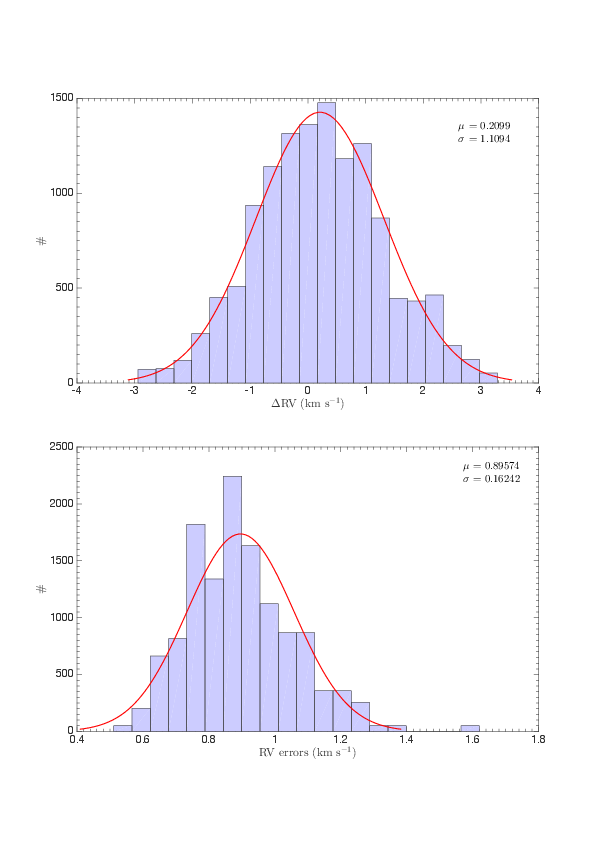}
\end{turn}
\caption{Example of Monte Carlo simulations for a B0.5\,V star with $v\sin\,i$ = 400 km s$^{-1}$ observed with $R$ = 50\,000. A total of 12\,750 trials were made. {\it Upper panel:} Deviations of the derived velocities with respect to the input values. {\it Lower panel:} Breakdown of the derived errors provided by RVSAO. The Gaussian that best represents each distribution is overplotted in red.}
\label{FigErrorRV}
\end{figure}

To check the RVSAO error values, we performed Monte Carlo simulations. Synthetic TLUSTY spectra of a typical B0.5\,V and O5\,V star were convolved with two rotational profiles ($v\sin\,i$ = 200 and 400 km s$^{-1}$, corresponding to the extreme values of our sample), blurred by noise (S/N $\sim$ 125, typical of our data), and shifted with different radial velocities (from --250 to 250 km s$^{-1}$ with a step of 10 km s$^{-1}$). Their RV was estimated as for real spectra and the dispersion of the difference between applied shifts and derived velocities examined. We found that the distributions of the velocity differences can be reasonably represented by Gaussians whose standard deviations agree well with the errors provided by RVSAO (e.g. $\sim$ 1 km s$^{-1}$ found in both cases for a B0.5\,V star with $v\sin\,i$ = 400 km s$^{-1}$ and observed with $R$ = 50\,000, see Fig. \ref{FigErrorRV}).

\subsection{$v\sin\,i$ }
The errors on the projected rotational velocities can be empirically estimated by comparing results obtained for a star observed with various instruments and analysed using different diagnostic lines. Taking \object{HD 149757} ($\zeta$\,Oph) as a prototypical example, we found dispersions of $v\sin\,i$ values of $\sim$ 8 km s$^{-1}$ when considering different lines (\ion{He}{I} 4026, 4471, 4922), but the same instrument. Alternatively, this translates to $\sim$ 12 km s$^{-1}$ for the same lines, but different instruments (ELODIE, FEROS, and HARPS). The overall dispersion considering all values amounts to 13 km s$^{-1}$. We therefore consider {a representative error} of $\sim$ 15 km s$^{-1}$ for our sample stars. 

\subsection{Atmospheric parameters and abundances} 
\subsubsection{Cooler stars}
\label{sectUncCoo}

To estimate the precision of our parameters ($T_{\rm{eff}}$, $\log g$) and abundances, we first examined the dispersion of the results obtained for different spectra (ELODIE, FEROS, and HARPS) of the same star (\object{HD 149757}). The differences are expected to mainly reflect the uncertainties related to the nature of the data and their treatment, especially errors in the normalisation to the continuum. In the case of \object{HD 149757}, our procedure also accounts for line-profile variations arising from non-radial pulsations (e.g. \citealt{kam97}). As a second step, we explored the impact of the choice of the microturbulence by repeating the analysis of \object{HD 149757} after adopting $\xi$ = 5 rather than 10 km s$^{-1}$. 

To accommodate both sources of errors, we quadratically summed the derived dispersions to get the final uncertainties that are quoted in Table \ref{sigmaUncert}. These uncertainties typically amount to 1000 K for $T_{\rm{eff}}$, 0.10 dex for $\log g$, 0.025 for $y$, and 0.12, 0.13, and 0.21 dex for the abundances of carbon, nitrogen, and oxygen, respectively. The errors on the nitrogen-to-carbon and nitrogen-to-oxygen abundance ratios ([N/C] and [N/O]{, defined as $\log [\mathcal{N}\rm{(N)} / {\mathcal{N}\rm{(C)}}]$ and $\log [\mathcal{N}\rm{(N)} / {\mathcal{N}\rm{(O)}}]$, respectively}) were then estimated to be 0.21 and 0.12 dex, respectively.

\subsubsection{Hotter stars}

Typical errors on $T_{\rm{eff}}$ and $\log g$ were assumed to be 1500 K and 0.15 dex, respectively, as generally adopted in CMFGEN analyses in the literature (\citealt{mar15b,rau16}). These values are higher than those considered for DETAIL/SURFACE since stronger winds have an impact on the resultant spectrum, making the analysis more challenging. The chosen errors are also comparable to the differences found when comparing our values of $T_{\rm{eff}}$ and $\log g$ with those in the literature (Appendix \ref{resB}), which supports our choice. \object{HD 41161}, which is representative of the sample of hotter stars, was chosen to determine the typical errors on He and CNO abundances. These errors were derived from the unnormalised $\chi^2$ function, considering values corresponding to $\Delta\,\chi^2=1$ above its minimum. This approach is different from that of \citet{mar15a} who first normalised the $\chi^2$ function such that the minimum is equal to one before considering $\Delta\,\chi^2=1$ (a procedure less valid than ours, statistically speaking). We caution that the errors on He and CNO abundances do not take the uncertainties on atmospheric parameters into account so that they are likely underestimated.

\section{Method validation}
\label{secMethVal}

\subsection{Comparison of atmospheric parameters and abundances with literature}
Half of our targets had been previously investigated in some detail (though usually CNO abundances are missing; Table \ref{tabResComp}). We note a good agreement overall, considering error bars. In particular, we underline the study of \citet{mar15a,mar15b}, which has 11 objects in common with our analysis. On average, differences in stellar parameters (ours minus Martins et al.) amount to $\Delta$$T_{\rm{eff}}$ = +282$\pm$627 K, $\Delta$$\log g$ = +0.02$\pm$0.12 dex, $\Delta$$y$ = --0.010$\pm$0.044, $\Delta$$\log \epsilon$(C) = 0.00$\pm$0.19 dex, $\Delta$$\log \epsilon$(N) = --0.10$\pm$0.13 dex, and $\Delta$$\log \epsilon$(O) = --0.12$\pm$0.18 dex, which are well within error bars. The largest differences are within, or close to, 2\,$\sigma$: \object{HD 46485} ($\Delta$$\log g$ = 0.25 dex), \object{HD 191423} ($\Delta$$y$ = --0.066), and \object{HD 13268} ($\Delta$$\log \epsilon$(O) = --0.39 dex). 

{Some differences are nevertheless worth mentioning. Our lower limit for the oxygen abundance in \object{HD 150574} is larger than the value derived by \citet{mar15b}. For \object{HD 191423}, we derive an upper limit for the carbon abundance that is lower than the value derived by \citet{vil02} and a nitrogen abundance that is lower than the lower limit reported by \citet{mar15b}. However, the differences for \object{HD 191423} are below 2\,$\sigma$, hence barely significant.  In addition,  this star has an extreme rotational velocity ($v\sin\,i$ = 420 km s$^{-1}$), which renders its analysis very difficult.}

\subsection{Comparison of DETAIL/SURFACE results with those previously obtained for well-studied slow rotators}
In order to validate the procedures used for the analysis of the cooler stars of our sample, the following four narrow-lined, well-studied objects were analysed \citep[see ][]{rau12,mor08}: $\xi^1$\,CMa (B0.5\,IV; $v$\,sin\,$i$ $\sim$ 10 km s$^{-1}$), $\tau$\,Sco (B0\,V; $v$\,sin\,$i$ $\sim$ 8 km s$^{-1}$), \object{HD 57682} (O9.2\,IV; $v$\,sin\,$i$ $\sim$ 25 km s$^{-1}$), and 10\,Lac (O9\,V; $v$\,sin\,$i$ $\sim$ 25 km s$^{-1}$). For $\xi^1$\,CMa, which is a well-known $\beta$ Cephei pulsator with slight variations of the physical parameters along the pulsation cycle \citep{mor06}, the HEROS exposure corresponding to the highest {effective} temperature was chosen.

A high rotation rate may bias our results because of, for example, blending issues or a more uncertain continuum placement. To assess the importance of these effects, we repeated the analysis after convolving the spectra with a rotational broadening function corresponding to 300 km s$^{-1}$, which is a value representative of our sample. 

\begin{figure*}
\centering
\begin{turn}{0}
\includegraphics[scale=0.43]{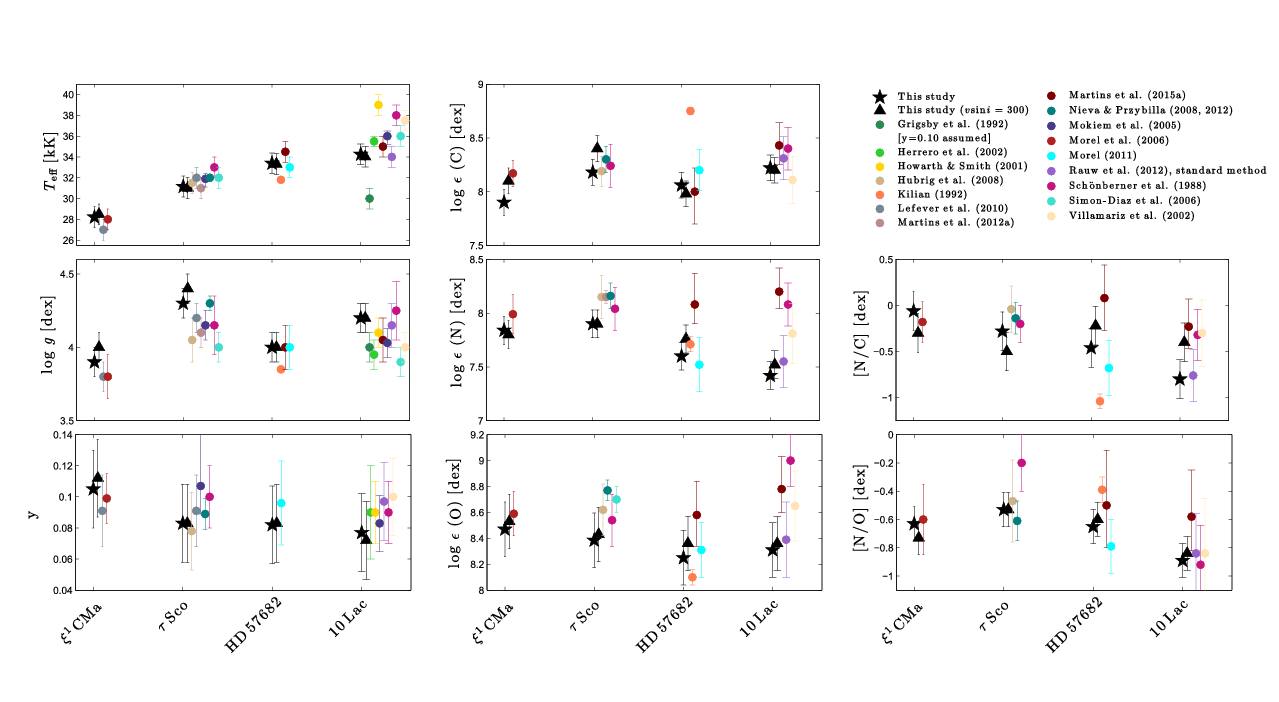}
\end{turn}
\caption{Comparison between our results for the slow rotators and those in the literature. For the $\beta$ Cephei $\xi^{1}$\,CMa, the results of \citet{lef10} are the values averaged along the pulsation cycle, while those of \citet{mor06} correspond to the highest temperature.}
\label{figMethVal}
\end{figure*}

Table \ref{tabMethVal} {presents} our results and {Fig. \ref{figMethVal} compares} them to literature values. Some study-to-study scatter exists, but there is an overall good agreement between our values and those in the literature. In particular, there is no evidence for systematic differences compared to previous results despite the different techniques employed; in fact, 10\,Lac displays a large dispersion in the literature values of $T_{\rm{eff}}$, hence provides a less significant comparison point. Furthermore, our results appear largely insensitive to the amount of rotational broadening, thereby validating our method.

\subsection{CMFGEN versus DETAIL/SURFACE}
\label{subsecCMF}
Previous studies have revealed a good agreement for main-sequence, early B-type stars between the parameters/abundances determined with DETAIL/SURFACE and the unified code FASTWIND (\citealt{lef10}; \citealt{nie11}). However, a full comparison of the results provided by DETAIL/SURFACE and CMFGEN was seldom performed. To our knowledge, only two stars have been analysed with both codes: $\tau$\,Sco (studied with DETAIL/SURFACE by \citealt{hub08} as well as by \citealt{nie12}, and with CMFGEN by \citealt{mar12a}) and \object{HD 57682} (studied with DETAIL/SURFACE by \citealt{mor11} and with CMFGEN by \citealt{mar15a}). In these cases, the results appear to agree within the errors. The only exception is the nitrogen abundance in \object{HD 57682}, but the origin of this discrepancy is unclear.

Since we made use of these two different line-formation codes for the analysis, our results for the subsamples of cool and hot objects could be affected by systematic errors. To be able to fully assess the magnitude of such differences, if any, it is necessary to study at least a few objects with both codes. To this end, three objects have been chosen: \object{HD 102415}, \object{HD 149757}, and \object{HD 163892}. The three stars were selected because they exhibit different degrees of nitrogen enrichment, spanning the range observed in our sample.

Table \ref{tabMethValCMFGEN} presents our results. The effective temperatures are in good agreement, within the error bars: the largest difference is $\Delta\,T_{\rm{eff}}=500$ K for \object{HD 149757}, which is still below the typical error bars of 1--1.5\,kK. The differences in gravities are also generally small (< 0.1 dex), although the largest difference (for \object{HD 102415}) reaches 0.24 dex, which is slightly larger than the errors (estimated to be 0.10--0.15 dex). The helium abundances agree well with the largest difference, $\Delta\,y=0.034$, found for \object{HD 102415}, being similar to the error bars. The CNO abundances yielded by the two codes also agree within the error bars. Therefore, we can conclude that there is no evidence for significant differences when analysing our targets with DETAIL/SURFACE or CMFGEN, ensuring that our overall results are to first order homogeneous.

\onecolumn
\begin{sidewaystable}
{\scriptsize
\caption[]{Stellar parameters derived for the stars in our sample {and assumed wind parameters for our hotter stars.}}
\label{tabResults}
\centering
\begin{tabular}{ccccccccccccccccccccccc}
\hline\hline 
\multirow{2}{*}{Star} &$v\sin\,i$              &$v_{\rm{mac}}$&\multirow{1}{*}{Multiplicity}&\multirow{2}{*}{Runaway?}&$T_{\rm{eff}}$&\multirow{2}{*}{$\log g$}&\multirow{2}{*}{$\log g_{\rm{C}}$}
&\multirow{2}{*}{{$\log(\dot{M})$}}&{$v_{\infty}$}&\multirow{2}{*}{{$\beta$}}
&\multirow{2}{*}{$y$}&\multirow{2}{*}{$\log \epsilon$(C) }&\multirow{2}{*}{$\log \epsilon$(N) }  &\multicolumn{3}{c}{$\log \epsilon$(O) }  & \multirow{2}{*}{[N/C]} &\multirow{2}{*}{[N/O]} \\ \cline{15-17} 
                     &[km s$^{-1}$]                 &[km s$^{-1}$]                  & from spectroscopy& &               [K]             &  &      
                     & & {[km s$^{-1}$]} &
                     & &                                & &4060--4082 $\AA$ & 4691--4709 $\AA$ & Adopted      & &\\       
\hline \multicolumn{19}{l}{\it Cooler stars (DETAIL/SURFACE)}\\
\object{ALS 864}                & 249                           &       <98             &...                 &No                             & 31\,500  & 4.00 & 4.05  &{...}&{...}&{...}   & 0.064  &$<$7.86 & 7.64          & 8.26  & 7.98 &8.12                    & $>$--0.22       &  --0.48       \\
\object{ALS 18675}            & 236                             &       <200            & ...             &No                             & 30\,300  & 3.90 & 3.95  &{...}&{...}&{...}  & 0.071    & 7.78        & 7.54    & 8.02  & 8.12 &8.07                 & --0.24                & --0.53        \\
\object{BD +60$^{\circ}$594}& 314                               &       <197            & RV var   &No                            & 33\,200  & 4.00 & 4.04  &{...}&{...}&{...}  & 0.131    & 7.66      & $<$8.24& 8.60  & 8.36 &8.48           & $<$0.58&   $<$--0.24     \\
\object{HD 28446A}          & 275                               &       <169         & { RV var}&No                  & 29\,400  & 3.70 & 3.78  &{...}&{...}&{...}  & 0.126        & 8.30        & 7.48     & 8.76  &      8.28&8.52               & --0.82          &    --1.04      \\
\object{HD 52266}              & 285                            &       <96           & SB1         &No                             & 30\,100  & 3.60 & 3.70  &{...}&{...}&{...}   & 0.187   & 7.78        & 7.74     & 8.08    & 7.92 &8.00           & --0.04                &   --0.26       \\
\object{HD 52533}              & 305                            &       <180         & SB1   &...                            & 34\,100  & 4.10 & 4.16  &{...}&{...}&{...}  & 0.065   & 7.76        & $<$7.78     & 8.08    & 8.22 &8.15   & $<$0.02 &$<$--0.37  \\
\object{HD 53755}$^a$    & 285                          &       <102            & ...             &No                             & 28\,100  & 3.60 & 3.70        &{...}&{...}&{...} & 0.135         & ...           & 7.32    & ...    & 8.38 &8.38                         & ...                     &  --1.06        \\
\object{HD 84567}              & 261                            &       <105         & RV var        &...                            & 27\,700        & 3.50& 3.59   &{...}&{...}&{...} & 0.204    & $<$7.84 & 7.98   & 8.62    & 8.16 &8.39           &$>$0.14        &--0.41      \\
\object{HD 90087}              & 276                            &       <63           & Pres. single &No                    & 29\,000  & 3.50 & 3.60  &{...}&{...}&{...} & 0.163    & 7.72        & 7.42     & 8.18    & 8.14 &8.16           & --0.30                & --0.74     \\
\object{HD 93521}              & 405                            &       <88           & Pres. single &...                   & 30\,000  & 3.60 & 3.78  &{...}&{...}&{...} & 0.166    & 7.68       & 8.10     & 8.50     & 8.16  &8.33          &  0.42                 &   --0.23       \\
\object{HD 102415}{ $\dag$} & 357                               &       <72          & { Pres. single}      &No             & 32\,900  & 4.10 & 4.19  &{...}&{...}&{...} & 0.158     & $<$7.54& 8.16     & 8.46    & 7.98 &8.22                 &  $>$0.62      &   --0.06       \\
\object{HD 149757}{ $\dag$}  & 378                      &       <104            & Pres. single&Yes                        & 31\,500       & 3.87 & 3.99  &{...}&{...}&{...} & 0.135         & 8.07        & 7.85    & 8.47    & 8.27 &8.37          & --0.22          & --0.52         \\
\object{HD 163892}{ $\dag$}  & 205                       &      <79           & SB1      &No                             & 32\,000  & 3.80 & 3.84  &{...}&{...}&{...} & 0.082   & 8.24        & 7.34    & 8.44      & 8.32 &8.38           & --0.90                &  --1.04    \\
\object{HD 172367}            & 266                             &       <107         & ...           &No                                     & 27\,600         & 3.60 & 3.69  &{...}&{...}&{...} & 0.140   & $<$8.09& 8.44    & 8.48      & 8.46 &8.47                  &  $>$0.35      & --0.03     \\
\object{HD 184915}            & 252                                     &         <62           & Pres. single &No                        & 27\,800 & 3.70 & 3.77  &{...}&{...}&{...}  & 0.183   &$<$8.18& 8.46    & $>$8.86 & 8.62  &8.62           &$>$0.28        &  --0.16\\
\object{HD 188439}            & 281                                     &         <197         & RV var   &...                            & 27\,700 & 3.70 & 3.79  &{...}&{...}&{...} & 0.122   & $<$8.09& 8.16   & $>$8.86  & 8.66  &8.66   &  $>$0.07      &  --0.50\\
\object{HD 198781}            & 222                                     &         <103         & Pres. single &No                 & 29\,100 & 3.90 & 3.94  &{...}&{...}&{...}      & 0.230  & $<$8.09  & 8.62   & 8.98    & 8.58  &8.78 &  $>$0.53  & --0.16    \\
{\it Typical errors}                     & {\it 15}                             &        ...             & ...           &...                            & {\it 1000}  &{\it 0.10}&{\it 0.10}&{...}&{...}&{...}&{\it 0.025}&{\it 0.12}&{\it 0.13}&...&...&{\it 0.21}&{\it 0.21} &{\it 0.12}\\
\hline \multicolumn{19}{l}{\it Hotter stars (CMFGEN)}\\
\object{BD +34$^{\circ}$1058}& 424                            &         <57             & ...             &No                             & 34\,400       & 3.67 & 3.85    &{--7.0}&{2035}&{1.03}& 0.119    & 7.90 & 8.14&... &...&8.68    &         0.24                    & --0.54 \\
\object{HD 13268}             & 301                           &         <86                 & RV var        &Yes                            & 32\,500       & 3.42 & 3.55   &{--6.6}&{2266}&{0.80}& 0.206     &$\le$7.50 &8.61&... &...&      8.10         &   $\ge$1.11   & 0.51 \\
\object{HD 14434}             & 408                           &         <44             & Pres. single &No                        & 40\,000       & 3.89 & 4.03   &{--6.2}&{1960}&{0.84}& 0.103 &7.96 &8.81       &... &...&$\le$8.10 &      0.85               &$\ge$0.71  \\
\object{HD 14442}$^b$     & 285                       &         <10             & Pres. single &No                        & 39\,200       & 3.69 & 3.78   &{--6.2}&{1960}&{0.84}& 0.097 &7.10 &8.61       &... &...&$\le$8.10 &    1.51        &$\ge$0.51  \\
\object{HD 15137}             & 267                           &         <76             & RV var  &Yes                            & 29\,500 & 3.18 & 3.31    &{--6.4}&{2575}&{0.89}& 0.112&7.63& 8.27        &... &...&          $\le$8.30 &   0.64   &$\ge$--0.03  \\
\object{HD 15642}             & 335                           &         <128         & Pres. single&No                       & 29\,700 & 3.28 & 3.47          &{--6.4}&{2575}&{0.89}& 0.150& $\le$7.55 &8.43 &... &...&       7.93         &       $\ge$0.88 &0.50  \\
\object{HD 41161}              & 303                          &         <81             & RV var  &Yes                            & 33\,900       & 3.67 & 3.77   &{--7.0}&{2035}&{1.03}  & 0.123    & 7.87  &8.09 &... &...&    8.67    &       0.22    & --0.58\\
\object{HD 41997}              & 247                          &         <118            & ...     &Yes                            & 34\,400       & 3.92 & 3.97   &{--6.8}&{2700}&{1.03}   & 0.110    & 8.59 & 8.21    &... &...&        8.79    &       --0.38  &--0.58\\
\object{HD 46056}              & 350                          &         <48             & Pres. single&No                 & 34\,500       & 3.90 & 4.00   &{--7.0}&{1305}&{1.03}  & 0.088    & 8.34 &7.78     &... &...&         8.32    &       --0.56           &--0.54 \\
\object{HD 46485}              & 315                          &         <149            & Pres. single&No                 & 37\,000       & 4.00 & 4.08   &{--6.6}&{1780}&{1.03}  & 0.076        & 8.46 & 7.95   &... &...&      8.72    &       --0.51  &--0.77\\
\object{HD 66811}              & 225                          &         <124         & Pres. single &Yes                     & 41\,000       & 3.55 & 3.62    &{--6.0}&{2485}&{0.92}  & 0.148 &$\le$7.00&8.94 &... &...&      8.20         &    $\ge$1.94   & { 0.74}\\
\object{HD 69106}              & 306                          &          <177    & Pres. single &No                      & 29\,500       & 3.45 & 3.58   &{--6.1}&{2700}&{1.03}  & 0.091    & 7.88 & 7.74 &... &...&8.47  &       --0.14         &--0.73\\
\object{HD 74920}              & 274                          &          <111    & Pres. single&No                       & 34\,100       & 3.85 & 3.92   &{--6.6}&{4410}&{1.20}   & 0.134    &7.78  &8.34         &... &...&      8.51         &       0.56                    &--0.17 \\
\object{HD 92554}              & 303                          &          <99             & { ...}  &Yes                          & 30\,000 & 3.41 & 3.54  &{--6.7}&{1260}&{0.99}  & 0.091 &7.57 &7.30     &... &...&      8.53         &       --0.27  & --1.23 \\
\object{HD 117490}            & 361                           &          <94             & Pres. single&No                       & 30\,000 & 3.55 & 3.70  &{--6.1}&{2970}&{0.95}  & 0.141 &$\le$7.39 &8.50&... &...&      8.15         &       $\ge$1.11       & 0.35  \\
\object{HD 124979}            & 246                           &          <215    & RV var        &Yes                            & 34\,100 & 3.85 & 3.91    &{--6.6}&{4410}&{1.20}        & 0.091         & 8.48 &7.92    &... &...&8.74       &       --0.56  & --0.82\\
\object{HD 150574}            & 233                           &         < 61              & Pres. single &No                      & 31\,500       & 3.32 & 3.41     &{--6.8}&{2960}&{0.90}& 0.172 & 7.48 &$\ge$9.08 &...&...&$\ge$8.93&$\ge$1.60 &{ ...} \\
\object{HD 175876}            & 265                          &           <100    & Pres. single&...                      & 35\,900  & 3.66 & 3.74   &{--6.3}&{2430}&{1.07}  & 0.110       &8.04  & 8.36   &... &...&      8.42         &        0.32           &--0.06  \\
\object{HD 191423}            & 420                          &           <29             & Pres. single&No                       & 30\,600  & 3.33 & 3.57   &{--6.4}&{600}&{0.99}  & 0.134 & $\le$7.24 &8.33       &... &...&$\le$8.33&$\ge$1.09 &$\ge$0.00 \\
\object{HD 192281}            & 276                          &          <99             & RV var    &No                   & 39\,000       & 3.64 & 3.73   &{--6.2}&{1960}&{0.84}  & 0.103   & 8.00 &8.76 &... &...&8.05 &           0.76                            & 0.71 \\
\object{HD 203064}            & 298                          &           <104    & SB1           &Yes                            & 35\,000  & 3.73 & 3.82   &{--6.6}&{2340}&{1.03} & 0.076    & 7.92& 8.23&... &...&8.46   &      0.31               & --0.23\\
\object{HD 210839}            & 214                         &            <82             & SB1   &Yes                            & 36\,000  & 3.50 & 3.56   &{--6.0}&{2600}&{1.03} & 0.113&7.83 &8.74 &... &...&8.13       &         0.91                            & 0.61 \\
\object{HD 228841}            & 305                         &            <189    & Pres. single&...                      & 34\,000  & 3.50  & 3.62  &{--7.1}&{2700}&{1.03}  &0.112&7.48 &8.74      &... &...&8.67  &       1.26                    & 0.07 \\
{\it Typical errors}                    & {\it 15}                          &            ...              & ...          &...            & {\it 1500}    &{\it 0.15}&{\it 0.15}                &{...}&{...}&{...}      &{\it 0.030}&{\it 0.27} &{\it 0.34}       &... &...& {\it 0.21}&{\it 0.43$^c$}&{\it 0.40$^c$}\\
\hline
\end{tabular}
\tablefoot{Because macroturbulent velocities cannot be determined reliably for fast rotators (Sect. \ref{secRotVel}), all values in column 3 are upper limits. Column 4 provides the multiplicity status (see Sect. \ref{sectRadVal} for the classification criterion and Appendix \ref{res} for the RV studies of each individual object). The runaway status is based on literature studies (references are given on a star-to-star basis in Appendix \ref{res}). {Columns 9, 10, and 11 list the assumed wind parameters.} For stars with the lowest temperatures (typically B0.5 stars), the carbon abundance cannot be firmly determined due to the weakness of the \ion{C}{III} lines at these temperatures. {Besides, \ion{N}{II} lines may also be very weak for the hottest stars studied with DETAIL/SURFACE. In these cases, we provide upper limits for both carbon and nitrogen abundances. They correspond to predicted lines becoming detectable, i.e. having a depth significantly exceeding the local noise. Similarly, CMFGEN fits may converge towards very high or very low CNO abundances. In both cases, the upper or lower limits were determined from the $\chi^2$ curves and correspond to the limit of their flat minimum. Since the CNO abundances are measured relative to the hydrogen content (assumed to be constant in our study), a correction should in principle be applied to the CNO abundances of stars that exhibit a very high helium abundance (and therefore have a reduced hydrogen abundance). However, we found this correction to be negligible ($\lesssim$0.1 dex) even for the most He-rich stars. The last row of each table section gives the typical 1\,$\sigma$ errors on the parameters. Stars studied with both DETAIL/SURFACE and CMFGEN (see Sect. \ref{subsecCMF}) are flagged with $\dag$.}

$^a$: Because the UCLES spectrum only covers the wavelength range $\sim$ 4350--6800 $\AA$, neither H$\epsilon$, H$\delta$, H$\gamma$, \ion{He}{I} 4026, nor the {C and O} lines in the 4060--4082 $\AA$ region were used. In order to check the reliability of our results, we determined the atmospheric parameters of \object{HD 172367} (whose spectral type is similar to \object{HD 53755}) considering either only one (H$\beta$) or four (H$\epsilon$, H$\delta$, H$\gamma$, H$\beta$) Balmer lines. No significant differences were found between the two sets of results, ensuring that our parameter derivation for \object{HD 53755} is secure. $^b$: Atmospheric parameters and surface abundances must be considered with caution, as they were derived from a low S/N spectrum that only ranges from 4075 to 4920 $\AA$ (i.e. with fewer diagnostic lines). $^c$: Quadratic sum of the CNO abundance uncertainties.}
}
\end{sidewaystable}

\twocolumn 

\begin{table}
{\begin{center}
\caption[]{Errors on the atmospheric parameters and abundances of \object{HD 149757} arising from the choice of the instrument and microturbulence value. The last column gives the adopted (combined) uncertainty.}
\label{sigmaUncert}
\centering
\begin{tabular}{lccccccccccccccccc}
\hline\hline
\multirow{1}{*}{Parameter}                      & $\sigma_{\rm{instr}}$ & $\sigma_{\rm{micro}}$ & $\sigma$ \\
\hline
$T_{\rm{eff}}$ [K]                                      & 630                           &320                                 &   1000\\ 
$\log g$                                                        & 0.06                          &0.07                        &0.10\\
$y$                                                             & 0.010                         &0.022                       &0.025\\
$\log \epsilon$(C)                                      & 0.10                          &0.06                        &0.12\\ 
$\log \epsilon$(N)                                      & 0.12                          &0.04                        &0.13\\
$\log \epsilon$(O)                                      & 0.13                          &0.16                        &0.21\\ 
 $[$N/C]                                                        & 0.21                          &0.02                        &0.21       \\
 $[$N/O]                                                        & 0.01                          &0.12                        &0.12       \\\hline
\end{tabular}
\tablefoot{$\sigma$ corresponds to $\sqrt{\sigma_{\rm{instr}}^2+\sigma_{\rm{micro}}^2}$.}
\end{center}}
\end{table}

\begin{table*}
{\scriptsize
\caption[]{Atmospheric parameters and metal abundances derived in this work for the slow rotators.}
\label{tabMethVal}
\centering
\begin{tabular}{ccccccccccc}
\hline\hline
\multirow{2}{*}{Star}&$T_{\rm{eff}}$&$\log g$& \multirow{2}{*}{$y$}&$\log \epsilon$(C)    &$\log \epsilon$(N)             &\multicolumn{3}{c}{$\log \epsilon$(O) }         &\multirow{2}{*}{[N/C]}  &\multirow{2}{*}{[N/O]}\\\cline{7-9} 
                                &[K]                 &            &                                &  4060--4082 $\AA$      & 4995--5011 $\AA$       & 4060--4082 $\AA$& 4691--4709 $\AA$&  Adopted   &                                 &\\\hline
\multirow{2}{*}{$\xi^1$\,CMa}&28\,200& 3.90&   0.105                    &  7.90                            & 7.84                             &     8.40                   & 8.54                        & 8.47     & --0.06                       & --0.63\\
                                &  \it  28\,500  & \it 4.00 &  \it  0.112                 & \it 8.10                         & \it 7.80                          &  \it    8.46              & \it 8.60                  &\it 8.53   & \it  --0.30                  & \it --0.73\\\hline
\multirow{2}{*}{$\tau$\,Sco}  &31\,200& 4.30 &   0.083                    &  8.18                            & 7.90                              &    8.27                  & 8.50                       &8.39          & --0.28                       & --0.49\\ 
                                &  \it 31\,000   & \it 4.40 &  \it 0.083                  &  \it 8.40                        &\it  7.90                          &  \it   8.24               &  \it 8.62                &\it 8.43   & \it --0.50                   & \it --0.53\\\hline
\multirow{2}{*}{\object{HD 57682}}   &33\,400& 4.00&   0.082                     &  8.06                            & 7.60                              &    8.24                   & 8.26                      &8.25          & --0.46                  & --0.65\\
                                &   \it 33\,300  & \it 4.00 &   \it 0.083                 &  \it 7.98                        & \it 7.76                          &  \it  8.42                & \it 8.30                  &\it 8.36   & \it --0.22                   & \it --0.60 \\\hline
\multirow{2}{*}{10\,Lac}&34\,300        & 4.20 &   0.077                     &  8.22                            & 7.42                             &     8.34                  & 8.28                           &8.31       & --0.80                     & --0.89\\
                                &   \it 34\,000  & \it 4.20 &  \it  0.072                 & \it 8.20                          & \it 7.80                         &   \it    8.48              & \it 8.24                 &\it 8.36   & \it --0.40                   & \it --0.56\\\hline
{\it Typical errors}    & {\it 1000}         &  {\it 0.10}&{\it 0.025}                &{\it 0.12}                            & {\it 0.13}                         &{...}                         &{...}                        &{\it 0.21}&{\it 0.21}                   &{\it 0.12}\\\hline                                                                
\end{tabular}
\tablefoot{For each star, the first row gives our nominal results, while the second row (in italics) provides the results obtained with spectra convolved with $v$\,sin\,$i$ $=$ 300 km s$^{-1}$. Note that solar [N/C] and [N/O] abundance ratios are --0.60 and --0.86, respectively \citep{asp09}.}
}
\end{table*}

\begin{figure}
\centering
\begin{turn}{0}
\includegraphics[scale=0.225]{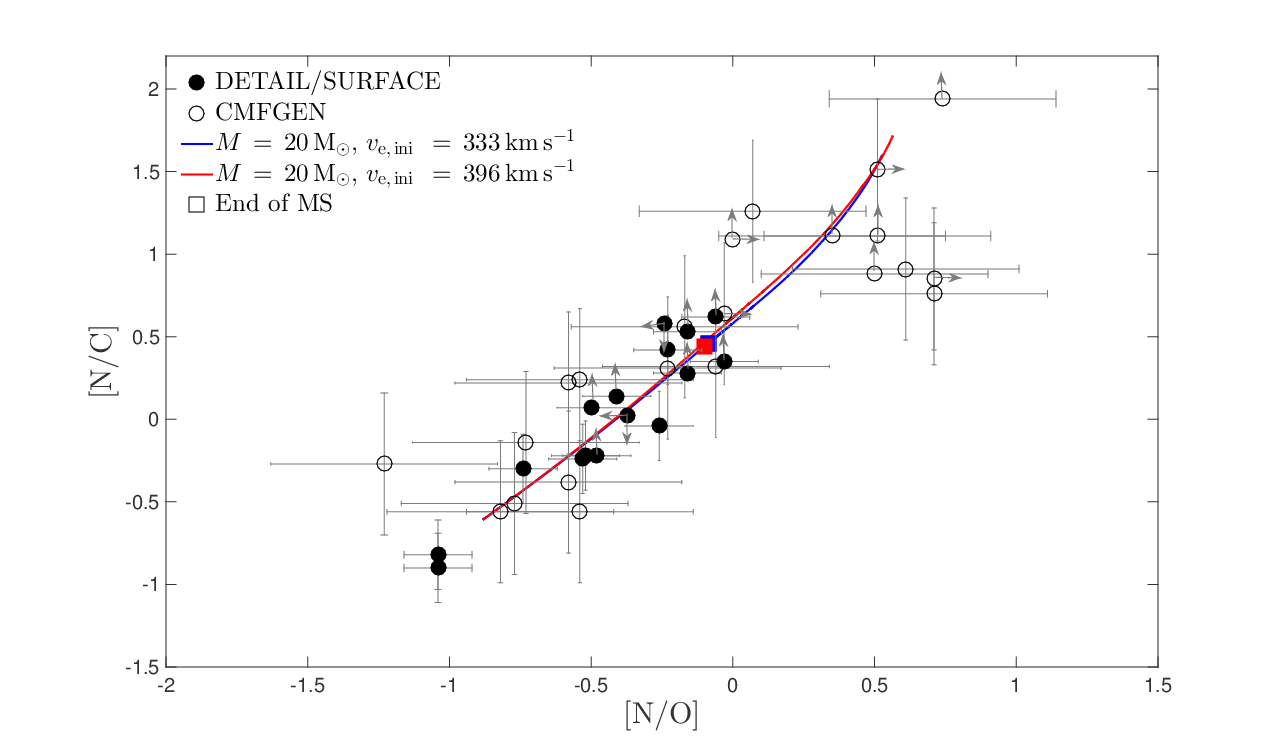}
\end{turn}
\caption{[N/C] as a function of [N/O] for the sample stars, along with theoretical predictions from Geneva models (solid lines for $Z$ = 0.014, Dr C. Georgy, private communication). Filled and open circles show values for the cool (studied with DETAIL/SURFACE) and hot (studied with CMFGEN) objects, respectively.}
\label{NC_NO_fig}
\end{figure}

\begin{table*}
{\scriptsize
\caption[]{Results obtained with DETAIL/SURFACE (columns D/S) and CMFGEN (columns CMF).}
\label{tabMethValCMFGEN}
\centering
\begin{tabular}{ccccccccccccccccccccc}
\hline\hline
\multirow{2}{*}{Star}  & \multicolumn{2}{c}{$T_{\rm{eff}}$ [K]} & \multicolumn{2}{c}{$\log g$ } & \multicolumn{2}{c}{$y$}&\multicolumn{2}{c}{$\log \epsilon$(C) }& \multicolumn{2}{c}{$\log \epsilon$(N) }& \multicolumn{2}{c}{$\log \epsilon$(O) } & \multicolumn{2}{c}{[N/C]} & \multicolumn{2}{c}{[N/O]} \\ 
                                        &D/S& CMF               &D/S& CMF &D/S& CMF & D/S& CMF          &D/S& CMF             &D/S& CMF &D/S& CMF           &D/S& CMF \\           
\hline
\object{HD 102415}              &32\,900&33\,000&4.10&3.86     &0.158&0.124 &$<$7.54&7.32  &8.16&8.51           &8.22&8.02   &$>$0.62&1.19  &--0.06&0.49          \\
\object{HD 149757} (FEROS)      &31\,800&32\,300&3.90&3.87     &0.124&0.096 & 8.06&8.19         & 7.92&7.54     &8.45& 8.27   &--0.14& --0.65       &--0.53&--0.73         \\ 
\object{HD 163892}              &32\,000&32\,400&3.80&3.80     &0.082&0.071 & 8.24&8.15         &7.34&7.44      &8.38&8.40    &--0.90&--0.71        &--1.04&--0.96   \\ 
{\it Typical errors}&{\it 1000}&{\it 1500}&{\it 0.10}&{\it 0.15}&{\it 0.025}&{\it 0.030}&{\it 0.12}&{\it 0.27}&{\it 0.13}&{\it 0.34}&{\it 0.21}&{\it 0.21}&{\it 0.21}&{\it 0.43}&{\it 0.12}&{\it 0.40} \\      
\hline
\end{tabular}}
\end{table*}

\subsection{Comparison with the CNO cycle predictions}
The abundance ratios [N/C] and [N/O] are very good indicators of rotational mixing in massive stars. The transformation of carbon into nitrogen is more efficient than that of oxygen into nitrogen for our sample stars. Hence, their surface carbon and nitrogen abundances should decrease and increase, respectively, whereas the surface abundance of oxygen should remain nearly constant as the star evolves. The loci in the [N/C] versus [N/O] diagram predicted by stellar evolution models reflect the efficiency of the mixing of the CNO material at equilibrium with the initial abundances \citep{prz11,mae14}. Fig. \ref{NC_NO_fig} shows very good consistency between our results and theoretical predictions for most of our targets. The consistent behaviour is preserved when comparing our results with predictions of models covering the full range of initial rotational velocities and masses spanned by our targets. Therefore, the abundances of fast rotators are in agreement with the predictions of CNO cycle nucleosynthesis. 

\subsection{Effect of stellar shape}
Rotation affects the stellar shape, increasing the equatorial radius while decreasing the polar one. This distortion implies that the equipotentials are closer in polar regions than near the equator. The local effective gravity, which is a measure of the gradient between equipotentials, is thus stronger at the pole than at the equator. As the energy passing through an equipotential is conserved in the absence of local energy production or destruction, polar regions are hotter than equatorial regions and more flux is emitted from the pole compared to the equator. This gravity darkening effect implies that the lines of a fast rotator can be created from different regions around the star; thus, \ion{He}{II} lines are preferentially formed near the poles, while \ion{He}{I} lines originate from a larger area of the stellar surface.

We used the Code of Massive Binary Spectral Computation \citep[CoMBISpeC;][]{pal12,pal13} to examine the effect of gravity darkening and stellar rotational flattening on the determination of stellar parameters. To this aim, two stars representing the extreme $v$\,sin\,$i$ values encountered in our sample (\object{HD 149757}, $v$\,sin\,$i$ $\sim$ 378 km s$^{-1}$ and \object{HD 163892}, $v$\,sin\,$i$ = 205 km s$^{-1}$) were considered.
 
\begin{figure}
\centering
\begin{turn}{0}
\includegraphics[scale=0.232]{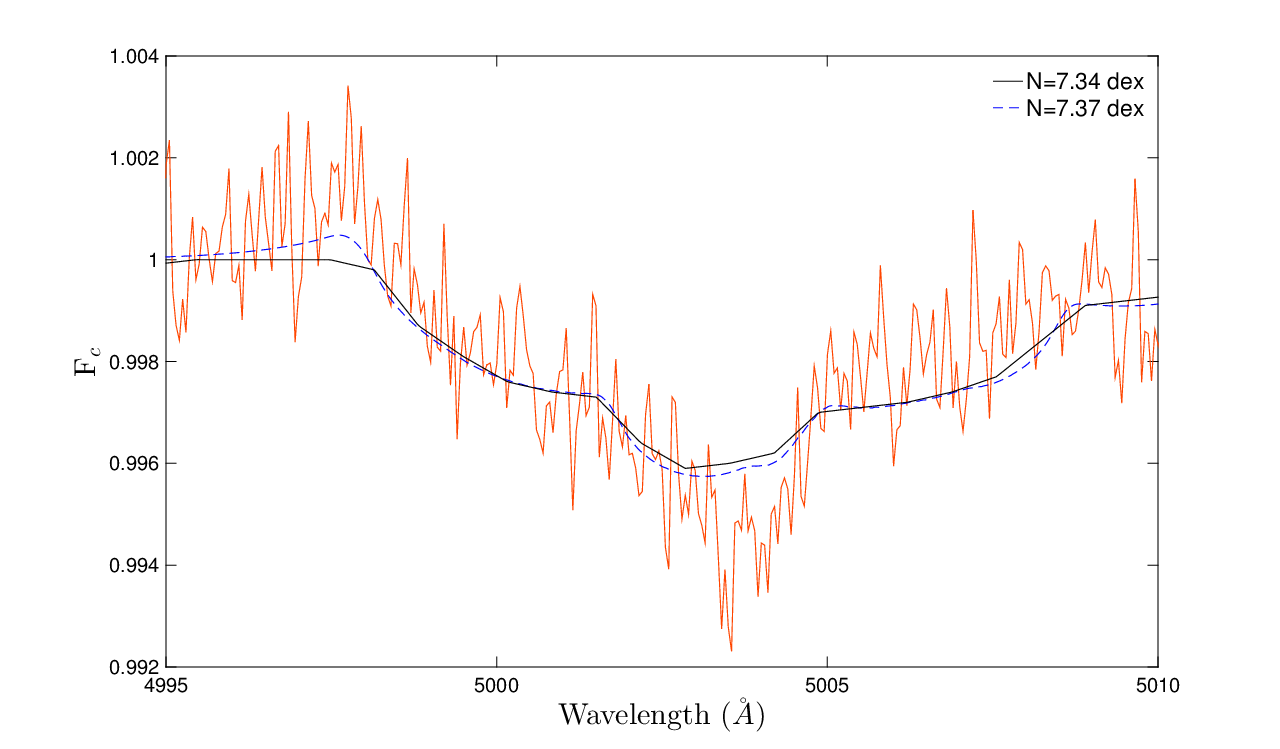}
\end{turn}
\caption{Observed \ion{N}{II} lines in \object{HD 163892} (solid red line) and best-fitting spectra for the spherical case (solid black line) or a flattened star seen under an inclination of $i=90^{\circ}$ (dashed blue line).}
\label{figHD163_COMBI}
\end{figure}

\begin{figure}
\centering
\begin{turn}{0}
\includegraphics[scale=0.232]{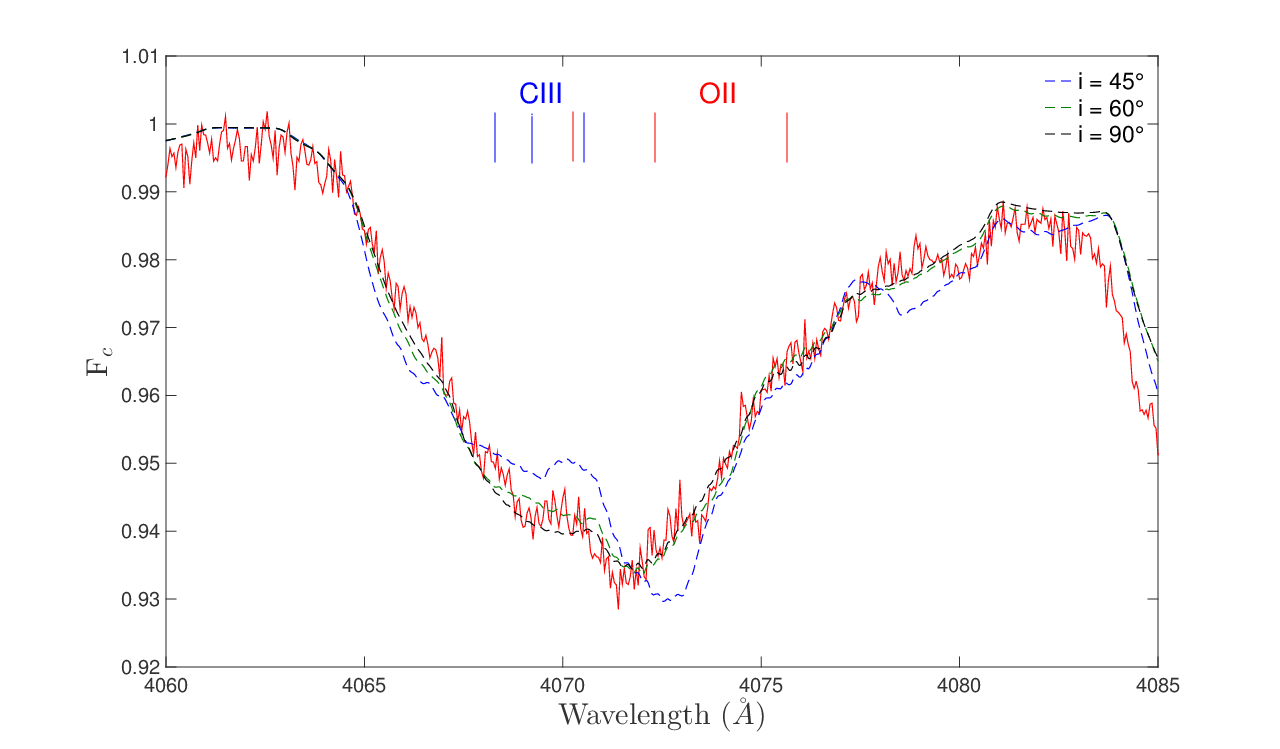}
\end{turn}
\caption{Example of the influence of the inclination and the related change of the rotational velocity on the strength of \ion{C}{III} and \ion{O}{II} lines for very fast rotators. The observed HARPS spectrum of \object{HD 149757} is shown as a solid red line, while model spectra for different inclinations are shown as dashed blue ($i=45^{\circ}$), green ($i=60^{\circ}$), and black ($i=90^{\circ}$) lines. The abundances of carbon and oxygen are set to 7.98 and 8.35 dex, respectively.}
\label{figCOMBI}
\end{figure}

We first determine the polar effective temperatures, $T_{\rm eff, p}$, and the polar radii, $R_{\rm{p}}$, of the stars depending on the inclination of the rotation axis, $i$, in such a way that $T_{\rm eff}$ and $\log g$ averaged over the visible hemisphere are equal to the values found with the method described in Sect. \ref{subSecMethTM}. This was carried out by fixing some parameters: the stellar mass was chosen to be 20\,M$_{\odot}$ since \object{HD 149757} and \object{HD 163892} are close to the corresponding evolutionary tracks in the $\log g_{\rm{C}}$--$\log\,T_{\rm eff}$ diagram (Paper II), the gravity darkening exponent was chosen to be 0.1875, as suggested by interferometric observations of rapidly rotating B stars \citep{krau12}, and finally the projected rotational velocities were fixed to the values that we derived (Table \ref{tabResults}). In these calculations, the $v$\,sin\,$i$ is held fixed. As a result, the true rotation rate varies as a function of $i$ (star intrinsically more rapidly rotating as $i$ decreases). Table \ref{tabCoMBI} presents the resulting parameters.
Once these parameters are known, we then explore how spectra of those flattened stars change with the CNO abundances. As both \ion{He}{I} and \ion{He}{II} line-formation zones are always seen, the helium abundance is correctly determined and we thus do not need to explore changes in $y$. Table \ref{tabCOMBI_incl} illustrates how the resulting abundances vary for the various cases considered. For \object{HD 163892}, we observe that different combinations of inclinations and rotation rates yield very similar best-fitting abundances (Fig. \ref{figHD163_COMBI}). Furthermore, the results are similar to those found with spherically symmetric models, yielding strong support to our methodology. The other star, \object{HD 149757}, is an apparently faster rotator. As expected, this translates into larger differences in the emerging spectrum. Fig. \ref{figCOMBI} shows an example of the variations affecting the C and O line profiles for a fixed abundance set. In fact, as inclination increases, cooler surface regions come into view and the true rotational velocity decreases (as $v\sin\,i$ is kept constant), modifying the strength of \ion{C}{III}, \ion{N}{II}, and \ion{O}{II} lines, which are our abundance diagnostics (see Sect. \ref{subSecMethTM}). For low inclinations, it appears that all CNO abundances of \object{HD 149757} are lower than those derived in the spherical case, while these abundances increase with inclination, reaching values similar to the spherical case when $i=90^{\circ}$; it should be noted that the carbon abundance is,
however, difficult to pinpoint precisely. Whatever the inclination, however, the differences remain well within the error bars and we therefore conclude that the spherically symmetric models used in this work are suitable to study our sample stars.

\subsection{Impact of binarity}
\label{sect_impact_binarity}
A few of our targets are firmly identified as SB1 systems. To examine the impact of the contamination of the spectrum by the secondary, we considered the system with by far the largest mass function hence the largest potential contamination (\object{HD 52533}; see Table \ref{orbSolHD}). Assuming an edge-on orbit and a primary mass of $\sim$ 20 M$_\odot$ (Paper II), we infer that the companion is a B1-B2 star. We repeated the analysis described in Sect. \ref{subSecMethTM} assuming for simplicity that the companion is on the zero age main sequence (ZAMS) and rotates at the same speed as the primary. We further adopted the following parameters: $T_{\rm eff}$ = 28\,000 K, $\log g$ = 4.3, $\xi$ = 10 km s$^{-1}$ and abundances typical of nearby B-type dwarfs (Table 6 of \citealt{mor08}). Grids of composite, synthetic spectra similar to those discussed in Sect. \ref{subSecMethTM} were computed assuming at each mesh point an appropriate flux ratio between the two components (typically $\sim$ 0.1-0.2 for the default parameters of the primary). 

As can be seen in Table \ref{tab_impact_binarity}, taking the cooler secondary in \object{HD 52533} into account would result in modest differences, close to or below the uncertainties. In any event, this more sophisticated approach strengthens the case for a lack of a He and N excess in this star. Furthermore, as the companions are less massive and much fainter for the other SB1 systems (Table \ref{orbSolHD}), even more negligible differences are expected for the parameters of these binaries. 

\section{Summary}
 \label{secCon}
The importance of rotational mixing was recently questioned after the discovery of a population of fast rotators with no or little evidence for a nitrogen enrichment. 

\begin{table}
{\begin{center}
\caption[]{Polar effective temperatures and radii of \object{HD 149757} and \object{HD 163892} as a function of the inclination of the rotation axis.}
\label{tabCoMBI}
\centering
\begin{tabular}{cccc}
\hline\hline
\multirow{2}{*}{Star}              & \multirow{1}{*}{$i$}               & \multirow{1}{*}{$T_{\rm eff, p}$} & \multirow{1}{*}{$R_{\rm{p}}$} \\
                                                   & \multirow{1}{*}{[$^{\circ}$]} & \multirow{1}{*}{[K]}                             & \multirow{1}{*}{[R$_{\odot}$]} \\                                          
\hline
\multirow{3}{*}{\object{HD 149757}} & 45                                                 & 34\,800                                            & 6.38  \\
                                            & 60                                                 & 33\,800                                            & 6.91 \\
 (HARPS)                                    & 90                                                 & 33\,300                                            & 7.20 \\
\hline
\multirow{3}{*}{\object{HD 163892}}  & 45                                                &  33\,300                                          & 7.89 \\
                                            & 60                                                 &  33\,000                                          & 8.12 \\
(FEROS)                              & 90                                                &  32\,800                                          & 8.24\\
\hline
\end{tabular}
\end{center}}
\end{table}

\begin{table}
{\begin{center}
\caption[]{Derived abundances for the fast rotators \object{HD 149757} and \object{HD 163892} for different models (spherical case vs. flattened star seen under different inclinations).}
\label{tabCOMBI_incl}
\centering
\begin{tabular}{lccccccccccccccccc}
\hline\hline
\multirow{3}{*}{Star}                   & \multirow{3}{*}{$i$ [$^{\circ}$]}                     &        \multicolumn{5}{c}{Abundances }\\ \cline{3-7}
                                                &                                                                       &\multirow{2}{*}{C}&\multirow{2}{*}{N}& \multicolumn{3}{c}{O}    \\ \cline{5-7}
                                                &                                                                       &                                         &                                       & (1)                   & (2)                   & Av. \\ 
\hline
                                                &                Sph. case                                      & 7.98                            & 7.92                          & 8.50                  & 8.36                    & 8.43 \\
\multirow{1}{*}{\object{HD 149757}}     &                \multirow{1}{*}{45}                            & 7.82                            & 7.82                          & 8.45                  & 8.30                    & 8.38 \\
 (HARPS)                                        &                \multirow{1}{*}{60}                            & 8.06                            & 7.91                          & 8.50                  & 8.35                    & 8.43 \\
                                                &                \multirow{1}{*}{90}                            & 7.98                            & 7.92                          & 8.50                  & 8.35                    & 8.43 \\
\hline
                                                &                Sph. case                                      & 8.24                            &7.34                           & 8.44                  & 8.32                    & 8.38 \\\multirow{1}{*}{\object{HD 163892}}    &                  \multirow{1}{*}{45}                            & 8.24                          & 7.37                            & 8.45                  & 8.33                  & 8.39 \\
(FEROS)                                 &                \multirow{1}{*}{60}                            & 8.24                            & 7.37                          & 8.45                  & 8.33                    & 8.39 \\
                                                &                \multirow{1}{*}{90}                            &8.24                           & 7.37                            & 8.45                  & 8.33                  & 8.39 \\                                         
\hline
\end{tabular}
\tablefoot{(1) and (2) refer to the spectral regions 4060--4082 and 4691--4709 $\AA$, respectively. `Av.' refers to the average of the oxygen abundances derived in the two regions.}
\end{center}} 
\end{table}

\begin{table}[h!]
\caption{Impact on parameters and abundances when taking the secondary in \object{HD 52533} into account.}
\label{tab_impact_binarity}
\centering
\begin{tabular}{lcc}
\hline \hline
                           & Difference & Typical error\\
\hline 
$\Delta$$T_{\rm eff}$ [K]    & +313& 1000\\
$\Delta$$\log g$                      & --0.10 & 0.10 \\
$\Delta$$y$                 & +0.007          & 0.025\\
$\Delta$$\log \epsilon$(C)  & +0.08     & 0.12\\
$\Delta$$\log \epsilon$(N)  & --0.24       & 0.13\\
$\Delta$$\log \epsilon$(O)  & --0.13     & 0.21\\
$\Delta$$[$N$/$C$]$         & --0.32      & 0.21\\
$\Delta$$[$N$/$O$]$         & --0.11      & 0.12\\
\hline
\end{tabular}
\tablefoot{
The differences are values considering the companion minus values not considering it (from Table \ref{tabResults}).}
\end{table}

We decided to revisit this issue by performing an in-depth study of the physical properties of a large sample of massive, fast rotators. Their properties were derived in several steps. First, the RVs were estimated with a cross-correlation technique, while a Fourier transform method yielded the projected rotational velocity. Then, a comparison with synthetic spectra, calculated either with DETAIL/SURFACE for the 17 late-type (B0.5-O9\,V-III) stars or with CMFGEN for the 23 objects with earlier types, was performed in a homogeneous way within the two subgroups. This provided the effective temperatures, surface gravities, and the He and CNO abundances for each object. 

We performed several checks to validate our method and, hence, its results. First, we studied a sample of well-known slow rotators and showed that our results are in good agreement with previous studies. Furthermore, after convolving the spectra of these stars to mimic a broadening typical of our sample stars, we again obtained similar results, demonstrating the limited impact of broadening on our derivation of physical parameters. Second, the synthetic spectra used in this work correspond to spherically symmetric stars, while fast rotators are flattened objects. We therefore compared our results with those obtained with CoMBISpeC, which takes the stellar deformation into account. Again, results were similar, within errors, further validating our method. Finally, a few targets could be analysed by both CMFGEN and DETAIL/SURFACE models, again showing a good agreement. Further confidence in our results comes from the fact that the [N/C] and [N/O] abundance ratios correlate along the theoretical locus expected for the CNO cycle.

This paper presents the stellar parameters and CNO abundances of 40 fast rotators, along with their multiplicity status, including {two} new and three revised orbital solutions (see Appendix \ref{res}). The second paper of this series will compare these results to predictions of evolutionary models of single stars or of interacting binaries with the aim to assess the impact of rotational mixing in hot stars. 

\begin{acknowledgements}
{We are very grateful to the referee for providing useful comments.} We thank John Pritchard from the User Support Department of the European Southern Observatory and the FIES team for their precious help in the reduction of data. We also thank Dr. Keith Butler, Dr. John Hillier, Dr. Hugues Sana, and Dr. Matthieu Palate for making their codes available to us.

We acknowledge the support from the Universities of Hamburg, Guanajuato, and Li\`ege for the TIGRE telescope. We thank the team that proposed, observed, and reduced the MIKE data: Marcelo Borges, Gustavo Bragan\c ca, Thomas Bensby, Katia Cunha, Katy Garmany, and John Glaspey. To get SOPHIE observations, the authors received funding from the European Community's Seventh Framework Programme (FP7/2013-2016) under grant agreement number 312430 (OPTICON). We thank Dr. Sergi Blanco-Cuaresma and Dr. Maroussia Roelens for obtaining CORALIE observations.
CC also acknowledges funding from `Patrimoine de l'ULg' for his stay at Rio de Janeiro and people at the Observat\'orio Nacional for their kind hospitality during his stay in Rio.

{This research has made use of the WEBDA database, operated at the Department of Theoretical Physics and Astrophysics of the Masaryk University.}

Computational resources have been provided by the Consortium des \'Equipements de Calcul Intensif (C\'ECI), funded by the Fonds de la Recherche Scientifique (F.R.S.-FNRS) under Grant No. 2.5020.11.

The Li\`ege team also acknowledges support from the Fonds National de la Recherche Scientifique (Belgium), the Communaut\'e Fran\c caise de Belgique, the PRODEX XMM and GAIA-DPAC contracts (Belspo), and an ARC grant for concerted research actions financed by the French community of Belgium (Wallonia-Brussels Federation). ADS and CDS were used for preparing this document. 
\end{acknowledgements}


\begin{appendix}
  
\section{Journal of observations and radial velocities of our targets}
\label{jourRV}  
{Table \ref{tabJour} provides the journal of observations and RVs of our sample stars. The RVs taken from the literature are not included.}
  
{\begin{table*}
\caption{Journal of observations and individual RV measurements. Heliocentric corrections were applied to both Julian dates and RVs. Spectra indicated in boldface were used to determine the stellar properties (multiple exposures were averaged).}
\label{tabJour}
\begin{tabular}{cccc}
\hline\hline
\multirow{2}{*}{Name}   & \multirow{2}{*}{Instrument}   & Mid-exp. (HJD       &  RV \\
                                         &                                               & -- 2,450,000 &  [km s$^{-1}$] \\
\hline
\multicolumn{4}{l}{\it Slow rotators}\\
\multirow{1}{*}{\object{HD 214680}}& \multirow{2}{*}{\bf HEROS}   & \multirow{2}{*}{6561.713}         & \multirow{2}{*}{--7.3$\pm$0.5} \\
(10\,Lac)                         &                                               &                              &                               \\ 
\hline
\multirow{1}{*}{\object{HD 46328}}& \multirow{2}{*}{\bf HEROS}     & \multirow{2}{*}{7017.855}    & \multirow{2}{*}{21.9$\pm$0.5}\\
($\xi^1$\,CMa)                     &                                            &                                 &                                                       \\
\hline
\multirow{5}{*}{\object{HD 57682}}& \multirow{5}{*}{\bf HEROS}  & 6565.951              & 28.3$\pm$0.3\\
\hline
\end{tabular}
\begin{tabular}{cccc}
\hline\hline
\multirow{2}{*}{Name}              & \multirow{2}{*}{Instrument} & Mid-exp. (HJD           &  RV \\
                                                &                                          & -- 2,450,000) &  [km s$^{-1}$] \\                                                                                                                                                                                                                        
\hline
\multirow{21}{*}{\object{HD 13268}}   & \multirow{18}{*}{AURELIE}          &        4421.499 & --90.1 $\pm$   6.6\\                                                                                     &                                             & 4421.549 & --94.3 $\pm$   6.4\\
                                                &                                                   & 4421.573 & --95.1 $\pm$   6.3\\
                                                &                                                   & 4421.598 & --91.0 $\pm$   6.8\\
                                                &                                                   & 4422.356 & --94.1 $\pm$   6.9\\
                                                &                                                   & 4422.373 & --100.0 $\pm$   7.1\\
\end{tabular}
\end{table*}}
{\begin{table*}
\addtocounter{table}{-1}
\caption{Continued.}
\label{tabJour2}
\begin{tabular}{cccc}
\hline\hline
\multirow{2}{*}{Name}              & \multirow{2}{*}{Instrument} & Mid-exp. (HJD           &  RV \\
                                                &                                          & -- 2,450,000) &  [km s$^{-1}$] \\                                                                                                                                                                                                                        
\hline          
\multirow{17}{*}{\object{HD 14442}}& \multirow{17}{*}{AURELIE} & 2163.526 &  --73.4$\pm$14.0 \\                                                                                                                                                                                                                
                                              &                                            & 2163.550 &  --72.2$\pm$15.6 \\                                                                                                                                                                                                                               
                                              &                                            & 2164.550 &  --41.9$\pm$15.1\\                                                                                                                                                                                                                                
                                              &                                            & 2164.572 &  --41.4$\pm$15.1 \\                                                                                                                                                                                                                               
                                              &                                            & 2165.532 &  --72.3$\pm$13.4 \\
\hline
\end{tabular}
\begin{tabular}{cccc}
\hline\hline
\multirow{2}{*}{Name}              & \multirow{2}{*}{Instrument} & Mid-exp. (HJD           &  RV \\
                                                &                                          & -- 2,450,000) &  [km s$^{-1}$] \\                                                                                                                                                                                                                        
\hline
\multirow{26}{*}{\object{HD 41161}}     & \multirow{1}{*}{ELODIE}& 3330.687 & --18.4$\pm$6.1\\ \cline{2-4}
                                              & \multirow{3}{*}{\bf FIES}          & 4779.634 & --23.3$\pm$5.3 \\ 
                                                &                                          & 5577.688 & --20.3$\pm$5.6  \\ 
                                                &                                          & 5815.715 & --16.4$\pm$5.5 \\ \cline{2-4}  
                                              & \multirow{22}{*}{HEROS}    & 6563.745 & --22.0$\pm$4.8 \\ 
                                                &                                          & 6574.779 & --23.0$\pm$5.9 \\
\hline                                 
\end{tabular}
\end{table*}
\begin{table*}
\addtocounter{table}{-1}
\caption{Continued.}
\begin{tabular}{cccc}
\hline\hline
\multirow{2}{*}{Name}   & \multirow{2}{*}{Instrument}   & Mid-expo. (HJD       &  RV \\
                                         &                                               & -- 2,450,000) &  [km s$^{-1}$] \\
\hline
\multirow{1}{*}{\object{HD 66811}}& \multirow{2}{*}{\bf ESPaDOnS}& 5971.798                 &--10.6$\pm$3.9\\ 
  ($\zeta$\,Pup)                  &                                                     & 6701.928                &--4.4$\pm$4.0\\ 
\hline                                          
\multirow{8}{*}{\object{HD 69106}}& Boller \& Chivens   & 5646.564              &8.2$\pm$4.9 \\ \cline{2-4}
                                        & \multirow{3}{*}{CORALIE}      & 7443.743                &14.1$\pm$5.3 \\                                        
                                        &                                               & 7447.562                &26.9$\pm$5.4\\ 
                                        &                                               & 7501.568                &4.9$\pm$5.5\\
\hline
\end{tabular}
\begin{tabular}{cccc}
\hline\hline
\multirow{2}{*}{Name}              & \multirow{2}{*}{Instrument} & Mid-expo. [(HJD          &  RV \\
                                                &                                          & -- 2,450,000) &  [km s$^{-1}$] \\   
\hline
\multirow{17}{*}{\object{HD 163892}}& \multirow{3}{*}{CORALIE}  & 7441.865              & 55.1$\pm$2.6\\ \
                                       &                                                & 7530.880                & --34.8$\pm$2.3\\ 
                                       &                                                & 7536.832                & 15.4$\pm$2.2\\ \cline{2-4}
                                      &\multirow{12}{*}{\bf FEROS}      & \bf 3546.882    & 30.6$\pm$3.5 \\
                                     &                                                   & 4600.887              & --49.0$\pm$3.5 \\ 
                                     &                                                   & 4626.849              & --1.5$\pm$3.0 \\
\hline                               
\end{tabular}
\end{table*}

\begin{table*}
\addtocounter{table}{-1}
\caption{Continued.}
\begin{tabular}{cccc}
\hline\hline
\multirow{2}{*}{Name}              & \multirow{2}{*}{Instrument} & Mid-exp. (HJD           &  RV \\
                                                &                                          & -- 2,450,000) &  [km s$^{-1}$] \\                                                                                                                                                                                                                        
\hline
\multirow{14}{*}{\object{HD 191423}}&Albireo                    & 5014.482              & 38.5$\pm$21.4\\ \cline{2-4}                     
                                     & \multirow{7}{*}{AURELIE} & 4711.496              & --34.8$\pm$20.3 \\  
                                     &                                                   & 4712.551              & --23.6$\pm$20.2 \\ 
                                     &                                                   & 4717.444              & --27.2$\pm$20.6 \\                                 
                                     &                                                   & 4718.588              & --18.3$\pm$21.8 \\                                 
                                     &                                                   & 4743.328              & --30.2$\pm$20.4 \\
\hline
\end{tabular}
\begin{tabular}{cccc}
\hline\hline
\multirow{2}{*}{Name}              & \multirow{2}{*}{Instrument} & Mid-exp. (HJD           &  RV \\
                                                &                                          & -- 2,450,000) &  [km s$^{-1}$] \\                                                                                                                                                                                                                        
\hline
\multirow{5}{*}{\object{HD 198781}}& \multirow{5}{*}{\bf HEROS}  & 6568.690                 & --17.0$\pm$7.3 \\ 
                                     &                                                   & 6580.597              & --10.1$\pm$7.4 \\ 
                                     &                                                   & 6592.576              & --22.0$\pm$7.9 \\                                  
                                     &                                                   & 6860.864              & --26.7$\pm$7.1 \\                                      
                                     &                                                   & 6860.885              & --25.5$\pm$6.9 \\ 
\hline

\multirow{8}{*}{\object{HD 203064}}& \multirow{1}{*}{\bf ESPaDOnS}      & 6200.773                & 32.6$\pm$8.0 \\
\hline                                      
\end{tabular}
\end{table*}

\section{Diagnostic lines used for the CNO abundance determinations with CMFGEN}  
\label{diagLinesCNO}
{Table \ref{tabLinesCNO} gives the spectral lines used to derive the carbon, nitrogen, and oxygen abundances with CMFGEN.}
    
\onecolumn
{
\end{sidewaystable}
}
\addtocounter{table}{-1}
\begin{sidewaystable}
\caption{Continued.}
\begin{tabular}{lcccccccccccccccccccccccccc}
\hline\hline
\multirow{2}{*}{Line}   &BD                             & \multicolumn{12}{c}{HD}\\ \cline{3-14} 
        &+34$^{\circ}$1058      &13268  &14434  &14442  &15137  &15642  &41161& 41997& 46056    & 46485 &66811  & 69106 &74920  \\\hline
\ion{N}{II}\,3995       &       &       &$\surd$&       &       &       &       &       &       &       &       &       &$\surd$\\
\ion{N}{III}\,4004      &       &$\surd$&$\surd$&       &$\surd $&      &$\surd$& $\surd$&        &       &       &       &$\surd$\\
\ion{N}{IV}\,4058       &       &$\surd$&$\surd$&       &       &       &$\surd$&       &       &$\surd$&       &       &$\surd$\\
\ion{N}{III}\,4095      &$\surd$&       &$\surd$&$\surd$&$\surd$&       &       &$\surd$&       &       &$\surd$&$\surd$&$\surd$\\
\ion{N}{III}\,4196      &       &$\surd$&$\surd$&$\surd$&$\surd$&$\surd$&       &       &       &       &       &$\surd$&$\surd$\\
\ion{N}{III}\,4216      &       &       &       &       &       &       &       &       &       &       &       &       &       \\
\end{tabular}
\end{sidewaystable}
\addtocounter{table}{-1}
\begin{sidewaystable}
\caption{Continued.}
\begin{tabular}{lcccccccccccccccccccccccccc}
\hline\hline
\multirow{2}{*}{Line}                                   & \multicolumn{13}{c}{HD}\\ \cline{2-14} 
                        &92554  &102415 &117490 &124979 &149757 &150574 &163892 &175876& 191423&192281&203064&210839    &228841\\\hline
\ion{N}{II}\,3995       &       &$\surd$&       &$\surd$&       &       &       &       &       &       &       &$\surd$&       \\
\ion{N}{III}\,4004      &       &$\surd$&$\surd$&$\surd$&       &$\surd$&       &$\surd$&       &$\surd$&       &$\surd$&$\surd$\\
\ion{N}{IV}\,4058       &$\surd$&       &       &       &       &$\surd$&       &       &       &$\surd$&       &       &$\surd$\\
\ion{N}{III}\,4095      &$\surd$&       &$\surd$&$\surd$&$\surd$&       &$\surd$&$\surd$&$\surd$&       &       &       &$\surd$\\
\ion{N}{III}\,4196      &$\surd$&$\surd$&       &$\surd$&       &$\surd$&       &$\surd$&$\surd$&       &       &$\surd$&$\surd$\\
\ion{N}{III}\,4216      &       &       &       &$\surd$&       &       &       &       &       &       &       &       &       \\
\end{tabular}
\end{sidewaystable}
}
\addtocounter{table}{-1}
\begin{sidewaystable}
\caption{Continued.}
\begin{tabular}{lcccccccccccccccccccccccccc}
\hline\hline
\multirow{2}{*}{Line}   &BD                             & \multicolumn{12}{c}{HD}\\ \cline{3-14} 
        &+34$^{\circ}$1058      &13268  &14434  &14442  &15137  &15642  &41161  & 41997   & 46056 & 46485 &66811  & 69106 &74920  \\
\hline
\ion{O}{II}\,3792       &       &$\surd$&$\surd$&       &       &       &       &       &       &$\surd$&$\surd$&$\surd$&       \\
\ion{O}{II}\,3913       &       &       &       &       &       &$\surd$&       &       &       &       &       &$\surd$&       \\
\ion{O}{II}\,3955       &       &       &       &       &$\surd$&$\surd$&       &       &       &       &       &$\surd$&       \\
\ion{O}{II}\,3963       &       &$\surd$&$\surd$&       &$\surd$&$\surd$&$\surd$&       &$\surd$&$\surd$&       &$\surd$&       \\
\ion{O}{II}\,4277--78   &       &       &       &       &       &       &       &       &       &$\surd$&       &       &       \\
\ion{O}{II}\,4284       &       &       &       &       &       &       &       &       &       &$\surd$&       &       &       \\
\hline
\end{tabular}
\end{sidewaystable}

\twocolumn 
             
\section{Binary and runaway status}
\label{res}

We determined the multiplicity status of our targets using our own RV measurements (Table \ref{tabJour}) complemented with literature information. Whenever possible, new or improved orbital elements are presented (Table \ref{orbSolHD}).

Table \ref{tab_multiplicity} summarises the detection status of visual companion(s) in the close vicinity of our targets, when available. The widely different field of view and sensitivity in terms of magnitude differences and angular separations may explain why some close companions are detected in some surveys, but not in others. It should be noted that the presence of such companions is not reflected in RV variations, considering our error bars, as they are too distant.

\subsection{\object{ALS 864}}
There is only a little information about this object in the literature and only one spectrum is available in our dataset. Therefore, we could not assess its multiplicity.

\subsection{\object{ALS 18675}}
The literature provides no additional information and we have only one spectrum of this object, hence its multiplicity status cannot be established.

\subsection{\object{BD +60$^{\circ}$594}}
\citet{hilw06} claimed that this star is probably a single-lined spectroscopic binary with an orbital period of the order of 20 days. Indeed, \citet{con77} gave a RV value of --50 km s$^{-1}$, while \citet{hilw06} quoted decreasing RVs (from --60 to --110 km s$^{-1}$) on a timescale of 4\,d. Our measurements further yield $-$51 and $-$15 km s$^{-1}$. Accordingly, we classify this star as RV variable. 

\subsection{\object{BD +34$^{\circ}$1058}}
As we have only a single spectrum of this star and no additional RV measurements are available in the literature, therefore, we cannot assess its multiplicity status. 
 
\subsection{\object{HD 13268}}
Low-amplitude, short-term (a few hours) periodic line-profile variations have been detected by \citet{deb08}. They attributed these variations to non-radial pulsations or structures associated with material in the circumstellar environment. \citet{bek74} further suggested that this star is a runaway with a peculiar velocity greater than 89 km s$^{-1}$ and \citet{ken96} proposed that it was ejected from within Per OB1. 

The variability test indicates no significant variation of the RVs in our dataset, but variations are clearly detected when literature values are added. Indeed, all recent RVs (HJD > 2,450,000) fall inside the interval $-90$ to $-120$ km s$^{-1}$, while much older data (HJD $\sim$ 2,440,500; \citealt{abt72}) provide RVs typically ranging from $-110$ to $-130$ km s$^{-1}$. As there are 78 RV measurements in total, we attempted a period determination. A search performed on all or only the recent data yields no clear periodicity, simply favouring variations occurring on long timescales, hence we thus simply -- and tentatively -- classify the star as RV variable.

\subsection{\object{HD 14434}} 
Significant line-profile variability of the \ion{He}{II} 4686 double-peaked emission and the H${\beta}$ absorption line has been reported by \citet{deb04}, but we found no significant RV variation for this star, even when considering literature values \citep{con77}. Hence we classify it as presumably single. 
 
\subsection{\object{HD 14442}}
Significant line-profile variability of the \ion{He}{II} 4686 double-peaked emission and of the H${\beta}$ absorption line has been interpreted as co-rotating features formed in the wind \citep{deb04}. In our data, we found no significant RV variation so we classify \object{HD 14442} as presumably single.

\subsection{\object{HD 15137}} 
\object{HD 15137} is a known runaway with a peculiar space velocity $V_{\rm{pec}}$ = 62.7$\pm$11.8 km s$^{-1}$ \citep{mcs07}. This star was proposed to be an SB1 system that was probably expelled from the open cluster NGC\,654 \citep{boy05}. The high eccentricity of the system ($e$ $\sim$ 0.5) can be explained by the widening of the orbit during the supernova event that also imparted the velocity kick. The mass of the companion star must be low \citep[1.4 M$_{\odot} \le M_{\rm{comp}} \le 3.0$ M$_{\odot}$;][]{mcs10}, as indicated by the mass function of the system. It may also be noted that this system is a faint X-ray emitter, although it could still be a {high mass X-Ray binary (HMXB) with a very low accretion rate \citep{boy05}}.

We re-investigated the system with our dataset, complemented by literature data (\citealt{con77}; \citealt{boy05}; \citealt{mcs07,mcs10}) {and found evidence for significant RV variations}. An error of 8.5 km s$^{-1}$ is considered for RV values from \citet{mcs07} and \citet{boy05}, as in \citet{mcs10}. Analysing the RVs with period search algorithms yields, however, no clear {periodicity, simply favouring variations occurring on long timescales; we do not find significant peaks at the periods proposed by these authors}. Furthermore, when we fold all data with the 28.61\,d period of \citet{boy05}, the RVs appear scattered. {Folding them with the 55.40\,d period of \citet{mcs10} results in slightly more coherent variations, although no convincing peak is seen in the periodograms at this orbital period.} Therefore, we simply classify the star as RV variable. Eliminating the oldest data point (HJD = 2,440,074.970; \citealt{con77}) does not modify our conclusions.
        
    
\subsection{\object{HD 15642}}    
Our RVs do not show any significant variations and no other information is available in the literature. Therefore, we classify this star as presumably single. 

\subsection{\object{HD 28446A}} 
\object{HD 28446} was first suggested to be a spectroscopic binary by \citet{fro26}. In this context, \citet{pla31} claimed that \object{HD 28446} is an SB2 with a large velocity amplitude ($K$ $\sim$ 140 km s$^{-1}$). However, the spectra of \citet{may94} show no trace of a secondary and these authors only found a RV variability with a period of 1.3\,d. More recently, \citet{str07} suggested that \object{HD 28446} is a triple system with three visual components surrounded by a \ion{H}{II} region of 1.5--2$^{\circ}$ diameter, while \citet{egg08} found only two components separated by 10\farcs \,\,For our HEROS observations, we made sure that the brightest component of this system was observed (\object{HD 28446A}). 

\citet{jer93} reported small variations in the photometric data of \object{HD 28446A}, which is consistent with a period of 0.22132 day and suggesting a $\beta$ Cephei nature for this star. \citet{sta05} rather proposed it to be a slowly pulsating B star (SPB). Our RVs do not significantly vary, but when combined with literature values \citep{may98}, evidence for variability is found. {Period searches on the full RV set yield peaks near 1.5 or 3\,d, with numerous close aliases. Phasing the RVs with such periods yields a noisy RV curve with variation amplitude of $\sim$ 10 km s$^{-1}$. It must, however, be noted that Mayer's values appear mostly below ours with an offset of $\sim$ 20 km s$^{-1}$ between both datasets. Furthermore, the existence of line-profile variations arising from the non-radial pulsations and the possible presence of a gravitationally bound tertiary might lead to such noisy RV curves. Therefore, more data are needed to clarify the source of the RV variations. We thus refrain from calculating an orbital solution, simply classifying the star as RV variable.}



\subsection{\object{HD 41161}}
\object{HD 41161} is a runaway star located about 355 pc above the Galactic plane \citep{dew05}. It is also a bow shock candidate \citep{per12}. 

\begin{figure}
\centering
\begin{turn}{0}
\includegraphics[scale=0.45]{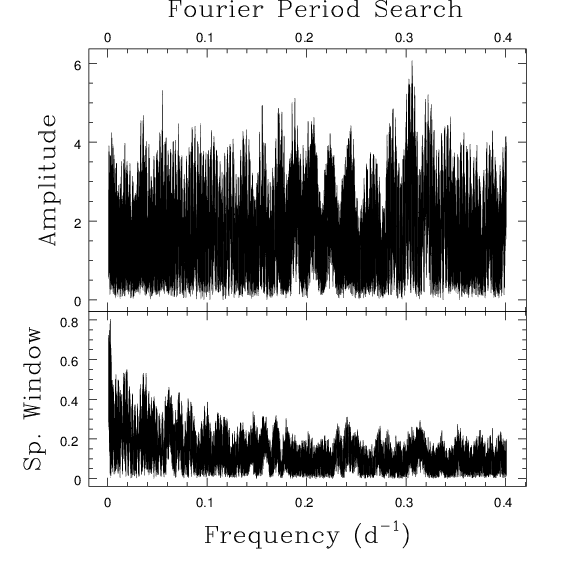}
\end{turn}
\caption{Fourier periodogram derived from the RVs (our work + literature) of \object{HD 41161}. Note the peak at 0.306\,d$^{-1}$.} 
\label{HD41161Fourier}
\end{figure}

Significant RV changes are found when all data (our work + literature; \citealt{con77}; \citealt{gar80}) are combined, with a maximum RV difference of 35 km s$^{-1}$ (corresponding to a 5\,$\sigma$ variation). {All periodograms have some peaks around a period of $\sim$ 3\,d} -- the best Fourier value is $P=3.26592\pm0.00006$\,d (Fig. \ref{HD41161Fourier}). The period error is certainly underestimated. In fact, there are numerous close aliases of that period because the data consist of widely separated observing blocks. {The amplitude of the variations, however, is very small,  i.e. only 6 km s$^{-1}$. This is not formally significant since peaks with this amplitude are typically found in periodograms calculated based on Monte Carlo simulations (using only the observing dates and noise). Besides, such a small amplitude could arise from line-profile variations}, and we thus do not attempt to calculate an orbital solution, waiting for more data taken with a more appropriate sampling, to solve the issue. In the meantime, we classify this object as RV variable.

\subsection{\object{HD 41997}} 
\object{HD 41997} is a runaway star with a peculiar RV of --40 km s$^{-1}$ \citep{car78}. Only one spectrum is available for this star and no further RV measurements are available in the literature, so its multiplicity cannot be assessed. 
  
\subsection{\object{HD 46056}}
\object{HD 46056} was suggested to be an SB1 \citep{wal73,und90}. \citet{mah09} rather found it to be single. These authors noticed variations of the line profiles, which could have led to a spurious detection of RV changes. In line with this result, we do not find any significant RV variation in the ESPaDOnS and ESPRESSO spectra, and we thus classify this star as presumably single. \citet{fea57} observed a large RV variation (from --21 to +65 km s$^{-1}$), but this needs to be confirmed because of the low precision of measurements on photographic plates. 


\subsection{\object{HD 46485}}  
We find a RV difference of $\sim$ 20 km s$^{-1}$ between our two spectra of this star, which are separated by about two months. A similar difference was reported by \citet{fea57} over two years. However, it corresponds only to a 1.9\,$\sigma$ variation, which is not significant. Therefore, we classify this star as presumably single. Adding a value from \citet{con77} does not change this multiplicity status (maximum RV difference of 27 km s$^{-1}$ corresponding to a 3.5\,$\sigma$ variation). 


\subsection{\object{HD 52266}} 
\label{subsecHD5226}
\begin{figure}
\centering
\begin{turn}{0}
\includegraphics[scale=0.45]{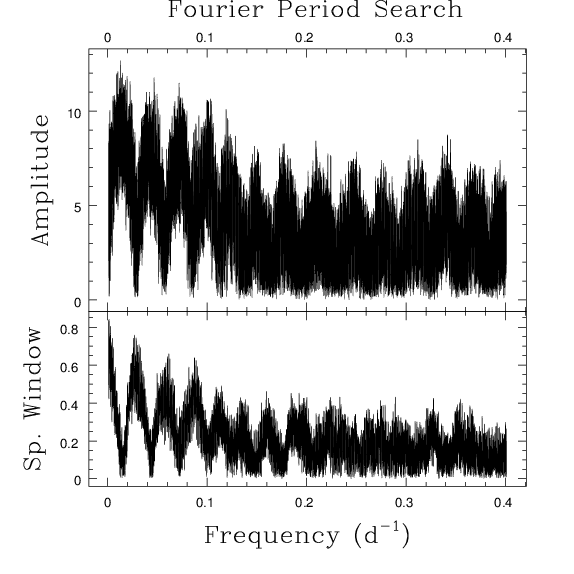}
\end{turn}
\caption{Fourier periodogram derived from the RVs (our work + literature) of \object{HD 52266}. Note the peak near 0.01\,d$^{-1}$.} 
\label{HD52266Fourier}
\end{figure}

\begin{figure}
\centering
\begin{turn}{0}
\includegraphics[scale=0.235]{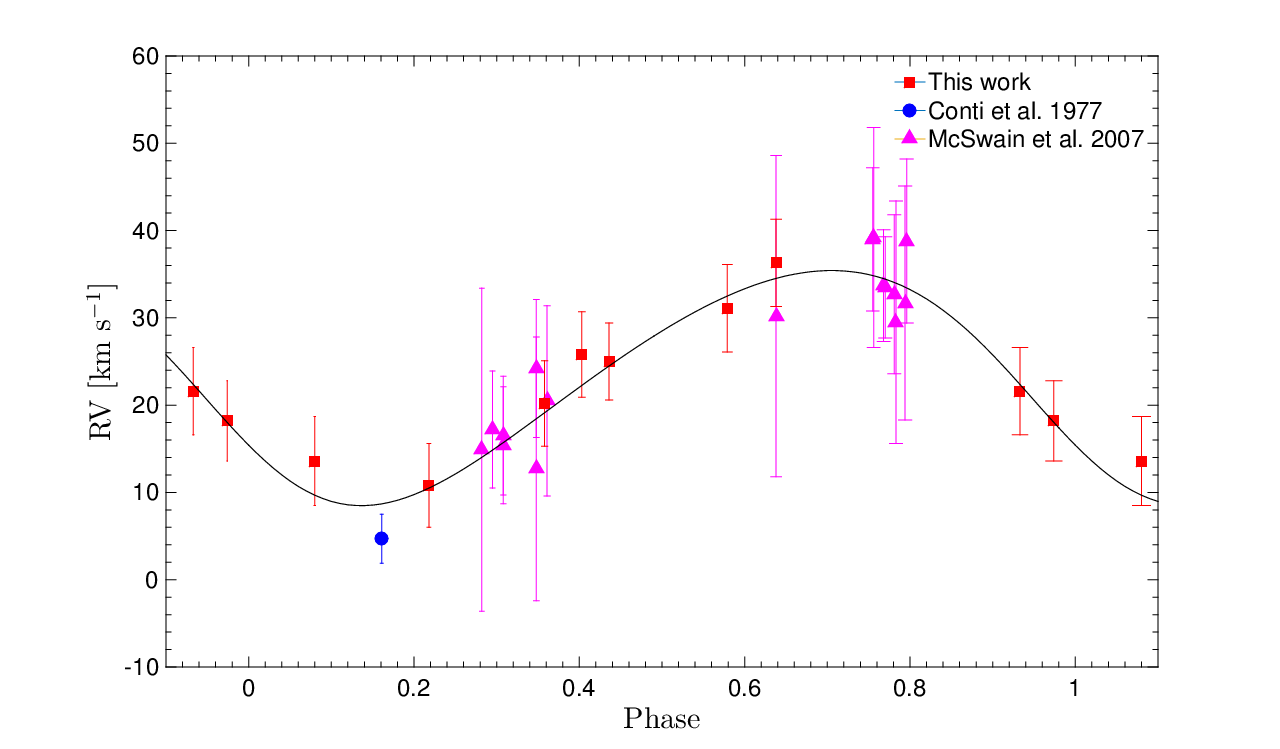}
\end{turn}
\caption{Phase diagram of the RV values of \object{HD 52266} folded with a 75.84\,d period. The best-fit orbital solution (Table \ref{orbSolHD}) is represented as a black curve.}
\label{HD52266_LOSP}
\end{figure}

The peculiar velocity of \object{HD 52266} is not very large (19.4$\pm$9.0 km s$^{-1}$) and it is thus presumably not a runaway star \citep{mcs07}. \citet{mcs07} further suggested that \object{HD 52266} is likely an SB1 system, but they could not determine an orbital period. They only constrained it to be longer than the time span of their data (i.e. RV variation from 12 to 39 km s$^{-1}$ over 40 days). We obtained many additional observations and our RV measurements span the range 14--36 km s$^{-1}$, confirming previous results. The RV changes are found to be significant when all data (our work + literature; \citealt{con77}; \citealt{mcs07}) are combined. We decided to search for a period and a clear signal was found: $P=75.84\pm0.04$\,d for the modified Fourier algorithm (Fig. \ref{HD52266Fourier}). The associated semi-amplitude is moderate (13\,km s$^{-1}$) and the period error is certainly underestimated; there are numerous close aliases of that period because the data consist of widely separated observing blocks. Adopting the period mentioned above, we nevertheless derived an orbital solution with LOSP (see Table \ref{orbSolHD} and Fig. \ref{HD52266_LOSP}). We caution, however, that the sampling is far from perfect, implying that this solution is still preliminary and requires confirmation.

\subsection{\object{HD 52533}}
\label{subsecHD5225}
\citet{gie86} found an SB1 solution for \object{HD 52533} with a 3.29\,d period, while \citet{mcs07} suggested that it might be an SB3. The \ion{He}{II} lines would originate from the primary O star, while a distant B companion would contribute to \ion{He}{I} and Balmer line profiles. \citet{mcs07} found a period of 22.1861$\pm$0.0002\,d from the lines associated with the O star, while the B-star lines appeared stationary. In addition, the peculiar velocity of \object{HD 52533} is 47.0$\pm$27.9 km s$^{-1}$, suggesting that it might be a runaway star \citep{mcs07}. In this context, the invisible companion of the O star could be a compact object, but its X-ray emission is typical of that of single O stars \citep{mot98}. A modest accretion rate could render the presence of a compact companion undetectable, however \citep{meu05}. A search for radio emission originating from a pulsar was unsuccessful \citep{phi96}.
 
\begin{figure}
\centering
\begin{turn}{0}
\includegraphics[scale=0.45]{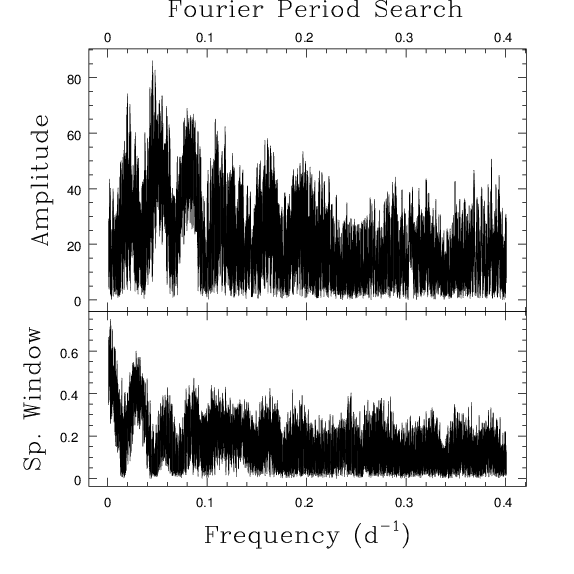}
\end{turn}
\caption{Fourier periodogram derived from the RVs (our work + literature) of \object{HD 52533}. Note the peak near 0.045\,d$^{-1}$.} 
\label{HD52533Fourier}
\end{figure}

\begin{figure}
\centering
\begin{turn}{0}
\includegraphics[scale=0.235]{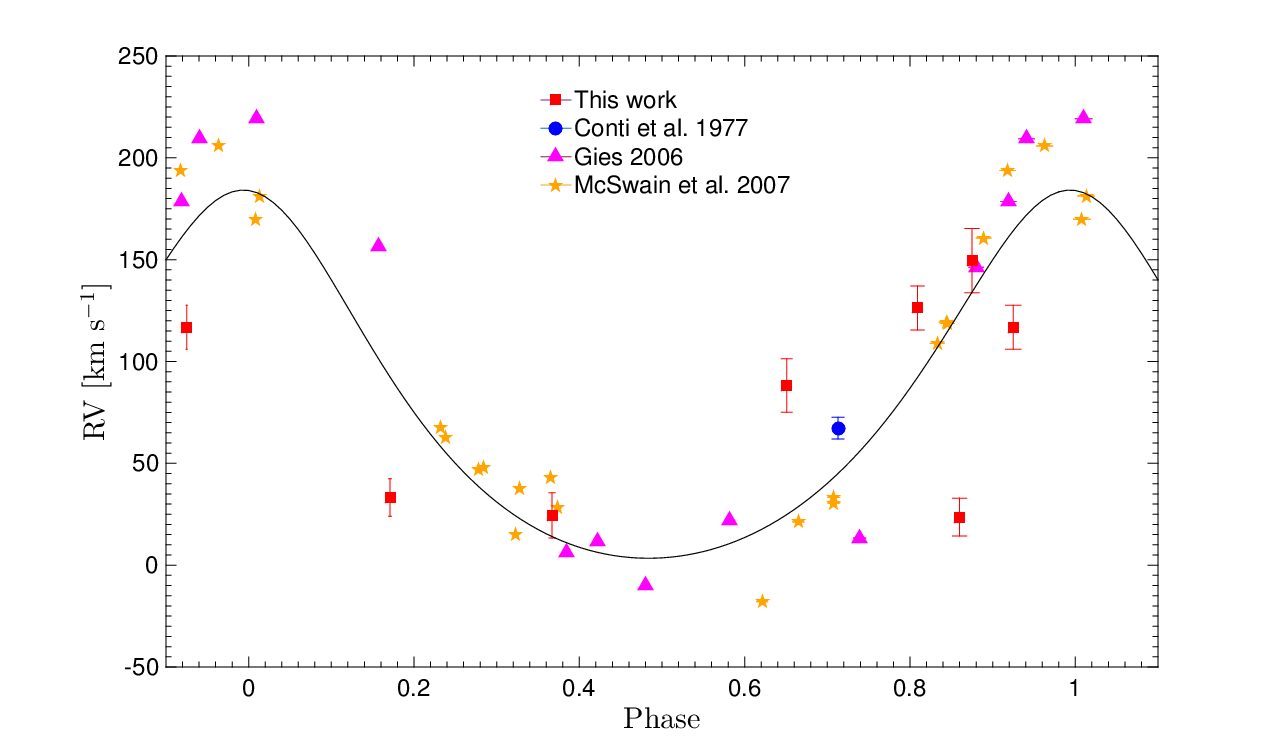}
\end{turn}
\caption{Phase diagram of the RV values of \object{HD 52533} folded with a 22.244\,d period. The best-fit orbital solution (Table \ref{orbSolHD}) derived is shown as a black curve.}
\label{HD52533_LOSP}
\end{figure}

Our RV values, which are significantly variable, indicate a phase shift relative to \citet{mcs07} ephemeris. This leads us to recalculate an orbital solution, adding values from literature (\citealt{con77}; D.R. Gies 2006 -- private communication, although these values were also used in \citealt{mcs07}). To this aim, we first use period search algorithms and found $P(\rm{Fourier})=22.243\pm0.003$\,d ({Fig. \ref{HD52533Fourier}}; again, because of the imperfect sampling with long intervals without observations, the period error is certainly underestimated). {The large amplitude of this peak makes it highly significant (significance level SL$\ll1$\%).} Using this period as first guess, we then derived an orbital solution thanks to the LOSP programme (Table \ref{orbSolHD}). This solution, illustrated in Fig. \ref{HD52533_LOSP}, agrees well with that of \citet{mcs07}, but it would certainly be improved by collecting data with a better phase coverage. 

\subsection{\object{HD 53755}}
\object{HD 53755} is a candidate $\beta$ Cephei according to \citet{sta05}. Our sole UCLES spectrum of \object{HD 53755}, along with the lack of other RV measurements in the literature, does not allow us to investigate the multiplicity of this star. 

\subsection{\object{HD 66811}}
This star is a runaway with a peculiar RV of $-40$ km s$^{-1}$ \citep{car78}. The RVs derived in our two spectra separated by about two years appear compatible within the error bars, but those reported by \citet{gar80} range from --11 to --28 km s$^{-1}$, which leads to a maximum RV difference of 24 km s$^{-1}$ corresponding to 3.5$\sigma$; the RV changes are thus only on the verge of being significant. Therefore, in view of current data, we are forced to keep a presumably single status for this star.


\subsection{\object{HD 69106}}
\citet{fea57} noticed that two Balmer lines (H${\delta}$ and H${\gamma}$) were double on one photographic plate, but no further study of this object was performed since then. In our data, we see no doubling of the lines and detect a maximum RV difference of 26 km s$^{-1}$, corresponding only to a 2.9\,$\sigma$ variation. \citet{fea57} provided additional RV measurements that are all in agreement with our data except one discrepant point at --31 km s$^{-1}$. Without further information, we discard this value as outlier and we tentatively classify the star as presumably single.


\subsection{\object{HD 74920}}
We found no significant RV variation for this star in our data. Hence we classify it as presumably single. 

\subsection{\object{HD 84567}}
\object{HD 84567} is a runaway star candidate \citep[$V_{\rm{pec}}$ = 33.4$_{-13.1}^{+10.9}$ km s$^{-1}$;][]{tez11}. A difference of 36 km s$^{-1}$ is found between our two RV measurements separated by $\sim$ 3 months, corresponding to a 5.5\,$\sigma$ variation. We hence classify this star as RV variable. 

\subsection{\object{HD 90087}}
The two sole RV measurements taken about three years apart (Table \ref{tabJour}) are compatible within the error bars. We therefore consider this star as presumably single. 

\subsection{\object{HD 92554}}
{As we only have two exposures of this star that are separated by less than one hour, and since no additional RV measurements are available in the literature, we cannot assess its multiplicity status. }

\subsection{\object{HD 93521}}
No significant RV variation is found for this star in our data and, while \citet{rau12} reported line-to-line RV variations and RV changes between different observing years, they attributed these changes to non-radial pulsations. Therefore, we concur with their classification of a presumably single object. This also agrees with the fact that no significant variation is found when examining values from \citet{gar80} and ours; {the maximum RV difference of 71 km s$^{-1}$ corresponds only to a 3.6\,$\sigma$ variation in view of the large error bar}s. The runaway status of this star is still uncertain, but no evidence for an accreting compact companion has been found in X-rays \citep{rau12}. 

\subsection{\object{HD 102415}}
Hints of RV variability were reported for this star by \citet{wal11}, \citet{sot14}, and \citet{mar15b}. {In our data, we found no significant RV variation so we classify \object{HD 102415} as presumably single.}


\subsection{\object{HD 117490}}
Some RV variability has previously been reported for this star \citep{mar15b}, but all our RV values are similar within the error bars, hence our choice of a presumably single status. 


\subsection{\object{HD 124979}}
\object{HD 124979} is a runaway star characterised by a peculiar velocity of 74.4$_{-8.3}^{+7.7}$ km s$^{-1}$ \citep{mas98, tez11}. It was suggested to be an SB2 \citep{pen96,bar10,sot14}. However, we do not observe the usual line doubling in our spectra. All recent data (our work and literature -- \citealt{wil11}) are very similar with RVs between $-$70 and $-$90 km s$^{-1}$; old RV measurements \citep{fea63,kil75} differ from these, reaching higher and lower values. However, even after discarding them, the RV changes are found to be significant hence we classify this star as RV variable. Period searches yield no clear periodicity; we therefore need more data to assess the timescale of this variability.

\subsection{\object{HD 149757}}
\object{HD 149757}, best known as $\zeta$\,Oph, is a runaway star \citep[$V_{\rm{pec}}$ = 25$_{-1.1}^{+2.9}$ km s$^{-1}$]{bla61, tez11} as testified by the bow shock in its vicinity (e.g. \citealt{vanB88}). It was claimed that this star was part of a binary and was ejected when its companion (now the pulsar PSR B1929+10) exploded as a supernova about 1 Myr ago \citep{van96,hoo01,tez10}, but \citet{kir15} recently refuted this hypothesis. In any case, it appears to be currently single; we found no significant RV variation in the data. 

\subsection{\object{HD 150574}} 
\citet{gar77,gar83} suggested \object{HD 150574} to be an SB2 based on the observation of double lines in the spectrum, although we do not detect any signature of a secondary in our high-resolution spectra. Furthermore, these new spectra do not reveal any significant RV variations, hence we classify this object as presumably single. 



\subsection{\object{HD 163892}}
\label{subSectHD163}
\begin{figure}
\centering
\begin{turn}{0}
\includegraphics[scale=0.45]{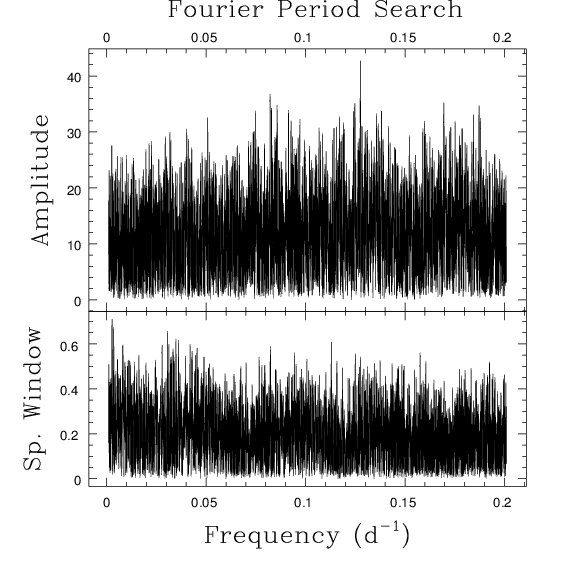}
\end{turn}
\caption{Fourier periodogram derived from the RVs (our work + literature) of \object{HD 163892}. Note the peak at 0.128\,d$^{-1}$.} 
\label{HD163892Fourier}
\end{figure}

\begin{figure}
\centering
\begin{turn}{0}
\includegraphics[scale=0.235]{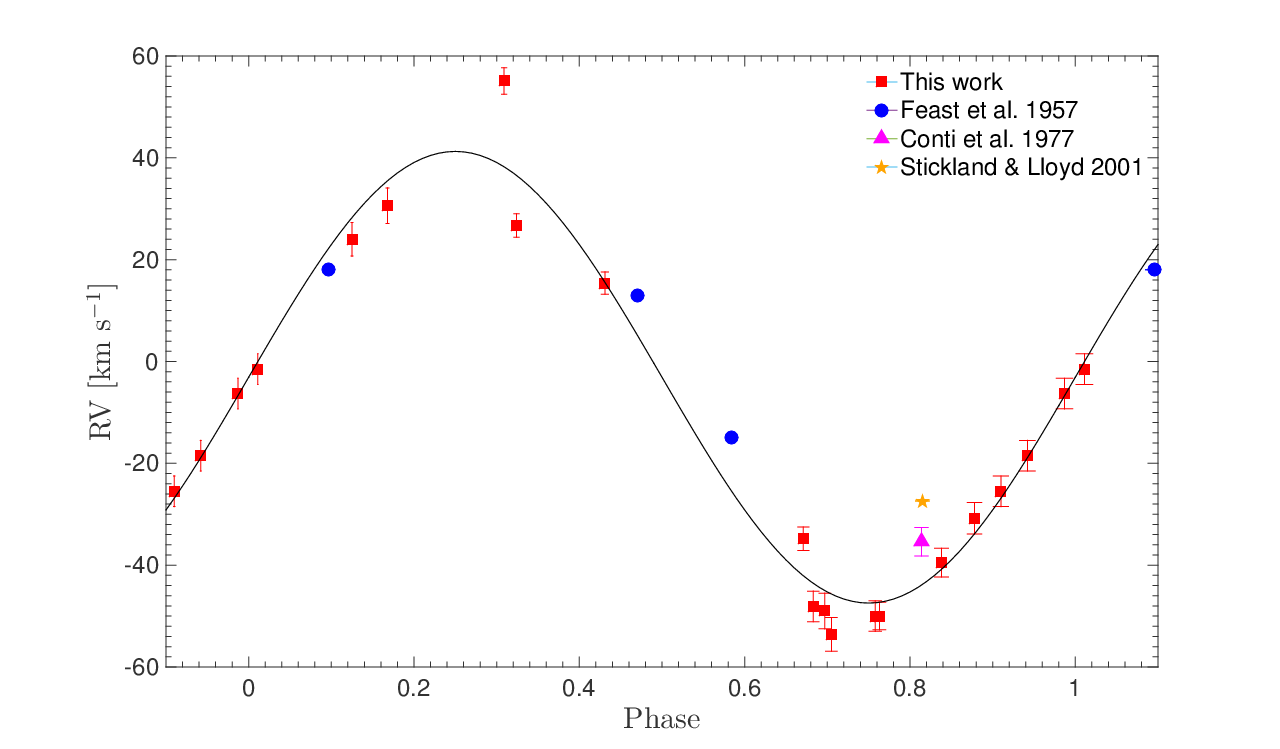}
\end{turn}
\caption{Phase diagram of the RV values of \object{HD 163892} folded with a 7.8348\,d period. The black curve shows the best-fit orbital solution derived with the LOSP programme (Table \ref{orbSolHD}).}
\label{HD163892_LOSP}
\end{figure}

This star is a member of the Sgr OB1 association \citep{hum78}. It has long been recognised as an SB1 system (\citealt{fea57}; \citealt{con77}), as recently confirmed by \citet{sti01} and the OWN Survey (\citealt{bar10}; \citealt{sot14}). An orbital solution for the SB1 was presented by \citet{may14} who found a 7.8\,d period. In the context of this work, we redetermined the RVs of the FEROS spectra used by Mayer et al. and complemented the set of RV values thanks to another FEROS spectrum, three CORALIE, and two FIES spectra (Table \ref{tabJour}). Our RVs appear systematically lower by $\sim$ 10 km s$^{-1}$ than those reported by \citet{may14}. This difference is not surprising as they used Gaussian fits of individual lines to derive their values, which can differ from correlation results by about 10 km s$^{-1}$, depending on the chosen rest wavelength of the fitted lines. It may further be noted that lowering by 10 km s$^{-1}$ the primary systemic velocity $V_{\gamma\rm{,\,pri}}$ given by \citet{may14} yields a value more consistent with the average RV of the Sgr OB1 members that they quote ($\sim$ --10 km s$^{-1}$).  Such a change in the orbital solution, however, implies that two out of the three RV measurements of \citet{fea63} and the \citet{sti01} measurement do not fit well the RV curve anymore; but these RVs were measured on photographic plates, hence have a larger error than ours based on high-resolution \'echelle spectra. We performed a period search on all available RVs and {one peak slightly stands out in the periodograms (e.g. Fig. \ref{HD163892Fourier})} with a period $P(\rm{Fourier})=7.8347\pm0.0003$\,d, although the long gaps without observations lead to the presence of numerous close aliases that increase the actual error on that value. {The large amplitude of this peak makes it highly significant (SL$\ll1$\%). Furthermore, when} folded with this period, RVs yield a clear sinusoidal variation with phase. The best-fit orbital solution was derived with the LOSP programme (Table \ref{orbSolHD} and Fig. \ref{HD163892_LOSP}){; this orbital solution was computed with an eccentricity fixed to zero after it was found to be compatible with this value within the error bars}. This orbital solution is in good agreement with the solution of \citet{may14} within the error bars, except for the primary systemic velocity ($V_{\gamma\rm{,\,pri}}=+2.8$ vs --3.1 km s$^{-1}$). 
 
\subsection{\object{HD 172367}}
We cannot investigate the multiplicity status because we only have one spectrum and no previous investigation of the RVs exists. 

\subsection{\object{HD 175876}}
\citet{tez11} suggested that \object{HD 175876} is a runaway star candidate; their value for the peculiar spatial velocity is 22.2 $_{-8.1}^{+5.9}$ km s$^{-1}$. The analysis of the RVs (ours complemented by literature values; \citealt{kil75}; \citealt{con77}; \citealt{boh78}; \citealt{gar80}) leads us to reject RV variability. We nevertheless investigated the extensive RV set with period search algorithms, but without conclusive results. We thus conclude that this star is presumably single.

\subsection{\object{HD 184915}} 
Weak emission is noticed in both wings of H${\alpha}$, but according to \citet{riv13} this emission is not produced in a circumstellar disc (as in Be stars), but arises from a stellar outflow. We found no significant RV variation for this star in our data. Hence we classify it as presumably single.  

\subsection{\object{HD 188439}}
\object{HD 188439}, or V819\,Cyg, is a runaway star candidate with a peculiar velocity of 61.9 $_{-4.3}^{+3.7}$ km s$^{-1}$ \citep{tez11}. According to \citet{sta05}, it is not a $\beta$ Cephei star even though its photometric variations have been associated with pulsational activity with periods of $\sim$ 0.3775\,d \citep{lyn59} or $\sim$ 0.7137\,d \citep{koe02}. \citet{lyn59} also suggested that \object{HD 188439} might be a very short-period binary in which the stars are partially merged. 

Our RV values do not present significant variations, but the full RV dataset (our work + literature; \citealt{gie86}) indicates the presence of significant changes. In addition, we performed a period search on our data combined to those of \citet{gie86}, excluding an outlier value (--50.1$\pm$4.6 km s$^{-1}$ at HJD = 2,444,803.737), but no significant period could be derived. We therefore classify the star as RV variable, requiring more data to constrain the periodicity.

\subsection{\object{HD 191423}} 
\object{HD 191423}, also known as ``Howarth's star'', is considered as one of the fastest rotators known amongst O stars since its rotation rate is believed to be close to critical ($\Omega$/$\Omega_{\rm{crit}}$ = 0.9; \citealt{how01}). From spectroscopic time series, \citet{mah13} argued that \object{HD 191423} is probably single. Excluding one deviant RV measured on the GOSSS spectrum, the RV differences in our data are not significant (maximum $\Delta$RV of 29 km s$^{-1}$ corresponding to a 0.8\,$\sigma$ difference because of the large error bars). {Furthermore, while no clear periodicity can be identified in the whole RV dataset, a potential variability timescale of about 2.1\,d is apparently detected in the Fourier periodogram when excluding the GOSSS measurement. However, the large RV uncertainties imply that this peak is totally insignificant after comparison with simulated data}. The sampling is not at all adapted to identify such a timescale and the phased RVs do not result in a convincing diagram, hence we keep the presumably single status until further information becomes available.

\subsection{\object{HD 192281}} 
Significant variability of the \ion{He}{II} 4686 double-peaked emission and of the H${\beta}$ absorption line was found by \citet{deb04}: they interpreted them as an effect of co-rotating features present in the wind. \citet{bara93} found RV variations with a period of 5.48\,d compatible with the presence of a low-mass companion. This was challenged by \citet{deb04}. These authors, after showing that this star is not a runaway, derived instead a 9.57\,d period for the RV variability, but with a so small amplitude that it was not considered significant. We do detect a significant variability in the RVs when combining our measurements with those in the literature. However, there is a clear outlier: one measurement by \citet{bara93} is positive while all others are clearly negative. Eliminating it, though, does not modify our conclusions, i.e. there is evidence for variability. However, there is no convincing detection of periodicity. We therefore classify this star as RV variable.


\subsection{\object{HD 198781}}
Our RVs do not display significant variations and no other measurements are available in the literature; we thus classify this star as presumably single. 

 \subsection{\object{HD 203064}} 
\begin{figure}
\centering
\begin{turn}{0}
\includegraphics[scale=0.45]{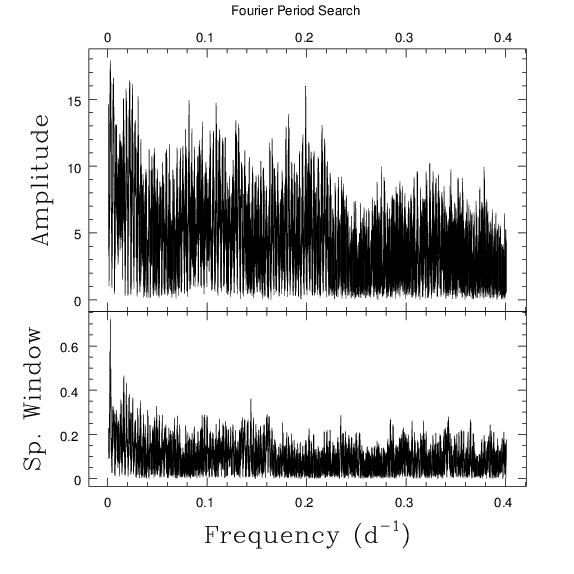}
\end{turn}
\caption{Fourier periodogram derived from the RVs (our work + literature) of \object{HD 203064}. Note the peak near 0.199\,d$^{-1}$.} 
\label{HD203064Fourier}
\end{figure}

\begin{figure}
\centering
\begin{turn}{0}
\includegraphics[scale=0.235]{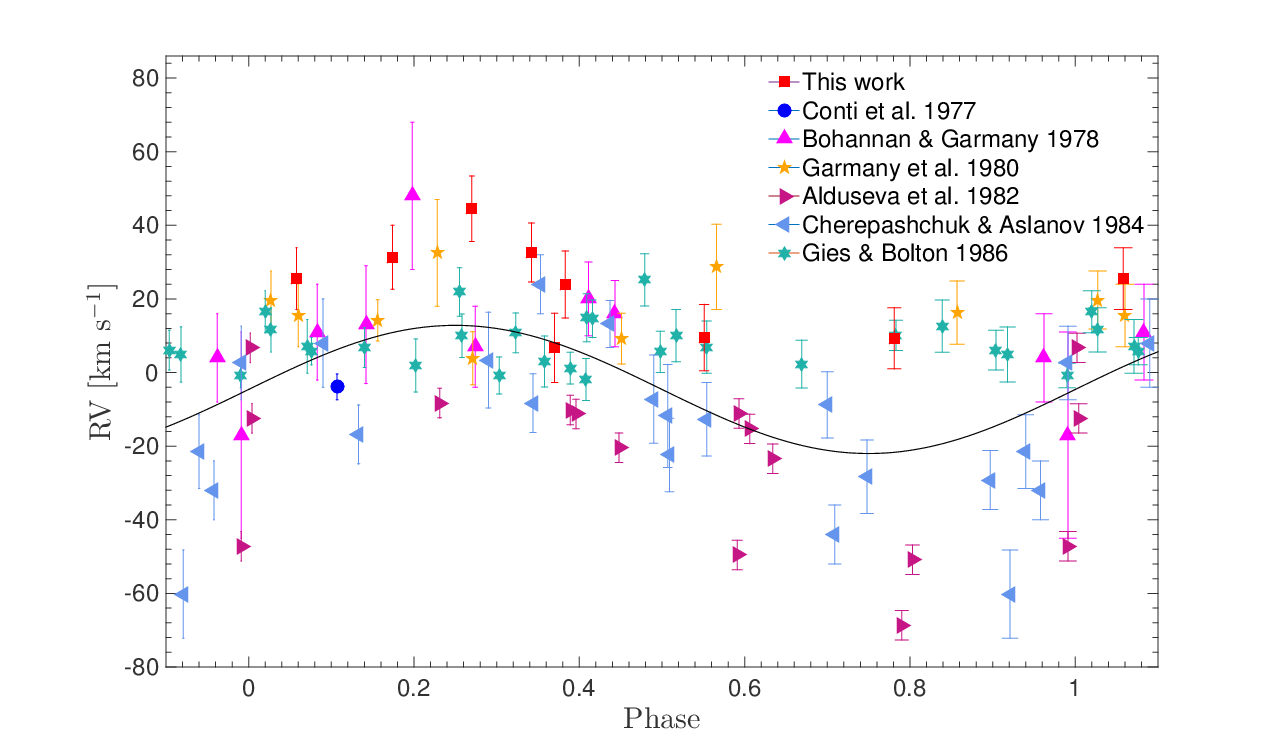}
\end{turn}
\caption{Phase diagram of the RV values of \object{HD 203064} folded with a 5.02292\,d period. The best-fit orbital solution (Table \ref{orbSolHD}) is shown as a black curve.}
\label{HD203064_LOSP}
\end{figure}

\object{HD 203064}, or 68\,Cyg, is a known runaway \citep[$V_{\rm{pec}}$ = 59.4$_{-23.2}^{+12.8}$ km s$^{-1}$;][]{gie86,tez11}. \citet{loz81} further detected a small photometric variability with a 3.34\,d period. It is also a known SB1 presenting discrete absorption components (DACs; \citealt{kap96}). {From the study of \ion{H}{$\delta$},} \citet[see also \citealt{che84}]{ald82} derived an orbital solution with a period of 5.1\,d. The amplitude of our RVs is large ($\Delta$RV $\sim$ 38 km s$^{-1}$), but this corresponds to a 2.9\,$\sigma$ variation only. However, a significant RV variability is detected when literature values (\citealt{con77}; \citealt{boh78}; \citealt{gar80}; {\citealt{ald82}; \citealt{che84};} \citealt{gie86}) are added; we find a maximum $\Delta$RV of {117 km s$^{-1}$ or a 14\,$\sigma$ difference}. {In our period searches, one peak slightly stands out from the Fourier periodogram (Fig. \ref{HD203064Fourier}), with a period $P$=5.02290$\pm$0.00016\,d. Its rather large amplitude, seldom reached in periodograms derived from Monte Carlo simulations, makes it highly significant (SL$\ll1$\%).} We tentatively calculated an orbital solution for this period using the LOSP programme. The derived orbital elements are presented in Table \ref{orbSolHD} and the orbital solution is shown in Fig. \ref{HD203064_LOSP}{; this orbital solution was computed with an eccentricity fixed to zero since it was found to be compatible with this value within the error bars}. {There is an indication for systematically lower RVs derived from \ion{H}{$\delta$}, which leads to the noisy appearance of the RV curve. Our derived orbital period is slightly shorter and our velocity amplitude is smaller than previous solutions \citep{ald82,che84}. However, since the sampling is far from being perfectly adequate for a 5\,d period, new data are required to confirm this tentative solution. }


\subsection{\object{HD 210839}}
\begin{figure}
\centering
\begin{turn}{0}
\includegraphics[scale=0.45]{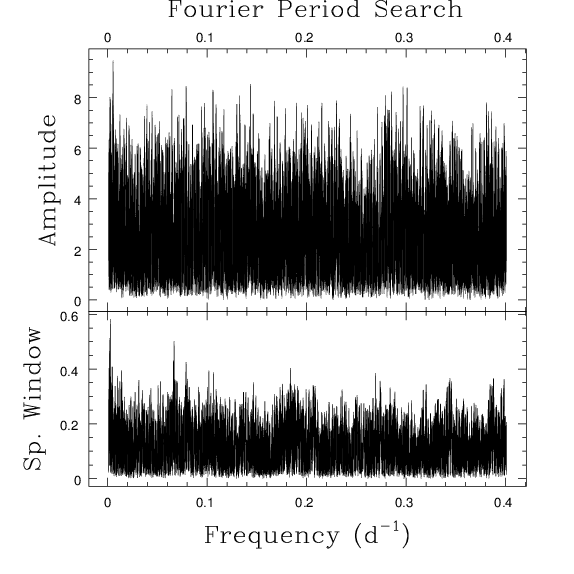}
\end{turn}
\caption{Fourier periodogram derived from the RVs (our work + literature) of \object{HD 210839}. Note the peak near 0.005\,d$^{-1}$} 
\label{HD210839Fourier}
\end{figure}

\begin{figure}
\centering
\begin{turn}{0}
\includegraphics[scale=0.235]{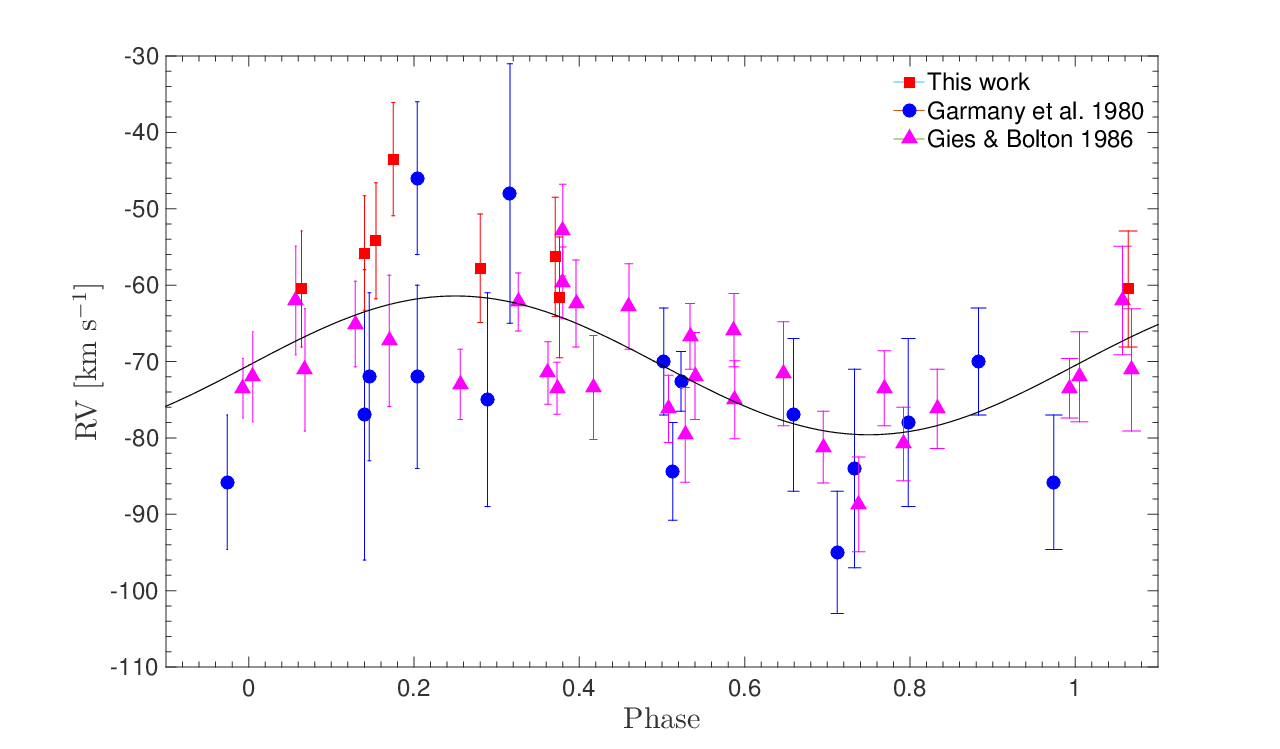}
\end{turn}
\caption{Phase diagram of the RV values of \object{HD 210839} folded with a 186.4\,d period. The best-fit orbital solution (Table \ref{orbSolHD}) is shown as a black curve.}
\label{HD210839_LOSP}
\end{figure}

\object{HD 210839} is a runaway star with a peculiar space velocity of 66.4$_{-2.3}^{+3.7}$ km s$^{-1}$ \citep{tez11}. No significant RV variation is detected in our dataset, but adding the literature values (\citealt{gar80}; \citealt{gie86}) results in a clear detection of RV variability; in this case, we find a maximum $\Delta$RV of 52 km s$^{-1}$ or a $4.7\sigma$ difference. Our period searches yield a small but significant peak in the periodograms at low frequencies, corresponding to $P(\rm{Fourier})$=186.4$\pm$0.2\,d (Fig. \ref{HD210839Fourier}). Although this period is tentative, we used LOSP to calculate a preliminary orbital solution (see Table \ref{orbSolHD} and Fig. \ref{HD210839_LOSP}){; this orbital solution was computed with an eccentricity fixed to zero since it was found to be compatible with this value within the error bars}. It requires new data to be confirmed. 

\subsection{\object{HD 228841}}
\citet{wil11} suggested that it is a runaway star (with a peculiar space velocity of 87 km s$^{-1}$). Following a short-term RV monitoring, \citet{mah13} found no significant variability and thus favoured a single status for \object{HD 228841}. Having a single spectrum of this star, we cannot assess this statement in detail, but our RV measurement agrees well with those of \citet{mah13} for \ion{He}{I}. Therefore, we simply adopt their presumably ``single'' status. 



\begin{sidewaystable}
\centering
\caption{Orbital solutions obtained with the LOSP programme for some confirmed or suspected single-lined spectroscopic binaries in our sample. As a first guess of the period, we used that corresponding to the Fourier periodogram peak, and then refinement was performed, if needed, within LOSP. {These solutions are tentative since the temporal coverage of the period is not satisfactory. Thus, intensive monitoring over at least one full period is needed to ascertain these orbital solutions.}}
\label{orbSolHD}
\begin{tabular}{*{22}{c}}
\hline\hline
Elements                                                &\object{HD 52266}                 & \object{HD 52533}     & \object{HD 163892}    & \object{HD 203064}                 & \object{HD 210839}            \\\hline
$P$ [d]                                         & 75.84$\pm$0.04                &22.244$\pm$0.003               &7.8348$\pm$0.0003      &{5.02292$\pm$0.00016}  &186.4$\pm$0.2\\
$T_{0}$ [HJD--2,400,000]                        &53560.8$\pm$6.9                &51130.7$\pm$1.7                &50842.6$\pm$0.1                &{56284.4$\pm$0.1}              &41191.1$\pm$5.4        \\
$e$                                                     &0.12$\pm$0.07          &0.23$\pm$0.13          &0 (Fixed)                         &{0 (Fixed)     }                       &0 (Fixed)                 \\
$\omega$ [$^{\circ}$]                   &119.0$\pm$33.2         &3.8$\pm$28.9                   &{ ...}                            &{ ...}                                 & ...\\
$K_{\rm{prim}}$ [km s$^{-1}$]           &13.5$\pm$0.8           &90.4$\pm$11.8          &44.4$\pm$2.0                   &{17.4$\pm$3.8}                 &9.1$\pm$1.5                    \\
$V_{\gamma\rm{,\,pri}}$ [km s$^{-1}$]&22.8$\pm$0.6              &73.3$\pm$7.3                   &--3.1$\pm$1.6                  &{--4.6$\pm$2.2}                        &--70.5$\pm$1.2\\
$f$(m) [M$_{\odot}$]                            &0.0188$\pm$0.0035      &1.57$\pm$0.63          &0.0708$\pm$0.0095      &{0.0027$\pm$0.0018}            &0.0145$\pm$0.0074\\
$a\,\sin\,i$ [R$_{\odot}$]                      &20.02$\pm$1.26                 &38.67$\pm$5.20         &6.86$\pm$0.31          &{1.73$\pm$0.38}                        &33.46$\pm$5.70\\
rms [km s$^{-1}$]                               &2.7                                     &36.6                           &6.7                                     &{19.7}                                 &8.1\\
\hline
\end{tabular}
\tablefoot{$T_{0}$ stands for the time of passage at periastron when $e\neq0$ and at conjunction (primary in front) otherwise.}
\end{sidewaystable}

\onecolumn
\begin{sidewaystable}
\caption{Results from high-resolution angular observations of our targets, with Y or N indicating whether close companion(s) have been detected or not ($\rho$ is the angular separation).}
\vspace*{0.3cm}
\label{tab_multiplicity}
\hspace*{-0.5cm}
\begin{tabular}{llllllllll} \hline\hline
     & \multicolumn{8}{c}{Reference}                &        \\
Star & [1] & [2] & [3] & [4] & [5] & [6] & [7] & [8] & Comments\\
\hline 
HD 14434     &   ...   &    ...  &     N    &     ...  &     ...  &      ...    &       ...  &    ...           &   \\   
HD 14442     &   ...   &    ...  &     N    &     ...  &     ...  &      ...    &       ...  &    ...           &   \\   
HD 15137     &   N     &    ...  &     N    &     ...  &     N    &      ...    &       ...  &    N             &   \\   
HD 15642     &   ...   &    ...  &     N    &     ...  &     ...  &      ...    &       ...  &    ...           &  \\    
HD 28446A    &   ...   &    ...  &     ...  &     Y    &     ...  &      ...    &       ...  &    ...           & $\rho$ = 10.32$''$ [4]     \\
HD 41161     &   Y     &    ...  &     N    &     ...  &     N    &      ...    &       ...  &    N             & $\rho$ = 9.8$''$, $P$ $\sim$ 243,000 yrs assuming $d$ = 1.5 kpc [1]\\   
HD 41997     &   ...   &    ...  &     N    &     ...  &     ...  &      ...    &       ...  &    N             &      \\
HD 46056     &   ...   &    ...  &     N    &     ...  &     ...  &      ...    &       ...  &    N             &      \\
HD 46485     &   ...   &    ...  &     N    &     ...  &     ...  &      ...    &       N    &    N             &      \\
HD 52266     &   N     &    ...  &     N    &     ...  &     N    &      ...    &       N    &    N             & 2MASS companion at 7.1$''$ [5]     \\
HD 52533     &   Y     &    ...  &     N    &     ...  &     ...  &      ...    &       Y    &    Y             & A-B with $\rho$ = 2.5$''$ ($P$ $\sim$ 40,000 yrs assuming $d$ = 2.0 kpc) + A-C with $\rho$ = 22.6$''$ [1]  \\ 
             &         &         &          &          &          &             &            &                  & Aa-Ab with $\rho$ = 0.64$''$ ($\Delta H$ $\sim$ 3.50 mag) + A-B with $\rho$ = 2.64$''$ ($\Delta H$ $\sim$ 5.02 mag) + \\
             &         &         &          &          &          &             &            &                  & A-G with $\rho$ = 2.86$''$ ($\Delta H$ $\sim$ 6.37 mag) [7]\\
             &         &         &          &          &          &             &            &                  & Aa-Ab with $\rho$ = 0.6259$''$ ($\Delta F5ND$ $\sim$ 3.812 mag) [8]\\
HD 53755     &   ...   &    Y    &     ...  &     ...  &     ...  &      ...    &       ...  &    ...           & A-B with $\rho$ = 6.42$''$ + A-C with $\rho$ = 35.77$''$ [2] \\
HD 66811     &   N     &    ...  &     N    &     ...  &     N    &      ...    &       N    &    ...           &     \\
HD 74920     &   ...   &    ...  &     ...  &     ...  &     ...  &      ...    &       N    &    ...           &   \\   
HD 90087     &   N     &    ...  &     N    &     ...  &     ...  &      ...    &       ...  &    ...           &  \\    
HD 93521     &   N     &    ...  &     N    &     ...  &     N    &      ...    &       ...  &    N             & \\      
HD 102415    &   ...   &    ...  &     N    &     ...  &     ...  &      ...    &       ...  &    N             & \\     
HD 117490    &   ...   &    ...  &     N    &     ...  &     ...  &      ...    &       ...  &    ...           & \\    
HD 124979    &   N     &    ...  &     N    &     ...  &     ...  &      ...    &       ...  &    ...           &  \\   
HD 149757    &   N     &    ...  &     N    &     ...  &     N    &      N      &       N    &    ...           &  \\    
HD 150574    &   ...   &    ...  &     N    &     ...  &     ...  &      ...    &       ...  &    ...           &  \\  
HD 163892    &   N     &    ...  &     N    &     ...  &     N    &      ...    &       Y    &    N             & A-B with $\rho$ = 2.01$''$ ($\Delta Ks$ $\sim$ 5.31 mag) + A-C with $\rho$ = 2.45$''$ ($\Delta Ks$ $\sim$ 5.87 mag) + \\
             &         &         &          &          &          &             &            &                  & A-D with $\rho$ = 6.45$''$ ($\Delta Ks$ $\sim$ 6.31 mag) + A-E with $\rho$ = 6.50$''$ ($\Delta Ks$ $\sim$ 5.05 mag) [7]\\
             &         &         &          &          &          &             &            &                  & 2MASS companion at 6.4$''$ [5]     \\
HD 175876    &   Y     &    ...  &     N    &     ...  &     N    &      ...    &       N    &    ...           & $\rho$ = 17.0$''$ [1] \\  
HD 184915    &   ...   &    ...  &     N    &     ...  &     ...  &      N      &       ...  &    ...           &      \\
HD 191423    &   ...   &    ...  &     N    &     ...  &     ...  &      ...    &       ...  &    ...           &     \\
HD 192281    &   N     &    ...  &     N    &     ...  &     N    &      ...    &       ...  &    N             &     \\
HD 203064    &   N     &    ...  &     N    &     ...  &     Y    &      ...    &       ...  &    N             &  $\rho$ = 3.84$''$ ($\Delta I$ $\sim$ 9.48 mag) [5] \\ 
HD 210839    &   N     &    ...  &     N    &     ...  &     N    &      ...    &       ...  &    ...           &    \\
HD 228841    &   ...   &    ...  &     N    &     ...  &     ...  &      ...    &       ...  &    ...           &     \\
\hline 
\end{tabular}
\begin{flushleft}
References: [1] \citet{mas98}; [2] \citet{mas04}; [3] \citet{mas09}; [4] \citet{mas11}; [5] \citet{tur08}; [6] \citet{tok10}; [7] \citet{san14}; [8] \citet{ald15}.
\end{flushleft}
\end{sidewaystable}

\twocolumn 

\section{Comparison with literature data}
\label{resB}

Table \ref{tabResComp} compares our stellar parameters with those in the literature. 

\onecolumn
\begin{sidewaystable}
{\scriptsize
\caption[]{Comparison between stellar parameters and abundances derived in this work and those in the literature.}
\label{tabResComp}
\centering
\begin{tabular}{cccclcccccclcccccccc}
\hline\hline 
\multirow{2}{*}{Star}&$v\sin\,i$&$T_{\rm{eff}}$&\multirow{2}{*}{$\log g$}&\multirow{2}{*}{$\log g_{\rm{C}}$}    &\multirow{2}{*}{$y$}&\multirow{2}{*}{$\log \epsilon$(C)}&\multirow{2}{*}{$\log \epsilon$(N)}&\multirow{2}{*}{$\log \epsilon$(O)}&\multirow{2}{*}{[N/C]}&\multirow{2}{*}{[N/O]}&\multirow{2}{*}{Source}\\
                              &[km s$^{-1}$]&     [K]      &                                &                                                    &                               &                                                    &                                                     &                     &       &               &    \\         
\hline \multicolumn{12}{l}{\it Cooler stars (DETAIL/SURFACE)}\\
\multirow{4}{*}{\object{HD 93521}}
&405$\pm$15     &30\,000$\pm$1000       &3.60$\pm$0.10  &3.78$\pm$0.10  &0.166$\pm$0.025                &7.68$\pm$0.12  &8.10$\pm$0.13          &8.33$\pm$0.21  &0.42$\pm$0.21  &--0.23$\pm$0.12        &This work\\
&400$\pm$25     &33\,500$\pm$1500       &...                       &3.80$\pm$0.20       &0.200$\pm$0.050                &...                            &...                                    &...                            &...                            &...                            &\citet{len91}\\
&435$\pm$20     &33\,500$\pm$1500       &...                       &...                         &0.180$\pm$0.030                &...                            &...                                    &...                            &...                            &...                            &\citet{how01}\\
&390$\pm$10     &30\,900$\pm$700        &3.67$\pm$0.12  &...                    &0.178$\pm$0.020        &7.56                   &7.97                           &8.24                   &0.41                   &{ --0.27}         &\citet{rau12}\\\hline

\multirow{2}{*}{\object{HD 102415}}
&357$\pm$15     &32\,900$\pm$1000       &4.10$\pm$0.10 &4.19$\pm$0.10   &0.158$\pm$0.025                &$<$7.54                &8.16$\pm$0.13          &8.22$\pm$0.21  &$>$0.62                 &--0.06$\pm$0.12        &This work\\
&376$\pm$10     &31\,000$\pm$1500       &3.50$\pm$0.15 &3.70                    &0.174$\pm$0.068                &$\le$7.78              &8.88$_{-0.21}^{+0.24}$ &$\le$8.48              &$\ge$1.10              &$\ge$0.40            &\citet{mar15b}\\\hline

\multirow{11}{*}{\object{HD 149757}}
&378$\pm$15     &31\,500$\pm$1000       &3.87$\pm$0.10 &3.99$\pm$0.10   &0.135$\pm$0.025                &8.07$\pm$0.12  &7.85$\pm$0.13          &8.37$\pm$0.21  &--0.22$\pm$0.21        &--0.52$\pm$0.12        &This work\\
&400                    &32\,500                        &3.70              &3.85                 &0.160                          &...                            &...                                    &...                            &...                            &...                           &\citet{pul96}\\
&399$\pm$20     &34\,300$\pm$1500       &...                       &...                         &0.200$\pm$0.030                &...                            &...                                    &...                            &...                            &...                            &\citet{how01}\\
&400                    &32\,500$\pm$1500       &3.50$\pm$0.10  &3.77                   &0.160$\pm$0.030                &...                            &...                                    &...                            &...                            &...                            &\citet{her02}\\
&400                    &32\,000$\pm$1000       &3.65$\pm$0.10 &3.85$_{-0.08}^{+0.10}$&0.145$\pm$0.022  &...                            &...                                    &...                            &...                            &...                            &\citet{rep04}\\
&340$\pm$25     &26\,400$\pm$700        &3.80$\pm$0.09 &4.05$\pm$0.07   &...                                    &...                            &...                                    &...                            &...                            &...                            &\citet{fre05}\\        
($\zeta$\,Oph)
&400                    &32\,100$\pm$700        &3.62              &3.83$_{-0.05}^{+0.16}$&0.099$_{-0.016}^{+0.032}$&...                &...                                    &...                            &...                            &...                            &\citet{mok05}\\        
&...                    &33\,500$\pm$1700       &...                       &3.85$\pm$0.10        &0.145                          &...                            &...                                    &...                            &...                            &...                            &\citet{rep05}$^b$\\    
&400$\pm$20     &34\,000$\pm$1000       &3.70$\pm$0.10 &...                             &0.110$\pm$0.028                &7.86$\pm$0.30  &8.34$\pm$0.30           &8.69$\pm$0.30  &0.48                   &--0.35                 &\citet{vil05}\\
&400                    &32\,000$\pm$2000       &3.60$\pm$0.20 &3.80                    &...                                    &...                            &...                                     &...                            &...                            &...                            &\citet{marc09}\\
&400$\pm$10     &31\,000$\pm$1000       &3.60$\pm$0.15  &...                    &...                                    &...                            &...                                    &...                            &...                            &...                           &\citet{mar15a}\\\hline

\multirow{3}{*}{\object{HD 184915}}
&252$\pm$15     &27\,800$\pm$1000       &3.70$\pm$0.10 &3.77$\pm$0.10   &0.183$\pm$0.025                &$<$8.18                &8.46$\pm$0.13          &8.62$\pm$0.21  &$>$0.28                &--0.16        &This work\\&270                    &26\,800                        &3.56              &...                          &0.160$\pm$0.011                &...                            &...                                    &...                            &...                            &...                           &\citet{lyu04}$^a$\\
($\kappa$\,Aql)
&229$\pm$13     &27\,100$\pm$500        &3.49$\pm$0.05  &3.53$\pm$0.06  &...                                    &...                            &...                                    &...                            &...                            &...                           &\citet{fre05}\\  
&249$\pm$7      &26\,700$\pm$750        &3.59$\pm$0.07  &...                    &...                                    &...                            &...                                    &...                            &...                            &...                           &\citet{hua08}\\\hline

\multirow{2}{*}{\object{HD 198781}}
&222$\pm$15     &29\,100$\pm$1000       &3.90$\pm$0.10 &3.94$\pm$0.10   &0.230$\pm$0.025                &$<$8.09                &8.62$\pm$0.13          &8.78$\pm$0.21  &$>$0.53                &--0.16$\pm$0.12        &This work\\
&224                    &24\,400                        &3.50             &...                           &0.148$\pm$0.011                &...                            &...                                    &...                            &...                            &...                           &\citet{lyu04}$^a$\\         

\hline \multicolumn{12}{l}{\it Hotter stars (CMFGEN)}\\ 

\multirow{7}{*}{\object{HD 13268}}
&301$\pm$15     &32\,500$\pm$1500       &3.42$\pm$0.15  &3.55$\pm$0.15  &0.206$\pm$0.030                &$\le$7.50              &8.61$\pm$0.34          &8.10$\pm$0.21  &$\ge$1.11              &0.51$\pm$0.40  &This work\\
&300                    &36\,000$\pm$2000       &3.70$\pm$0.30 &...                             &...                                    &...                            &...                                    &...                             &...                            &...                           &\citet{ken96}\\
&320                    &35\,000                        &3.30              &3.50                 &0.200                          &...                            &...                                    &...                            &...                            &...                           &\citet{pul96}\\  
&320                    &35\,000$\pm$1500 &3.30$\pm$0.10 &3.42                  &$\ge$0.200                     &...                            &...                                    &...                            &...                            &...                           &\citet{her92}\\
&300                    &33\,000$\pm$1000 &3.25$\pm$0.10 &3.48$_{-0.08}^{+0.11}$        &0.200$\pm$0.019        &...                            &...                                    &...                            &...                            &...                           &\citet{rep04}\\
&...                    &33\,000$\pm$1650       &...                       &3.48$\pm$0.10        &0.200                          &...                            &...                                    &...                            &...                            &...                         &\citet{rep05}$^b$\\
&310$\pm$10     &32\,000$\pm$1500 &3.50$\pm$0.15  &3.63                 &0.167$\pm$0.069                &$\le$7.70              &8.70$_{-0.17}^{+0.24}$ &8.49$_{-0.20}^{+0.35}$&$\ge$1.00               &0.21                 &\citet{mar15b}\\\hline

\multirow{2}{*}{\object{HD 14434}}
&408$\pm$15     &40\,000$\pm$1500       &3.89$\pm$0.15  &4.03$\pm$0.15  &0.103$\pm$0.030                &7.96$\pm$0.27  &8.81$\pm$0.34          &$\le$8.10              &0.85$\pm$0.43  &$\ge$0.71              &This work\\
&380                    &43\,000$\pm$2000       &3.80$\pm$0.20 &...                             &...                                    &...                            &...                                    &...                             &...                            &...                         &\citet{ken96}\\    
\hline

\multirow{2}{*}{\object{HD 14442}}
&285$\pm$15     &39\,200$\pm$1500       &3.69$\pm$0.15  &3.78$\pm$0.15  &0.097$\pm$0.030                &7.10$\pm$0.27  &8.61$\pm$0.34          &$\le$8.10              &1.51$\pm$0.43  &$\ge$0.51    &This work\\
&260                    &43\,000$\pm$2000       &3.60$\pm$0.20 &...                             &...                                    &...                            &...                                    &...                            &...                            &...                  &\citet{ken96}\\   
\hline

\multirow{3}{*}{\object{HD 15137}}
&267$\pm$15     &29\,500$\pm$1500       &3.18$\pm$0.15  &3.31$\pm$0.15  &0.112$\pm$0.030                &7.63$\pm$0.27  &8.27$\pm$0.34          &$\le$8.30              &0.64$\pm$0.43  &$\ge$--0.03    &This work\\
&234$\pm$10     &29\,700$\pm$700        &3.50$\pm$0.10 &...                             &...                                    &...                            &...                                    &...                            &...                            &...                  &\citet{mcs07}\\
&258$\pm$20     &29\,700$\pm$1700 &3.50$\pm$0.25 &...                           &...                                    &...                            &...                                    &...                            &...                            &...                  &\citet{mcs10}\\   \hline

\multirow{2}{*}{\object{HD 46056}}
&350$\pm$15     &34\,500$\pm$1500       &3.90$\pm$0.15  &4.00$\pm$0.15  &0.088$\pm$0.030                &8.34$\pm$0.27  &7.78$\pm$0.34          &8.32$\pm$0.21  &--0.56$\pm$0.43        &--0.54$\pm$0.40        &This work\\
&330$\pm$10     &34\,500$\pm$1000       &3.75$\pm$0.15  &...                    &...                                    &8.28$\pm$0.07  &7.78$\pm$0.14          &8.45$\pm$0.28  &--0.50                 &--0.67               &\citet{mar15a}\\\hline

\multirow{2}{*}{\object{HD 46485}}
&315$\pm$15     &37\,000$\pm$1500       &4.00$\pm$0.15  &4.08$\pm$0.15  &0.076$\pm$0.030                &8.46$\pm$0.27  &7.95$\pm$0.34          &8.72$\pm$0.21  &--0.51$\pm$0.43        &--0.77$\pm$0.40        &This work\\
&300$\pm$10     &36\,000$\pm$1000       &3.75$\pm$0.15  &...                    &...                                    &8.43$_{-0.08}^{+0.11}$ &7.95$\pm$0.10  &8.64$_{-0.20}^{+0.22}$&--0.48          &--0.69                 &\citet{mar15a}\\
\hline 
\end{tabular}
}
\end{sidewaystable}

\begin{sidewaystable}
{\scriptsize
\addtocounter{table}{-1}
\caption[]{Continued.}
\label{tabResComp1}
\centering
\begin{tabular}{cccclcccccclcccccccc}
\hline\hline 
\multirow{2}{*}{Star}&$v\sin\,i$&$T_{\rm{eff}}$&\multirow{2}{*}{$\log g$}&\multirow{2}{*}{$\log g_{\rm{C}}$}    &\multirow{2}{*}{$y$}&\multirow{2}{*}{$\log \epsilon$(C)}&\multirow{2}{*}{$\log \epsilon$(N)}&\multirow{2}{*}{$\log \epsilon$(O)}&\multirow{2}{*}{[N/C]}&\multirow{2}{*}{[N/O]}&\multirow{2}{*}{Source}\\
                              &[km s$^{-1}$]&     [K]      &                                &                                                    &                               &                                                    &                                                     &                     &               &       &    \\         
\hline
\multirow{5}{*}{\object{HD 66811}}
&225$\pm$15     &41\,000$\pm$1500       &3.55$\pm$0.15  &3.62$\pm$0.15  &0.148$\pm$0.030                        &$\le$7.00                      &8.94$\pm$0.34          &8.20$\pm$0.21  &$\ge$1.94              & { 0.74}      &This work\\       
&220                    &42\,000                         &3.50             &3.60                 &0.107                                  &...                                    &...                                    &...                            &...                            &...                  &\citet{pul96}\\
&220                    &39\,000$\pm$1500       &3.55$\pm$0.10   &3.59$\pm$0.09 &0.167$\pm$0.021                        &...                                    &...                                    &...                            &...                            &...                  &\citet{rep04}\\
($\zeta$\,Pup)
&...                    &39\,000$\pm$1950       &...                       &3.59$\pm$0.10        &0.145                                  &...                                    &...                                    &...                            &...                            &...                  &\citet{rep05}$^b$\\
&210                    &40\,000$\pm$1000       &...                       &3.64$\pm$0.10        &0.140                                  &6.60$\pm$0.25          &9.10$\pm$0.17          &8.13$\pm$0.30  &2.50                   &0.97        &\citet{bou12}\\    
&210$\pm$10     &40\,000$\pm$1000       &3.64$\pm$0.15  &...                    &...                                            &6.60$\pm$0.22          &9.10$\pm$0.17          &8.13$\pm$0.29  &2.50                   &0.97    &\citet{mar15a}\\\hline                                                  
\multirow{2}{*}{\object{HD 69106}}
&306$\pm$15     &29\,500$\pm$1500       &3.45$\pm$0.15  &3.58$\pm$0.15  &0.091$\pm$0.030                        &7.88$\pm$0.27          &7.74$\pm$0.34          &8.47$\pm$0.21  &--0.14$\pm$0.43         &--0.73$\pm$0.40&This work\\
&320$\pm$10     &29\,000$\pm$1000       &3.40$\pm$0.15  &...                    &...                                            &7.60$_{-0.11}^{+0.22}$ &$\le$8.00                      &8.40$_{-0.14}^{+0.21}$&$\le$0.40               &$\le$--0.40    &\citet{mar15a}\\\hline

\multirow{2}{*}{\object{HD 117490}}
&361$\pm$15     &30\,000$\pm$1500       &3.55$\pm$0.15  &3.70$\pm$0.15  &0.141$\pm$0.030                        &$\le$7.39                      &8.50$\pm$0.34          &8.15$\pm$0.21  &$\ge$1.11              &0.35$\pm$0.40 &This work\\
&375$\pm$10     &30\,500$\pm$1500       &3.50$\pm$0.15  &3.66                   &0.138$_{-0.037}^{+0.067}$      &$\le$7.48                      &8.88$_{-0.15}^{+0.17}$ &8.40$_{-0.35}^{+0.45}$&$\ge$1.40               &0.48     &\citet{mar15b}\\\hline

\multirow{2}{*}{\object{HD 150574}}
&233$\pm$15     &31\,500$\pm$1500       &3.32$\pm$0.15  &3.41$\pm$0.15  &0.172$\pm$0.030                        &7.48$\pm$0.27          &$\ge$9.08                      &$\ge$8.93              &$\ge$1.60                 &{ ...} &This work\\
&240$\pm$10     &31\,000$\pm$1500       &3.40$\pm$0.15  &3.49                   &0.187$\pm$0.040                        &$\le$7.70                      &$\ge$9.00                      &8.78$_{-0.17}^{+0.21}$&{ $\ge$1.30}              &{ $\ge$0.22}     &\citet{mar15b}\\ \hline

\multirow{9}{*}{\object{HD 191423}}
&420$\pm$15     &30\,600$\pm$1500       &3.33$\pm$0.15  &3.57$\pm$0.15  &0.134$\pm$0.030                        &$\le$7.24                      &8.33$\pm$0.34          &$\le$8.33              & $\ge$1.09               & $\ge$0.00     &This work\\
&450                    &34\,000$\pm$1500 &3.40$\pm$0.10 &3.68                  &0.200$_{-0.030}^{+0.050}$      &...                                    &...                                    &...                            &...                            &...                    &\citet{her92}\\
&450                    &34\,000                        &3.40              &3.70                 &0.200                                  &...                                    &...                                    &...                            &...                            &...                     &\citet{pul96}\\
&435                    &34\,300$_{-800}^{+700}$&...               &...                         &0.190$\pm$0.030                        &...                                    &...                                    &...                            &...                            &...                    &\citet{how01}\\
&450                    &35\,000$\pm$1000       &3.40$\pm$0.10   &...                   &0.120$\pm$0.030                        &7.57$\pm$0.24          &8.36$\pm$0.17          &8.23$\pm$0.48  &0.79                   &0.13   &\citet{vil02}$^c$\\    
&400                    &32\,500$\pm$1000       &3.35$\pm$0.10   &3.60$_{-0.08}^{+0.11}$&0.167$\pm$0.021                &...                                    &...                                    &...                            &...                            &...                    &\citet{rep04}\\
&...                    &32\,000$\pm$1600       &...                       &3.56$\pm$0.10        &0.167                                  &...                                    &...                                    &...                            &...                            &...                    &\citet{rep05}$^b$\\    
&410                    &30\,600$\pm$1000       &3.50$\pm$0.10   &3.67                  &...                                            &...                                    &8.73$\pm$0.20          &...                            &...                            &...                   &\citet{mah15}\\
&445$\pm$10     &31\,500$\pm$1500       &3.50$\pm$0.15  &3.72                   &0.200$\pm$0.051                        &...                                    &$\ge$8.70                      &...                            &...                            &... &\citet{mar15b}\\\hline

\multirow{1}{*}{\object{HD 192281}}
&276$\pm$15     &39\,000$\pm$1500       &3.64$\pm$0.15  &3.73$\pm$0.15  &0.103$\pm$0.030                        &8.00$\pm$0.27          &8.76$\pm$0.34          &8.05$\pm$0.21  &0.76$\pm$0.43         & 0.71$\pm$0.40 &This work\\
(V819\,Cyg)
&245$\pm$10     &39\,000$\pm$1000       &3.65$\pm$0.15  &...                    &...                                            &8.11$_{-0.33}^{+0.37}$ &8.92$_{-0.22}^{+0.41}$ &8.15$\pm$0.09  &0.81                   &0.77   &\citet{mar15a}\\\hline

\multirow{5}{*}{\object{HD 203064}}
&298$\pm$15     &35\,000$\pm$1500&3.73$\pm$0.15  &3.82$\pm$0.15 &0.076$\pm$0.030                        &7.92$\pm$0.27          &8.23$\pm$0.34          &8.46$\pm$0.21  &0.31$\pm$0.43         & --0.23$\pm$0.40&This work\\
&315                    &37\,500                        &3.50              &3.65                 &0.123                                  &...                                    &...                                    &...                            &...                            &...                     &\citet{pul96}\\
&315                    &37\,500$\pm$1000 &3.50$\pm$0.10 &3.62                  &0.120$\pm$0.030                        &...                                    &...                                    &...                            &...                            &...                   &\citet{her92}\\
(68\,Cyg)
&300                    &34\,500$\pm$1000&3.50$\pm$0.10   &3.60$_{-0.08}^{+0.09}$&0.091$\pm$0.017               &...                                    &...                                    &...                            &...                            &...                   &\citet{rep04}\\
&...                    &34\,500$\pm$1700       &...                       &3.60$\pm$0.10        &0.167                                  &...                                    &...                                    &...                            &...                            &...                   &\citet{rep05}$^b$\\      
&300$\pm$10     &34\,000$\pm$1000       &3.60$\pm$0.15  &...                    &...                                            &8.20$\pm$0.11          &8.20$\pm$0.19          &...                            &0.00                   &...  &\citet{mar15a}\\\hline 

\multirow{5}{*}{\object{HD 210839}}
&214$\pm$15     &36\,000$\pm$1500       &3.50$\pm$0.15  &3.56$\pm$0.15  &0.113$\pm$0.030                        &7.83$\pm$0.27          &8.74$\pm$0.34          &8.13$\pm$0.21  & 0.91$\pm$0.43   & 0.61$\pm$0.40 &This work\\
&100                    &38\,000                         &3.60             &3.65                 &0.091                                  &...                                    &...                                    &...                            &...                            &...                   &\citet{pul96}\\  
&250                    &37\,000$\pm$1500       &3.55$\pm$0.10   &...                   &0.250$\pm$0.030                        &...                                    &...                                    &...                            &...                            &...                   &\citet{her00}\\($\lambda$Cep)    &200                    &36\,000$\pm$1500       &3.55$\pm$0.10   &3.58$\pm$0.09        &0.091$\pm$0.017                        &...                                    &...                                    &...                            &...                            &...                   &\citet{rep04}\\  

&210                    &36\,000$\pm$1000       &...                       &3.54$\pm$0.10        &0.107                                  &8.22$\pm$0.21          &8.70$\pm$0.15          &8.48$\pm$0.14  &0.48                   &0.22           &\citet{bou12}\\        
&210$\pm$10     &36\,000$\pm$1000       &3.50$\pm$0.15  &...                    &...                                            &7.78$\pm$0.15          &8.78$_{-0.09}^{+0.14}$ &8.40$\pm$0.35  &1.00                   &0.38           &\citet{mar15a}\\\hline

\multirow{2}{*}{\object{HD 228841}}
&305$\pm$15     &34\,000$\pm$1500       &3.50$\pm$0.15  &3.62$\pm$0.15  &0.112$\pm$0.030                        &7.48$\pm$0.27          &8.74$\pm$0.34          &8.67$\pm$0.21  &1.26$\pm$0.43  & 0.07$\pm$0.40&This work\\&317                    &34\,500$\pm$1000       &3.50$\pm$0.10   &3.62                  &...                                            &...                                    &8.73$\pm$0.22          &...                            &...                            &...                   &\citet{mah15}\\

\hline
\end{tabular}
\tablefoot{$^a$: The value of $y$ corresponding to $\xi$ derived from \ion{He}{I} lines is chosen. $^b$: Only values from the infrared analysis are indicated. $^c$: Values not corrected for Galactic chemical gradient.}
} 
\end{sidewaystable}

\twocolumn 

\section{Comparison with CMFGEN spectra}
\label{secCompaCMFGENFig}

This appendix provides a comparison between the observations of the hotter stars and their best-fit CMFGEN models. {Lines useful for the abundance derivations are indicated (see Sect. \ref{subSecMethCMFGEN} for details on the fitting procedure and Table \ref{tabLinesCNO} for the actual list of lines used for each star). Finally, in the caption we mention the remaining fitting imperfections for each star. In this context, we recall that the wind parameters were not derived, explaining why wind-sensitive lines (e.g. \ion{N}{III}\,4634--4643, \ion{He}{II}\,4686) may not be perfectly fitted.}

\onecolumn
\begin{figure}
\begin{center}
\begin{turn}{0}
\includegraphics[scale=0.9]{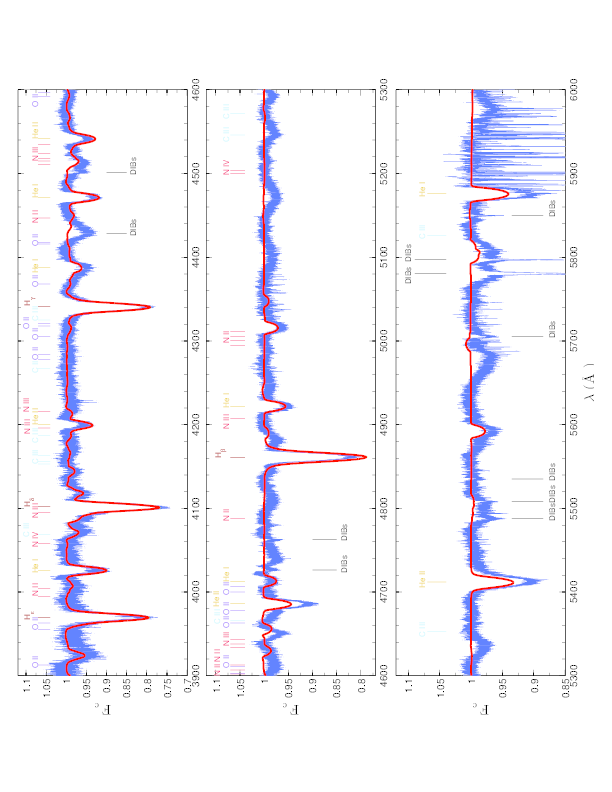}
\end{turn}
\caption{Best-fit CMFGEN model (red) compared to the {observed} spectrum {(blue)} of \object{BD +34$^{\circ}$1058} {(O8nn; $v\sin\,i$ = 424 km s$^{-1}$)}. Diagnostic lines are indicated. {The \ion{He}{I}\,5876 line appears too strong compared to the best-fit model, while the fit of \ion{He}{II}\,4686 and \ion{He}{II}\,5412 is imperfect, probably because of normalisation problems.}}
\label{cmfBD}
\end{center}
\end{figure} 

\begin{figure}
\begin{center}
\begin{turn}{0}
\includegraphics[scale=0.9]{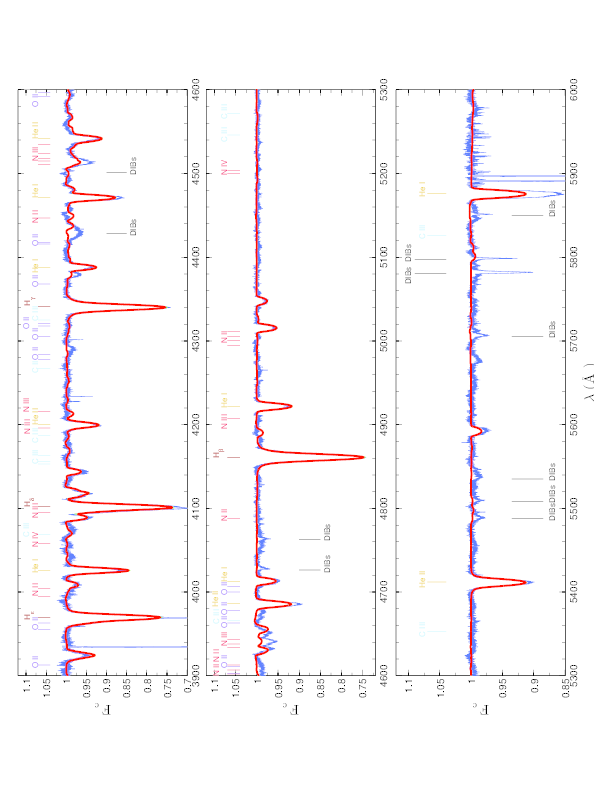}
\end{turn}
\caption{Same as Fig. \ref{cmfBD}, but for \object{HD 13268} {(ON8.5IIIn; $v\sin\,i$ = 301 km s$^{-1}$)}. {\ion{He}{I} lines are generally well fitted, except \ion{He}{I}\,5876. Some nitrogen lines, e.g. \ion{N}{III}\,4634--4643, are not perfectly fitted.}}
\end{center}
\end{figure} 

\begin{figure}
\begin{center}
\begin{turn}{0}
\includegraphics[scale=0.9]{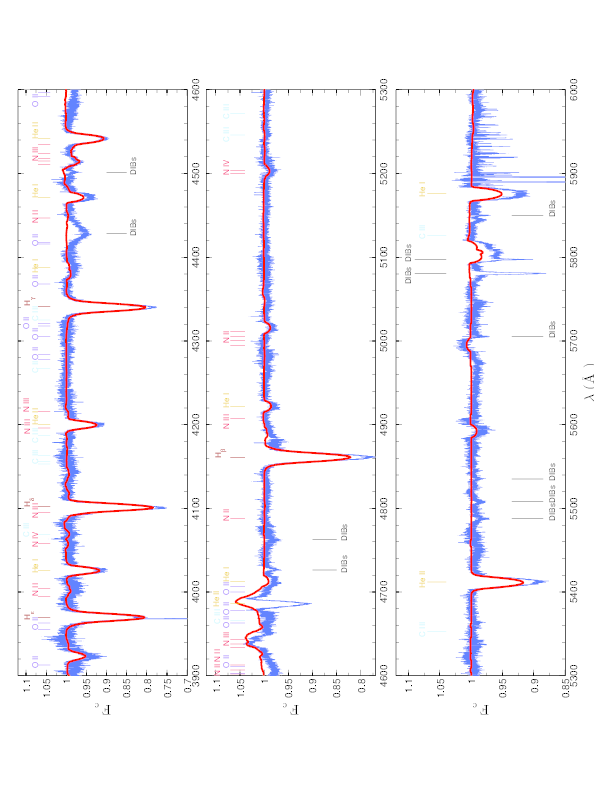}
\end{turn}
\caption{Same as Fig. \ref{cmfBD}, but for \object{HD 14434} {(O5.5Vnn((f))p; $v\sin\,i$ = 408 km s$^{-1}$)}. {\ion{He}{I} lines are generally well fitted, except \ion{He}{I}\,5876; some local normalisation problems explain the apparently imperfect fit to \ion{H}{$\beta$} and \ion{He}{II}\,5412. Because wind parameters were not derived, the wind-sensitive line \ion{He}{II}\,4686 is not well reproduced as too much emission is seen for the best-fit model.}}
\end{center}
\end{figure} 

\begin{figure}
\begin{center}
\begin{turn}{0}
\includegraphics[scale=0.9]{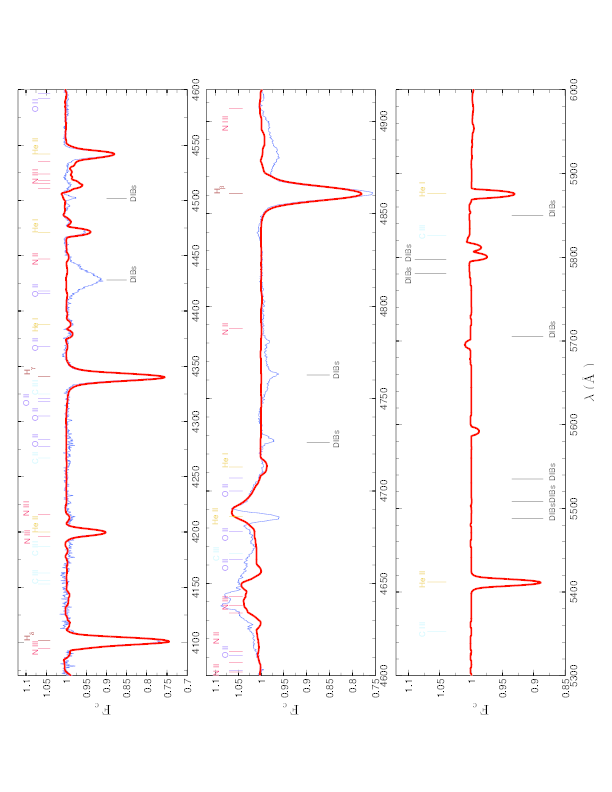}
\end{turn}
\caption{Same as Fig. \ref{cmfBD}, but for \object{HD 14442} {(O5n(f)p; $v\sin\,i$ = 285 km s$^{-1}$)}. {Because wind parameters were not derived, wind-sensitive lines are not well reproduced: too much emission is seen in \ion{He}{II}\,4686 for the best-fit model, but too little emission for the neighbouring \ion{N}{III}\,4634--4643 lines.}}
\end{center}
\end{figure} 

\begin{figure}
\begin{center}
\begin{turn}{0}
\includegraphics[scale=0.9]{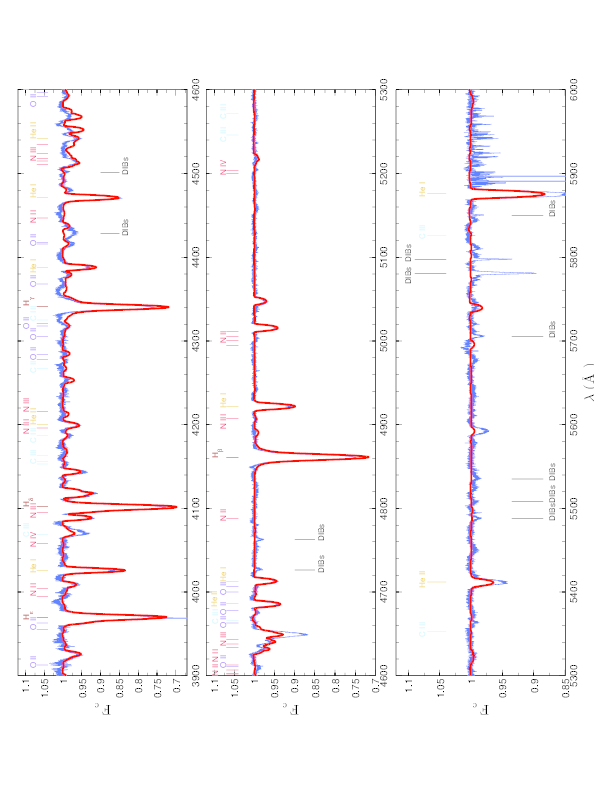}
\end{turn}
\caption{Same as Fig. \ref{cmfBD}, but for \object{HD 15137} {(O9.5II-IIIn; $v\sin\,i$ = 267 km s$^{-1}$)}. {The wings of \ion{H}{$\gamma$} are affected by a normalisation problem. \ion{He}{II}\,5412 and \ion{He}{I}\,5876 appear somewhat too weak in the model.}}
\end{center}
\end{figure} 

\begin{figure}
\begin{center}
\begin{turn}{0}
\includegraphics[scale=0.9]{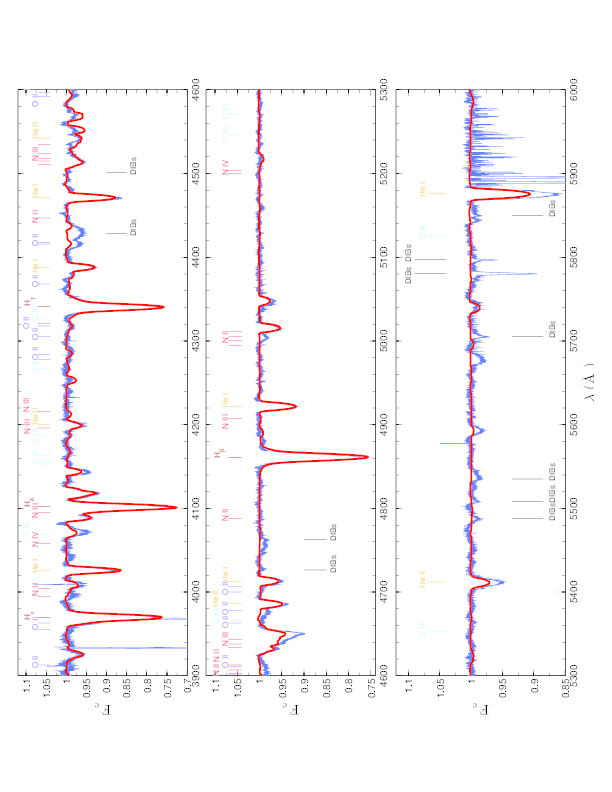}
\end{turn}
\caption{Same as Fig. \ref{cmfBD}, but for \object{HD 15642} {(O9.5II-IIIn; $v\sin\,i$ = 335 km s$^{-1}$)}. {While other He lines appear well fitted, \ion{He}{II}\,5412 and \ion{He}{I}\,5876 are too weak in the model.}}
\end{center}
\end{figure} 

\begin{figure}
\begin{center}
\begin{turn}{0}
\includegraphics[scale=0.9]{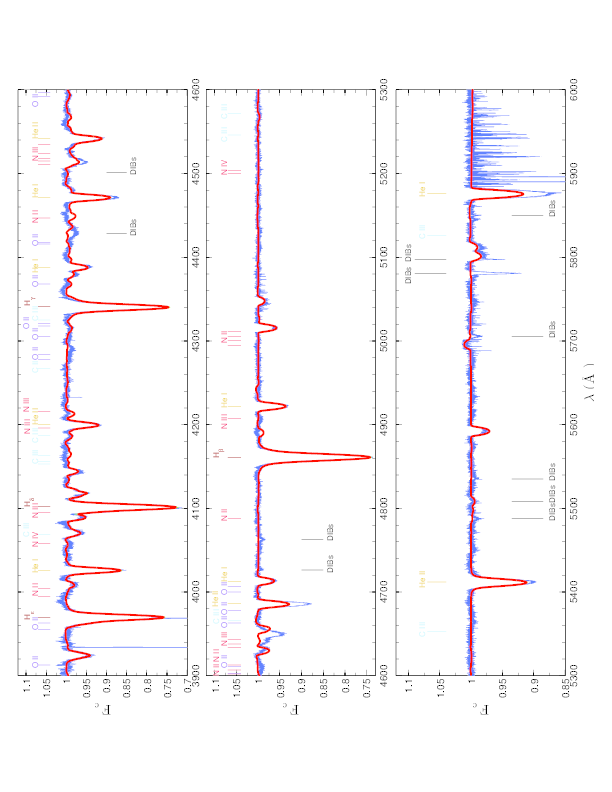}
\end{turn}
\caption{Same as Fig. \ref{cmfBD}, but for \object{HD 41161} {(O8Vn; $v\sin\,i$ = 303 km s$^{-1}$)}. {While other He lines appear well fitted, \ion{He}{II}\,4686 and \ion{He}{I}\,5876 are too weak in the model.}}
\end{center}
\end{figure} 

\begin{figure}
\begin{center}
\begin{turn}{0}
\includegraphics[scale=0.9]{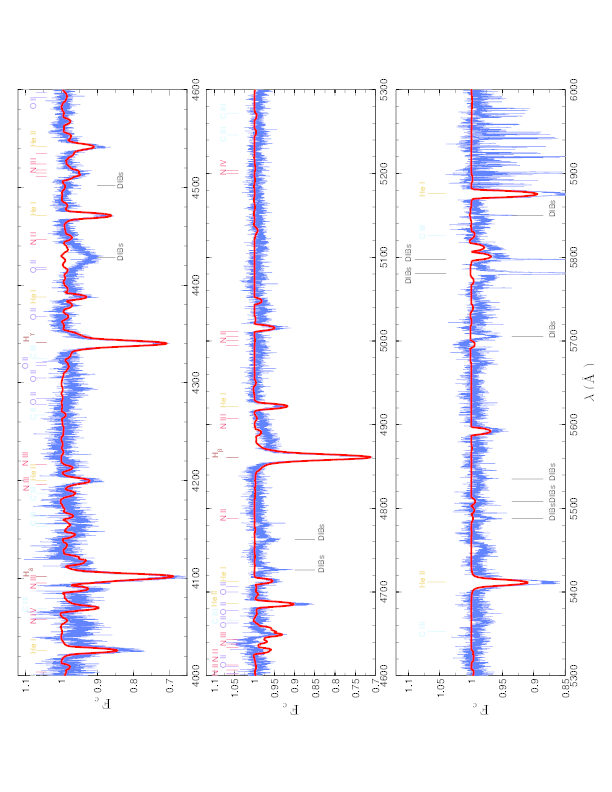}
\end{turn}
\caption{Same as Fig. \ref{cmfBD}, but for \object{HD 41997} {(O7.5Vn((f)); $v\sin\,i$ = 247 km s$^{-1}$)}. {Our sole ELODIE spectrum of this star has a low S/N, rendering the fitting more uncertain. While other He lines appear well fitted, \ion{He}{II}\,5412 and \ion{He}{I}\,5876 are too weak in the model.}}
\end{center}
\end{figure} 

\begin{figure}
\begin{turn}{0}
\includegraphics[scale=0.9]{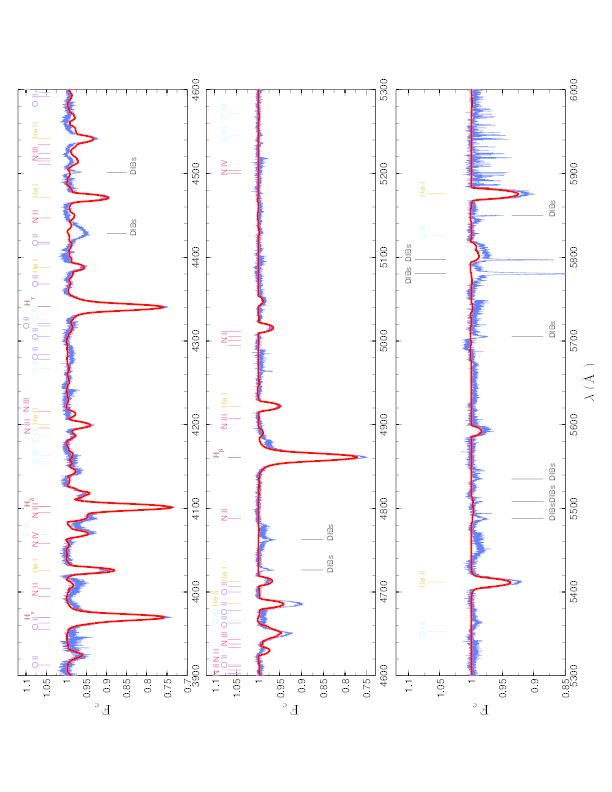}
\end{turn}
\caption{Same as Fig. \ref{cmfBD}, but for \object{HD 46056} {(O8Vn; $v\sin\,i$ = 350 km s$^{-1}$)}. {While other He lines appear well fitted, \ion{He}{II}\,4686 is too weak in the model, which is explained by the fact that wind parameters were not derived (see Sect. \ref{subSecMethCMFGEN}).}}
\end{figure} 

\begin{figure}
\begin{turn}{0}
\includegraphics[scale=0.9]{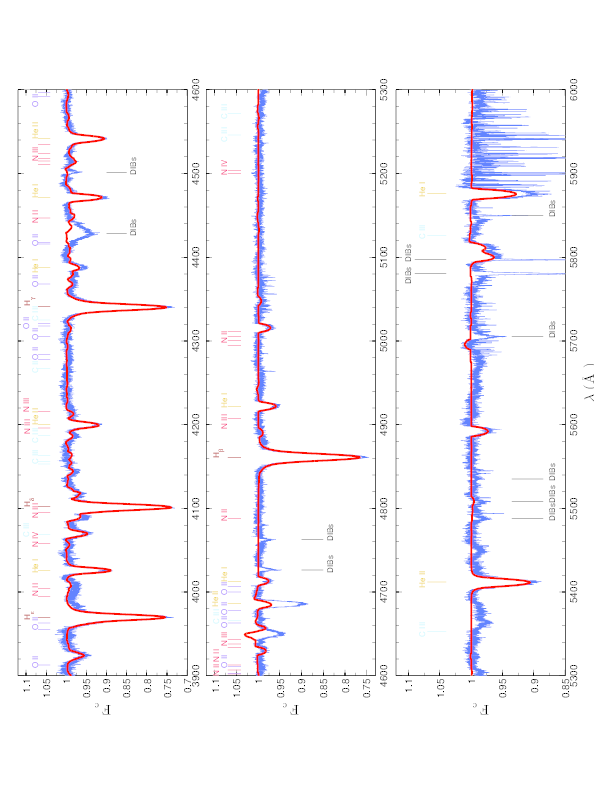}
\end{turn}
\caption{Same as Fig. \ref{cmfBD}, but for \object{HD 46485} {(O7V((f))nz; $v\sin\,i$ = 315 km s$^{-1}$)}. {The \ion{He}{I}\,5876 and \ion{He}{II}\,4686 lines appear too strong compared to the best-fit model.}  }
\end{figure} 

\begin{figure}
\begin{turn}{0}
\includegraphics[scale=0.9]{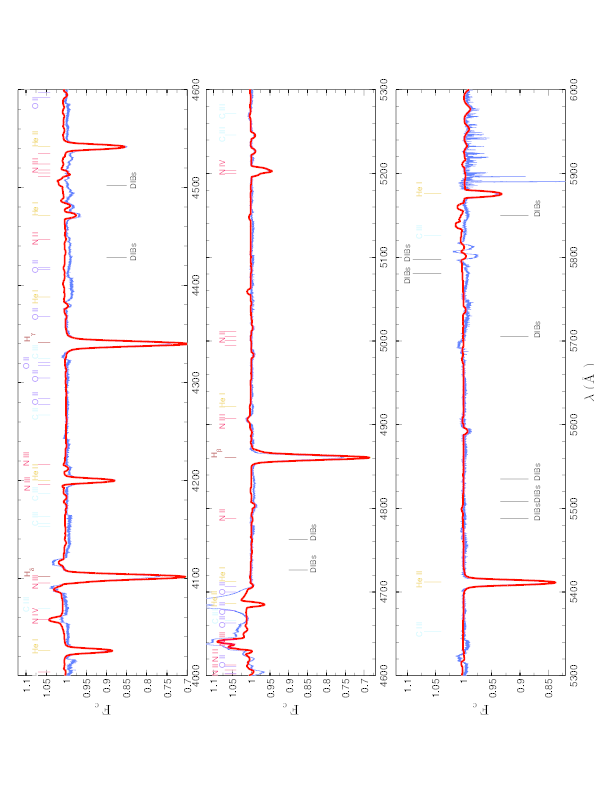}
\end{turn}
\caption{Same as Fig. \ref{cmfBD}, but for \object{HD 66811} {(O4I(n)fp; $v\sin\,i$ = 225 km s$^{-1}$)}. {The normalisation around \ion{H}{$\beta$} is imperfect, leading to some slight mismatch between the observation and the fit; the emission of the \ion{He}{II}\,4686 line is not reproduced in the model, but we recall that wind parameters were not derived (see Sect. \ref{subSecMethCMFGEN})}}
\end{figure} 

\begin{figure}
\begin{turn}{0}
\includegraphics[scale=0.9]{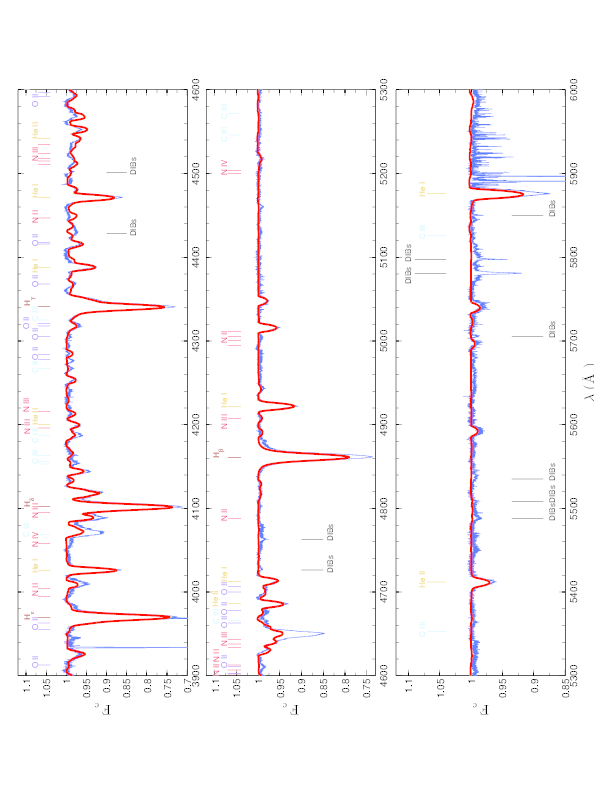}
\end{turn}
\caption{Same as Fig. \ref{cmfBD}, but for \object{HD 69106} {(O9.7IIn; $v\sin\,i$ = 306 km s$^{-1}$)}. {The H${\beta}$ and \ion{He}{I}\,5876 lines appear too strong compared to the best-fit model.}  }
\end{figure}
 
\begin{figure}
\begin{turn}{0}
\includegraphics[scale=0.9]{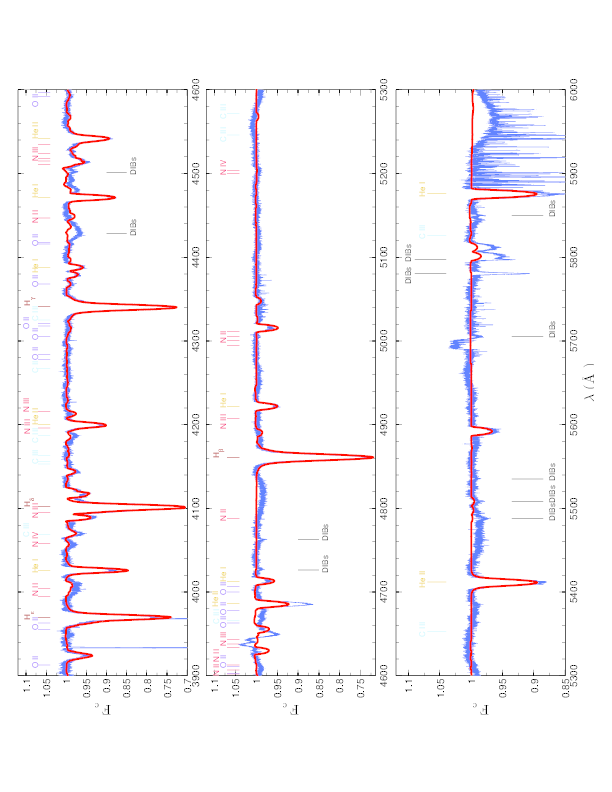}
\end{turn}
\caption{Same as Fig. \ref{cmfBD}, but for \object{HD 74920} {(O7.5IVn((f)); $v\sin\,i$ = 274 km s$^{-1}$)}. {The \ion{He}{II}\,4686 line appears too strong compared to the best-fit model, but this mismatch may be due to a normalisation issue.}  }
\end{figure} 

\begin{figure}
\begin{turn}{0}
\includegraphics[scale=0.9]{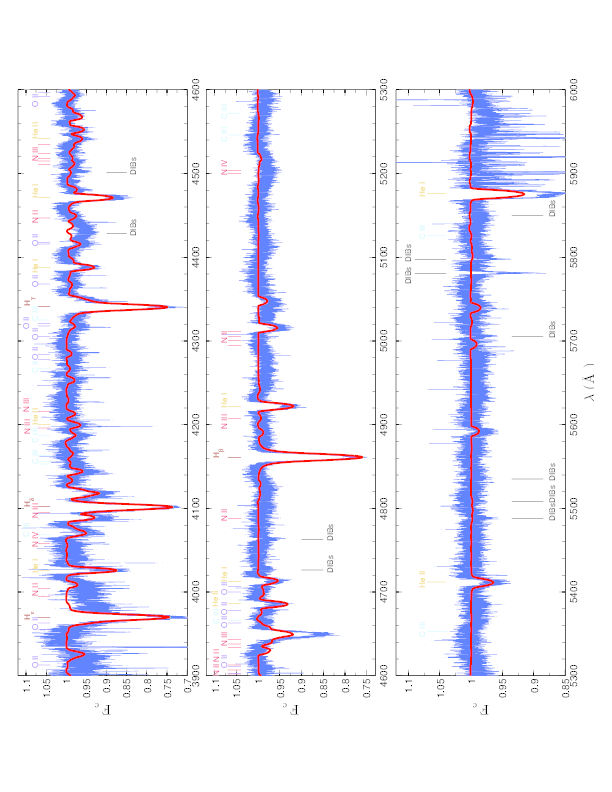}
\end{turn}
\caption{Same as Fig. \ref{cmfBD}, but for \object{HD 92554} {(O9.5III; $v\sin\,i$ = 303 km s$^{-1}$)}. {Our CORALIE spectra of this star have a low S/N, rendering the fitting more uncertain. Despite a good fit of the He lines, \ion{He}{I}\,5876 appears weaker than observed.}}
\end{figure} 

\begin{figure}
\begin{center}
\begin{turn}{0}
\includegraphics[scale=0.9]{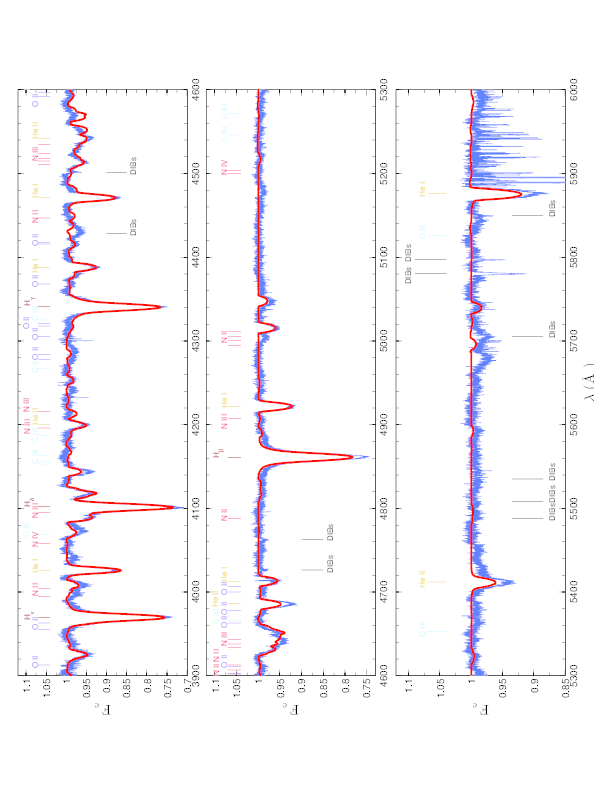}
\end{turn}
\caption{Same as Fig. \ref{cmfBD}, but for \object{HD 117490} {(ON9.5IIInn; $v\sin\,i$ = 361 km s$^{-1}$)}. {Some small mismatches between the model and observation for \ion{H}{$\beta$} and \ion{He}{II}\,5412 are mainly due to normalisation imperfections.}}
\end{center}
\end{figure} 

\begin{figure}
\begin{center}
\begin{turn}{0}
\includegraphics[scale=0.9]{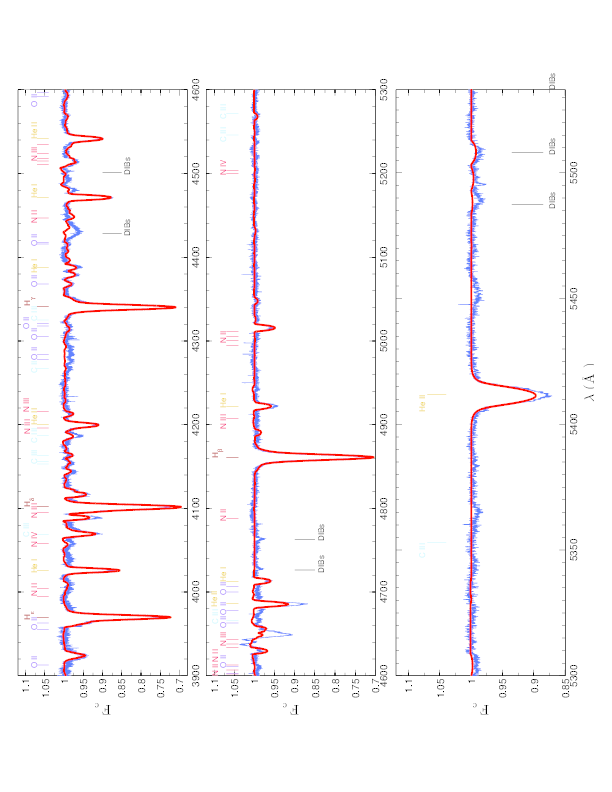}
\end{turn}
\caption{Same as Fig. \ref{cmfBD}, but for \object{HD 124979} {(O7.5IV(n)((f)); $v\sin\,i$ = 246 km s$^{-1}$)}. {Despite a good fit of the He lines, \ion{He}{II}\,4686 and \ion{He}{II}\,5412 appear weaker than observed; however, the normalisation is imperfect near 4686 $\AA$.}}
\end{center}
\end{figure} 

\begin{figure}
\begin{center}
\begin{turn}{0}
\includegraphics[scale=0.9]{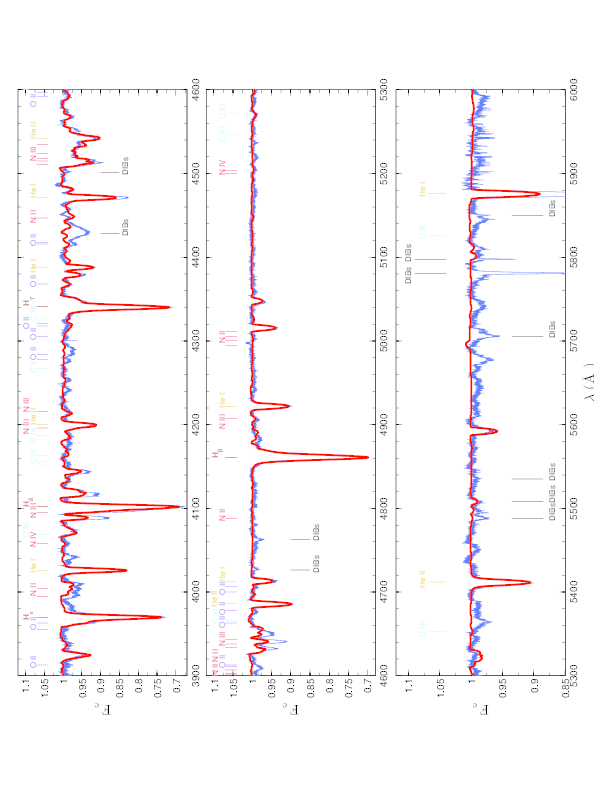}
\end{turn}
\caption{Same as Fig. \ref{cmfBD}, but for \object{HD 150574} {(ON9III(n); $v\sin\,i$ = 233 km s$^{-1}$)}. {The observed \ion{He}{I}\,5876 line appears too strong compared to the model, despite a good fit of the other He lines.}}
\end{center}
\end{figure} 

\begin{figure}
\begin{center}
\begin{turn}{0}
\includegraphics[scale=0.9]{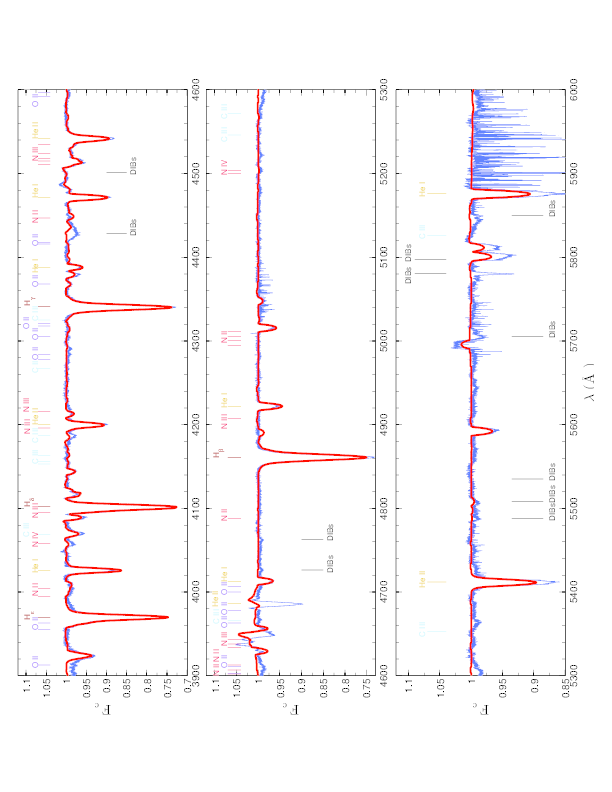}
\end{turn}
\caption{Same as Fig. \ref{cmfBD}, but for \object{HD 175876} {(O6.5III(n)(f); $v\sin\,i$ = 265 km s$^{-1}$)}. {There remains some mismatches in the wind-sensitive lines (but see Sect. \ref{subSecMethCMFGEN}), in particular \ion{He}{II}\,5412 (though an imperfect normalisation may have an impact) and \ion{He}{I}\,5876.}}
\end{center}
\end{figure} 

\onecolumn
\begin{figure}
\begin{center}
\begin{turn}{0}
\includegraphics[scale=0.9]{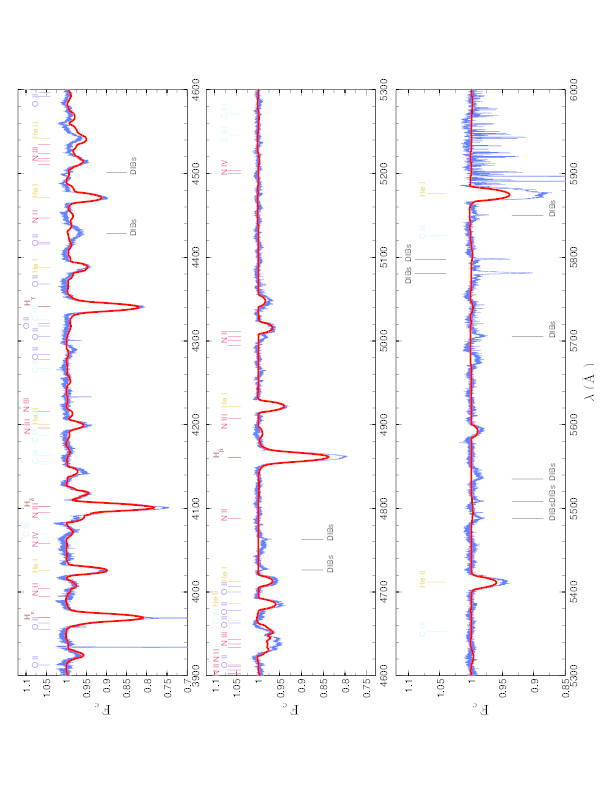}
\end{turn}
\caption{Same as Fig. \ref{cmfBD}, but for \object{HD 191423} {(ON9II-IIInn; $v\sin\,i$ = 420 km s$^{-1}$)}. {Despite an overall good fit of the H and He lines, some mismatches remain for \ion{H}{$\beta$} and  \ion{He}{I}\,5876.}}
\end{center}
\end{figure} 

\begin{figure}
\begin{turn}{0}
\includegraphics[scale=0.9]{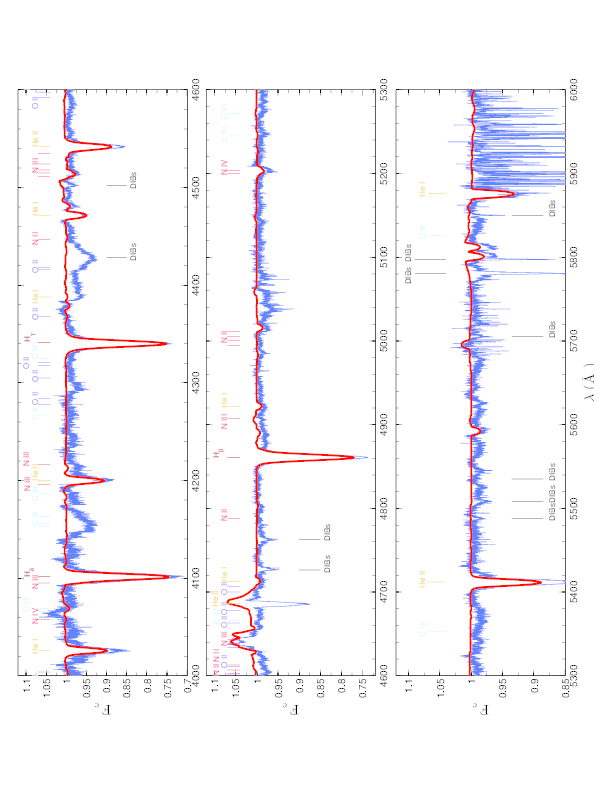}
\end{turn}
\caption{Same as Fig. \ref{cmfBD}, but for \object{HD 192281} {(O4.5V(n)((f)); $v\sin\,i$ = 276 km s$^{-1}$)}. {Our ELODIE spectra of this star have a rather low S/N, rendering the fitting more uncertain. Some mismatches remain, in particular for \ion{He}{II}\,5412 and the \ion{C}{III}\,4153--4163 complex. Because wind parameters were not derived, wind-sensitive lines are not well reproduced: too much emission is seen in \ion{He}{II}\,4686 and \ion{N}{III}\,4634--4643 for the best-fit model.}}
\end{figure} 

\begin{figure}
\begin{center}
\begin{turn}{0}
\includegraphics[scale=0.9]{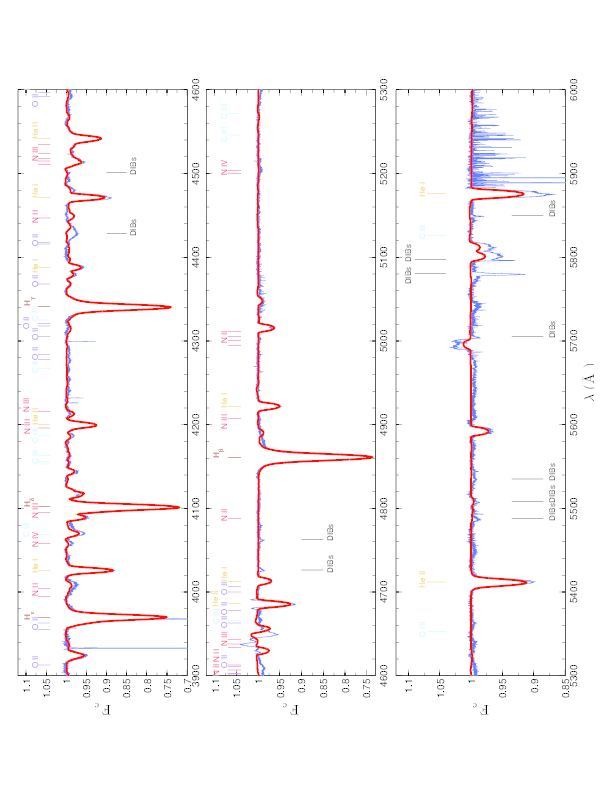}
\end{turn}
\caption{Same as Fig. \ref{cmfBD}, but for \object{HD 203064} {(O7.5IIIn((f)); $v\sin\,i$ = 298 km s$^{-1}$)}. {Despite an overall good fit of the He lines, a mismatch remains for \ion{He}{I}\,5876; some fitting imperfections are also spotted in the region of wind-sensitive lines (4600--4700 $\AA$, see Sect. \ref{subSecMethCMFGEN}).}}
\end{center}
\end{figure} 

\begin{figure}
\begin{turn}{0}
\includegraphics[scale=0.9]{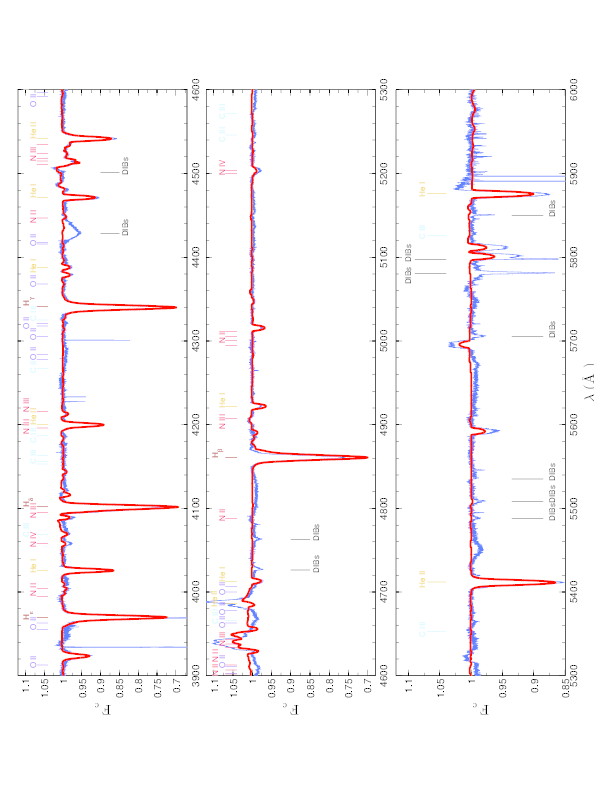}
\end{turn}
\caption{Same as Fig. \ref{cmfBD}, but for \object{HD 210839} {(O6.5I(n)fp; $v\sin\,i$ = 214 km s$^{-1}$)}. {Some normalisation imperfections remain, notably near \ion{H}{$\beta$}, and the spectral domain encompassing wind-sensitive lines is not well fitted (4600-4700 $\AA$, see Sect. \ref{subSecMethCMFGEN}).}}
\end{figure} 

\begin{figure}
\begin{turn}{0}
\includegraphics[scale=0.9]{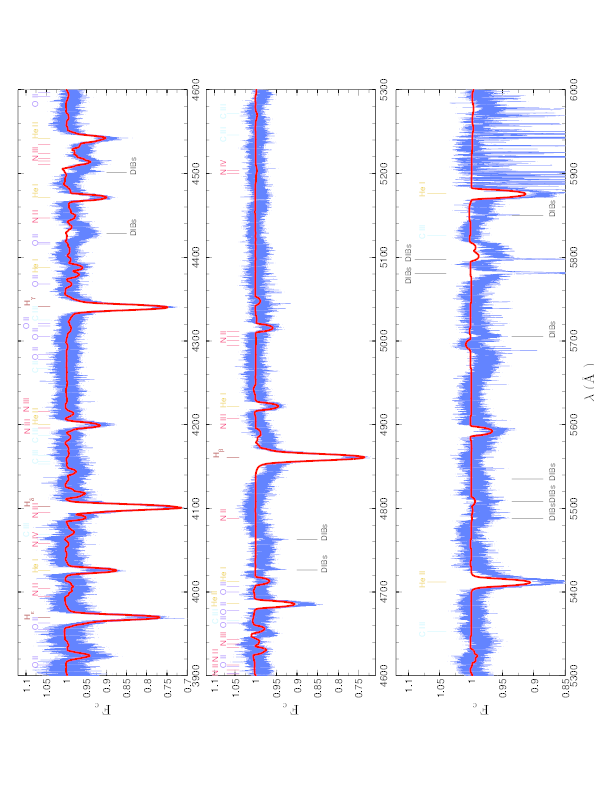}
\end{turn}
\caption{Same as Fig. \ref{cmfBD}, but for \object{HD 228841} {(O6.5Vn((f)); $v\sin\,i$ = 305 km s$^{-1}$)}. {Our sole SOPHIE spectrum of this star has a low S/N, rendering the fitting more uncertain; a good match is found overall except maybe for \ion{He}{II}\,5412, which could be affected by normalisation problems.}}
\end{figure}

\end{appendix}


\begin{thebibliography}{}

\bibitem[Abt et al.(1972)]{abt72} Abt, H.~A., Levy, S.~G., \& Gandet, T.~L.\ 1972, \aj, 77, 138 
\bibitem[Aldoretta et al.(2015)]{ald15} Aldoretta, E.~J., Caballero-Nieves, S.~M., Gies, D.~R., et al.\ 2015, \aj, 149, 26 
\bibitem[Alduseva et al.(1982)]{ald82} Alduseva, V.~I., Aslanov, A.~A., Kolotilov, E.~A., \& Cherepashchuk, A.~M.\ 1982, Soviet Astronomy Letters, 8, 717 
\bibitem[Asplund et al.(2009)]{asp09} Asplund, M., Grevesse, N., Sauval, A.~J., \& Scott, P.\ 2009, \araa, 47, 481
\bibitem[Bates et al.(1992)]{bat92} Bates, B., Wood, K.~D., Catney, M.~G., \& Gilheany, S.\ 1992, \mnras, 254, 221 
\bibitem[Baranne et al.(1996)]{bar96} Baranne, A., Queloz, D., Mayor, M., et al.\ 1996, \aaps, 119, 373
\bibitem[Barannikov(1993)]{bara93} Barannikov, A.~A.\ 1993, Astronomy Letters, 19, 420
\bibitem[Barb{\'a} et al.(2010)]{bar10} Barb{\'a}, R.~H., Gamen, R., Arias, J.~I., et al.\ 2010, Revista Mexicana de Astronomia y Astrofisica Conference Series, 38, 30
\bibitem[Bastiaansen(1992)]{bas92} Bastiaansen, P.~A.\ 1992, \aaps, 93, 449 
\bibitem[Bekenstein \& Bowers(1974)]{bek74} Bekenstein, J.~D., \& Bowers, R.~L.\ 1974, \apj, 190, 653
\bibitem[Blaauw(1961)]{bla61} Blaauw, A.\ 1961, \bain, 15, 265
\bibitem[Bohannan \& Garmany(1978)]{boh78} Bohannan, B., \& Garmany, C.~D.\ 1978, \apj, 223, 908 
\bibitem[Bouret et al.(2012)]{bou12} Bouret, J.-C., Hillier, D.~J., Lanz, T., \& Fullerton, A.~W.\ 2012, \aap, 544, A67
\bibitem[Bouret et al.(2013)]{bou13} Bouret, J.-C., Lanz, T., Martins, F., et al.\ 2013, \aap, 555, A1
\bibitem[Boyajian et al.(2005)]{boy05} Boyajian, T.~S., Beaulieu, T.~D., Gies, D.~R., et al.\ 2005, \apj, 621, 978
\bibitem[Bragan{\c c}a et al.(2012)]{bra12} Bragan{\c c}a, G.~A., Daflon, S., Cunha, K., et al.\ 2012, \aj, 144, 130
\bibitem[Brott et al.(2011b)]{bro11b} Brott, I., Evans, C.~J., Hunter, I., et al.\ 2011b, \aap, 530, A116
\bibitem[Butler \& Giddings(1985)]{but85} Butler, K., \& Giddings, J.~R.\ 1985, in Newsletter of Analysis of Astronomical Spectra, No. 9 (Univ. London)
\bibitem[Carrasco \& Creze(1978)]{car78} Carrasco, L., \& Creze, M.\ 1978, \aap, 65, 279 
\bibitem[Cazorla et al.(2007)]{caz17} Cazorla, C., Naz{\'e}, Y., Morel, T., Georgy, G., Godart, M., \& Langer, N.\ 2017, \aap, submitted (Paper II)
\bibitem[Cherepashchuk \& Aslanov(1984)]{che84} Cherepashchuk, A.~M., \& Aslanov, A.~A.\ 1984, \apss, 102, 97 
\bibitem[Cincotta et al.(1999)]{cin99} Cincotta, P.~M., Helmi, A., Mendez, M., Nunez, J.~A., \& Vucetich, H.\ 1999, \mnras, 302, 582
\bibitem[Cincotta(1999)]{cin99b} Cincotta, P.~M.\ 1999, \mnras, 307, 941
\bibitem[Conti et al.(1977)]{con77} Conti, P.~S., Leep, E.~M., \& Lorre, J.~J.\ 1977, \apj, 214, 759 
\bibitem[Cuypers(1987)]{cuy87} Cuypers, J.\ 1987, \aaps, 69, 445
\bibitem[Daflon et al.(2007)]{daf07} Daflon, S., Cunha, K., de Ara{\'u}jo, F.~X., Wolff, S., \& Przybilla, N.\ 2007, \aj, 134, 1570 
\bibitem[De Becker \& Rauw(2004)]{deb04} De Becker, M., \& Rauw, G.\ 2004, \aap, 427, 995
\bibitem[De Becker et al.(2008)]{deb08} De Becker, M., Linder, N., \& Rauw, G.\ 2008, Information Bulletin on Variable Stars, 5841, 1
\bibitem[de Mink et al.(2009)]{dem09} de Mink, S.~E., Cantiello, M., Langer, N., et al.\ 2009, \aap, 497, 243
\bibitem[de Mink et al.(2013)]{dem13} de Mink, S.~E., Langer, N., Izzard, R.~G., Sana, H., \& de Koter, A.\ 2013, \apj, 764, 166
\bibitem[Dervi{\c s}o{\v g}lu et al.(2010)]{der10} Dervi{\c s}o{\v g}lu, A., Tout, C.~A., \& Ibano{\v g}lu, C.\ 2010, \mnras, 406, 1071
\bibitem[de Wit et al.(2005)]{dew05} de Wit, W.~J., Testi, L., Palla, F., \& Zinnecker, H.\ 2005, \aap, 437, 247 
\bibitem[Dufton et al.(2011)]{duf11} Dufton, P.~L., Dunstall, P.~R., Evans, C.~J., et al.\ 2011, \apjl, 743, L22 
\bibitem[Eggleton \& Tokovinin(2008)]{egg08} Eggleton, P.~P., \& Tokovinin, A.~A.\ 2008, \mnras, 389, 869
\bibitem[Evans et al.(2008)]{eva08} Evans, C., Hunter, I., Smartt, S., et al.\ 2008, The Messenger, 131, 25 
\bibitem[Feast et al.(1957)]{fea57} Feast, M.~W., Thackeray, A.~D., \& Wesselink, A.~J.\ 1957, \memras, 68, 1 
\bibitem[Feast \& Thackeray(1963)]{fea63} Feast, M.~W., \& Thackeray, A.~D.\ 1963, \memras, 68, 173
\bibitem[Firnstein \& Przybilla(2012)]{fir12} Firnstein, M., \& Przybilla, N.\ 2012, \aap, 543, A80 
\bibitem[Fitzgerald \& Moffat(1975)]{fit75} Fitzgerald, M.~P., \& Moffat, A.~F.~J.\ 1975, \aaps, 20, 289
\bibitem[Fr{\'e}mat et al.(2005)]{fre05} Fr{\'e}mat, Y., Zorec, J., Hubert, A.-M., \& Floquet, M.\ 2005, \aap, 440, 305
\bibitem[Frost et al.(1926)]{fro26} Frost, E.~B., Barrett, S.~B., \& Struve, O.\ 1926, \apj, 64, 1
\bibitem[Gaia Collaboration et al.(2016)]{gai16} Gaia Collaboration, Prusti, T., de Bruijne, J.~H.~J., et al.\ 2016, \aap, 595, A1 
\bibitem[Garmany et al.(1980)]{gar80} Garmany, C.~D., Conti, P.~S., \& Massey, P.\ 1980, \apj, 242, 1063 
\bibitem[Garmany et al.(2015)]{gar15} Garmany, C.~D., Glaspey, J.~W., Bragan{\c c}a, G.~A., et al.\ 2015, \aj, 150, 41
\bibitem[Garrison et al.(1977)]{gar77} Garrison, R.~F., Hiltner, W.~A., \& Schild, R.~E.\ 1977, \apjs, 35, 111 
\bibitem[Garrison et al.(1983)]{gar83} Garrison, R.~F., Schild, R.~E., \& Hiltner, W.~A.\ 1983, \apjs, 52, 1
\bibitem[Gies \& Bolton(1986)]{gie86} Gies, D.~R., \& Bolton, C.~T.\ 1986, \apjs, 61, 419
\bibitem[Giddings(1981)]{gid81} Giddings, J.~R.\ 1981, Ph.D.~Thesis
\bibitem[Gillet et al.(1994)]{gil94} Gillet, D., Burnage, R., Kohler, D., et al.\ 1994, \aaps, 108, 181
\bibitem[Gosset et al.(2001)]{gos01} Gosset, E., Royer, P., Rauw, G., Manfroid, J., \& Vreux, J.-M.\ 2001, MNRAS, 327, 435
\bibitem[Graham et al.(2013)]{gra13} Graham, M.~J., Drake, A.~J., Djorgovski, S.~G., Mahabal, A.~A., \& Donalek, C.\ 2013, \mnras, 434, 2629
\bibitem[Grigsby et al.(1992)]{gri92} Grigsby, J.~A., Morrison, N.~D., \& Anderson, L.~S.\ 1992, \apjs, 78, 205 
\bibitem[Grin et al.(2016)]{grin16} Grin, N.~J., Ramirez-Agudelo, O.~H., de Koter, A., et al.\ 2016, arXiv:1609.00197 
\bibitem[Gray(2005)]{gra05} Gray, D.~F.\ 2005, ''The Observation and Analysis of Stellar Photospheres", 3rd Edition, ISBN 0521851866, Cambridge University Press
\bibitem[Heck et al.(1985)]{hmm} Heck, A., Manfroid, J., \& Mersch, G.\ 1985, A\&AS, 59, 63
\bibitem[Herrero et al.(2000)]{her00} Herrero, A., Puls, J., \& Villamariz, M.~R.\ 2000, \aap, 354, 193 
\bibitem[Herrero et al.(1992)]{her92} Herrero, A., Kudritzki, R.~P., Vilchez, J.~M., et al.\ 1992, \aap, 261, 209
\bibitem[Herrero et al.(2002)]{her02} Herrero, A., Puls, J., \& Najarro, F.\ 2002, \aap, 396, 949 
\bibitem[Hillier \& Miller(1998)]{hil98} Hillier, D.~J., \& Miller, D.~L.\ 1998, \apj, 496, 407
\bibitem[Hillwig et al.(2006)]{hilw06} Hillwig, T.~C., Gies, D.~R., Bagnuolo, W.~G., Jr., et al.\ 2006, \apj, 639, 1069
\bibitem[Hiltner et al.(1969)]{hil69} Hiltner, W.~A., Garrison, R.~F., \& Schild, R.~E.\ 1969, \apj, 157, 313 
\bibitem[Hoogerwerf et al.(2001)]{hoo01} Hoogerwerf, R., de Bruijne, J.~H.~J., \& de Zeeuw, P.~T.\ 2001, \aap, 365, 49 
\bibitem[Howarth et al.(1997)]{how97} Howarth, I.~D., Siebert, K.~W., Hussain, G.~A.~J., \& Prinja, R.~K.\ 1997, \mnras, 284, 265
\bibitem[Howarth \& Smith(2001)]{how01} Howarth, I.~D., \& Smith, K.~C.\ 2001, \mnras, 327, 353
\bibitem[Huang \& Gies(2008)]{hua08} Huang, W., \& Gies, D.~R.\ 2008, \apj, 683, 1045-1051 
\bibitem[Hubrig et al.(2008)]{hub08} Hubrig, S., Briquet, M., Morel, T., et al.\ 2008, \aap, 488, 287
\bibitem[Humphreys(1973)]{hum73} Humphreys, R.~M.\ 1973, \aaps, 9, 85 
\bibitem[Humphreys(1978)]{hum78} Humphreys, R.~M.\ 1978, \apjs, 38, 309
\bibitem[Hunter et al.(2007)]{hun07} Hunter, I., Dufton, P.~L., Smartt, S.~J., et al.\ 2007, \aap, 466, 277
\bibitem[Hunter et al.(2009)]{hun09} Hunter, I., Brott, I., Langer, N., et al.\ 2009, \aap, 496, 841
\bibitem[Hut(1981)]{hut81} Hut, P.\ 1981, \aap, 99, 126
\bibitem[Jerzykiewicz(1993)]{jer93} Jerzykiewicz, M.\ 1993, \aaps, 97, 421
\bibitem[Jurkevich(1971)]{jur71} Jurkevich, I.\ 1971, \apss, 13, 154
\bibitem[Kaper et al.(1996)]{kap96} Kaper, L., Henrichs, H.~F., Nichols, J.~S., et al.\ 1996, \aaps, 116, 257 
\bibitem[Kambe et al.(1997)]{kam97} Kambe, E., Hirata, R., Ando, H., et al.\ 1997, \apj, 481, 406 
\bibitem[Kendall et al.(1996)]{ken96} Kendall, T.~R., Dufton, P.~L., \& Lennon, D.~J.\ 1996, \aap, 310, 564 
\bibitem[Kilian(1992)]{kil92} Kilian, J.\ 1992, \aap, 262, 171 
\bibitem[Kilkenny \& Hill(1975)]{kil75} Kilkenny, D., \& Hill, P.~W.\ 1975, \mnras, 172, 649 
\bibitem[Kirsten et al.(2015)]{kir15} Kirsten, F., Vlemmings, W., Campbell, R.~M., Kramer, M., \& Chatterjee, S.\ 2015, \aap, 577, A111 
\bibitem[Koen \& Eyer(2002)]{koe02} Koen, C., \& Eyer, L.\ 2002, \mnras, 331, 45
\bibitem[Kraus et al.(2012)]{krau12} Kraus, S., Monnier, J.~D., Che, X., et al.\ 2012, \apj, 744, 19
\bibitem[Kurtz \& Mink(1998)]{kur98} Kurtz, M.~J., \& Mink, D.~J.\ 1998, \pasp, 110, 934
\bibitem[Lafler \& Kinman(1965)]{lafkin} Lafler, J., \& Kinman, T.~D.\ 1965, \apjs, 11, 216
\bibitem[Langer et al.(2008)]{lan08} Langer, N., Cantiello, M., Yoon, S.-C., et al.\ 2008, Massive Stars as Cosmic Engines, 250, 167
\bibitem[Lanz \& Hubeny(2003)]{lan03} Lanz, T., \& Hubeny, I.\ 2003, \apjs, 146, 417
\bibitem[Langer et al.(2003)]{lan03b} Langer, N., Wellstein, S., \& Petrovic, J.\ 2003, A Massive Star Odyssey: From Main Sequence to Supernova, 212, 275 
\bibitem[Lanz \& Hubeny(2007)]{lan07} Lanz, T., \& Hubeny, I.\ 2007, \apjs, 169, 83
\bibitem[Lefever et al.(2010)]{lef10} Lefever, K., Puls, J., Morel, T., et al.\ 2010, \aap, 515, A74
\bibitem[Lennon et al.(1991)]{len91} Lennon, D.~J., Dufton, P.~L., Keenan, F.~P., \& Holmgren, D.~E.\ 1991, \aap, 246, 175 
\bibitem[Lesh(1968)]{les68} Lesh, J.~R.\ 1968, \apjs, 17, 371 
\bibitem[Lozinskaya \& Lyuty(1981)]{loz81} Lozinskaya, T.~A., \& Lyuty, V.~M.\ 1981, Astronomicheskij Tsirkulyar, 1196, 1 
\bibitem[Lynds(1959)]{lyn59} Lynds, C.~R.\ 1959, \apj, 130, 577
\bibitem[Lyubimkov et al.(2004)]{lyu04} Lyubimkov, L.~S., Rostopchin, S.~I., \& Lambert, D.~L.\ 2004, \mnras, 351, 745
\bibitem[Maeder(1995)]{mae95} Maeder, A.\ 1995, IAU Colloq.~155: Astrophysical Applications of Stellar Pulsation, 83, 1 
\bibitem[Maeder \& Meynet(1996)]{mae96} Maeder, A., \& Meynet, G.\ 1996, \aap, 313, 140
\bibitem[Maeder et al.(2009)]{mae09} Maeder, A., Meynet, G., Ekstr{\"o}m, S., \& Georgy, C.\ 2009, Communications in Asteroseismology, 158, 72 
\bibitem[Maeder et al.(2014)]{mae14} Maeder, A., Przybilla, N., Nieva, M.-F., et al.\ 2014, \aap, 565, A39
\bibitem[Maeder \& Meynet(2015)]{mae15} Maeder, A., \& Meynet, G.\ 2015, IAU Symposium, 307, 9
\bibitem[Mahy et al.(2009)]{mah09} Mahy, L., Naz{\'e}, Y., Rauw, G., et al.\ 2009, \aap, 502, 937
\bibitem[Mahy et al.(2013)]{mah13} Mahy, L., Rauw, G., De Becker, M., Eenens, P., \& Flores, C.~A.\ 2013, \aap, 550, A27
\bibitem[Mahy et al.(2015)]{mah15} Mahy, L., Rauw, G., De Becker, M., Eenens, P., \& Flores, C.~A.\ 2015, \aap, 577, A23
\bibitem[Ma{\'{\i}}z Apell{\'a}niz et al.(2008)]{mai08} Ma{\'{\i}}z Apell{\'a}niz, J., Alfaro, E.~J., \& Sota, A.\ 2008, arXiv:0804.2553 
\bibitem[Ma{\'{\i}}z Apell{\'a}niz et al.(2011)]{mai11} Ma{\'{\i}}z Apell{\'a}niz, J., Sota, A., Walborn, N.~R., et al.\ 2011, Highlights of Spanish Astrophysics VI, 467 
\bibitem[Marcolino et al.(2009)]{marc09} Marcolino, W.~L.~F., Bouret, J.-C., Martins, F., et al.\ 2009, \aap, 498, 837 
\bibitem[Martins et al.(2005)]{mar05} Martins, F., Schaerer, D., \& Hillier, D.~J.\ 2005, \aap, 436, 1049 
\bibitem[Martins et al.(2012a)]{mar12a} Martins, F., Escolano, C., Wade, G.~A., et al.\ 2012a, \aap, 538, A29 
\bibitem[Martins et al.(2012b)]{mar12b} Martins, F., Mahy, L., Hillier, D.~J., \& Rauw, G.\ 2012b, \aap, 538, A39 
\bibitem[Martins et al.(2015a)]{mar15a} Martins, F., Herv{\'e}, A., Bouret, J.-C., et al.\ 2015a, \aap, 575, A34
\bibitem[Martins et al.(2015b)]{mar15b} Martins, F., Sim{\'o}n-D{\'{\i}}az, S., Palacios, A., et al.\ 2015b, \aap, 578, A109
\bibitem[Mason et al.(1998)]{mas98} Mason, B.~D., Gies, D.~R., Hartkopf, W.~I., et al.\ 1998, \aj, 115, 821
\bibitem[Mason et al.(2004)]{mas04} Mason, B.~D., Hartkopf, W.~I., Wycoff, G.~L., et al.\ 2004, \aj, 128, 3012 
\bibitem[Mason et al.(2009)]{mas09} Mason, B.~D., Hartkopf, W.~I., Gies, D.~R., Henry, T.~J., \& Helsel, J.~W.\ 2009, \aj, 137, 3358
\bibitem[Mason et al.(2011)]{mas11} Mason, B.~D., Hartkopf, W.~I., \& Wycoff, G.~L.\ 2011, \aj, 141, 157 
\bibitem[Mayer et al.(1994)]{may94} Mayer, P., Chochol, D., Hanna, M.~A.-M.~., \& Wolf, M.\ 1994, Contributions of the Astronomical Observatory Skalnate Pleso, 24, 65
\bibitem[Mayer et al.(1998)]{may98} Mayer, P., Hanna, M.~A., Wolf, M., \& Chochol, D.\ 1998, \apss, 262, 163 
\bibitem[Mayer et al.(2014)]{may14} Mayer, P., Drechsel, H., \& Irrgang, A.\ 2014, \aap, 565, A86
\bibitem[McSwain et al.(2007)]{mcs07} McSwain, M.~V., Boyajian, T.~S., Grundstrom, E.~D., \& Gies, D.~R.\ 2007, \apj, 655, 473 
\bibitem[McSwain et al.(2010)]{mcs10} McSwain, M.~V., De Becker, M., Roberts, M.~S.~E., et al.\ 2010, \aj, 139, 857
\bibitem[Meurs et al.(2005)]{meu05} Meurs, E.~J.~A., Fennell, G., \& Norci, L.\ 2005, \apj, 624, 307 
\bibitem[Meynet \& Maeder(2000)]{mey00} Meynet, G., \& Maeder, A.\ 2000, \aap, 361, 101
\bibitem[Meynet et al.(2011)]{mey11} Meynet, G., Eggenberger, P., \& Maeder, A.\ 2011, \aap, 525, L11
\bibitem[Mokiem et al.(2005)]{mok05} Mokiem, M.~R., de Koter, A., Puls, J., et al.\ 2005, \aap, 441, 711 
\bibitem[Morel et al.(2006)]{mor06} Morel, T., Butler, K., Aerts, C., Neiner, C., \& Briquet, M.\ 2006, \aap, 457, 651 
\bibitem[Morel et al.(2008)]{mor08} Morel, T., Hubrig, S., \& Briquet, M.\ 2008, \aap, 481, 453
\bibitem[Morel(2011)]{mor11} Morel, T.\ 2011, Bulletin de la Societe Royale des Sciences de Liege, 80, 405
\bibitem[Morgan et al.(1955)]{mor55} Morgan, W.~W., Code, A.~D., \& Whitford, A.~E.\ 1955, \apjs, 2, 41
\bibitem[Morton(1975)]{mor75} Morton, D.~C.\ 1975, \apj, 197, 85 
\bibitem[Motch et al.(1998)]{mot98} Motch, C., Guillout, P., Haberl, F., et al.\ 1998, \aaps, 132, 341
\bibitem[Muijres et al.(2012)]{mui12} Muijres, L.~E., Vink, J.~S., de Koter, A., M{\"u}ller, P.~E., \& Langer, N.\ 2012, \aap, 537, A37 
\bibitem[Murphy(1969)]{mur69} Murphy, R.~E.\ 1969, \aj, 74, 1082
\bibitem[Nieva \& Przybilla(2007)]{nie07} Nieva, M.~F., \& Przybilla, N.\ 2007, \aap, 467, 295 
\bibitem[Nieva \& Przybilla(2008)]{nie08} Nieva, M.~F., \& Przybilla, N.\ 2008, \aap, 481, 199 
\bibitem[Nieva \& Przybilla(2012)]{nie12} Nieva, M.-F., \& Przybilla, N.\ 2012, \aap, 539, A143 
\bibitem[Nieva \& Sim{\'o}n-D{\'{\i}}az(2011)]{nie11} Nieva, M.-F., \& Sim{\'o}n-D{\'{\i}}az, S.\ 2011, \aap, 532, A2 
\bibitem[Packet(1981)]{pac81} Packet, W.\ 1981, \aap, 102, 17
\bibitem[Palate \& Rauw(2012)]{pal12} Palate, M., \& Rauw, G.\ 2012, \aap, 537, A119
\bibitem[Palate et al.(2013)]{pal13} Palate, M., Rauw, G., Koenigsberger, G., \& Moreno, E.\ 2013, \aap, 552, A39
\bibitem[Penny(1996)]{pen96} Penny, L.~R.\ 1996, \apj, 463, 737
\bibitem[Peri et al.(2012)]{per12} Peri, C.~S., Benaglia, P., Brookes, D.~P., Stevens, I.~R., \& Isequilla, N.~L.\ 2012, \aap, 538, A108
\bibitem[Petrovic et al.(2005a)]{pet05a} Petrovic, J., Langer, N., Yoon, S.-C., \& Heger, A.\ 2005a, \aap, 435, 247
\bibitem[Petrovic et al.(2005b)]{pet05b} Petrovic, J., Langer, N., \& van der Hucht, K.~A.\ 2005b, \aap, 435, 1013
\bibitem[Philp et al.(1996)]{phi96} Philp, C.~J., Evans, C.~R., Leonard, P.~J.~T., \& Frail, D.~A.\ 1996, \aj, 111, 1220 
\bibitem[Piskunov \& Valenti(2002)]{pis02} Piskunov, N.~E., \& Valenti, J.~A.\ 2002, \aap, 385, 1095 
\bibitem[Plaskett \& Pearce(1931)]{pla31} Plaskett, J.~S., \& Pearce, J.~A.\ 1931, Publications of the Dominion Astrophysical Observatory Victoria, 5, 1
\bibitem[Podsiadlowski et al.(1992)]{pod92} Podsiadlowski,P., Joss, P.~C., \& Hsu, J.~J.~L.\ 1992, \apj, 391, 246
\bibitem[Pols et al.(1991)]{pol91} Pols, O.~R., Cote, J., Waters, L.~B.~F.~M., \& Heise, J.\ 1991, \aap, 241, 419
\bibitem[Potter et al.(2012)]{pot12} Potter, A.~T., Chitre, S.~M., \& Tout, C.~A.\ 2012, \mnras, 424, 2358 
\bibitem[Prinja et al.(1990)]{pri90} Prinja, R.~K., Barlow, M.~J., \& Howarth, I.~D.\ 1990, \apj, 361, 607
\bibitem[Proffitt \& Quigley(2001)]{pro01} Proffitt, C.~R., \& Quigley, M.~F.\ 2001, \apj, 548, 429 
\bibitem[Przybilla et al.(2011)]{prz11} Przybilla, N., Nieva, M.-F., \& Butler, K.\ 2011, Journal of Physics Conference Series, 328, 012015 
\bibitem[Puls et al.(1996)]{pul96} Puls, J., Kudritzki, R.-P., Herrero, A., et al.\ 1996, \aap, 305, 171 
\bibitem[Raucq et al.(2016)]{rau16} Raucq, F., Rauw, G., Gosset, E., et al.\ 2016, \aap, 588, A10 
\bibitem[Rauw et al.(2003)]{rau03} Rauw, G., De Becker, M., \& Vreux, J.-M.\ 2003, \aap, 399, 287 
\bibitem[Rauw \& De Becker(2004)]{rau04} Rauw, G., \& De Becker, M.\ 2004, \aap, 421, 693
\bibitem[Rauw et al.(2012)]{rau12} Rauw, G., Morel, T., \& Palate, M.\ 2012, \aap, 546, A77
\bibitem[Renson(1978)]{renson} Renson, P.\ 1978, \aap, 63, 125
\bibitem[Repolust et al.(2004)]{rep04} Repolust, T., Puls, J., \& Herrero, A.\ 2004, \aap, 415, 349 
\bibitem[Repolust et al.(2005)]{rep05} Repolust, T., Puls, J., Hanson, M.~M., Kudritzki, R.-P., \& Mokiem, M.~R.\ 2005, \aap, 440, 261 
\bibitem[Rivero Gonz{\'a}lez et al.(2012a)]{riv12a} Rivero Gonz{\'a}lez, J.~G., Puls, J., Najarro, F., \& Brott, I.\ 2012, \aap, 537, A79 
\bibitem[Rivero Gonz{\'a}lez et al.(2012b)]{riv12b} Rivero Gonz{\'a}lez, J.~G., Puls, J., Massey, P., \& Najarro, F.\ 2012, \aap, 543, A95 
\bibitem[Rivinius et al.(2013)]{riv13} Rivinius, T., Carciofi, A.~C., \& Martayan, C.\ 2013, \aapr, 21, 69
\bibitem[Sana(2013)]{san13} Sana, H.\ 2013, Astrophysics Source Code Library, 1309.003
\bibitem[Sana et al.(2013)]{san13b} Sana, H., de Koter, A., de Mink, S.~E., et al.\ 2013, \aap, 550, A107 
\bibitem[Sana et al.(2014)]{san14} Sana, H., Le Bouquin, J.-B., Lacour, S., et al.\ 2014, \apjs, 215, 15
\bibitem[Schmitt et al.(2014)]{sch14} Schmitt, J.~H.~M.~M., Schr{\"o}der, K.-P., Rauw, G., et al.\ 2014, Astronomische Nachrichten, 335, 787 
\bibitem[Sch\"onberner et al.(1988)]{sch88} Sch\"onberner, D., Herrero, A., Becker, S., et al.\ 1988, \aap, 197, 209 
\bibitem[Schwarzenberg-Czerny(1989)]{sch89} Schwarzenberg-Czerny, A.\ 1989, \mnras, 241, 153
\bibitem[Sim{\'o}n-D{\'{\i}}az et al.(2006)]{sim06} Sim{\'o}n-D{\'{\i}}az, S., Herrero, A., Esteban, C., \& Najarro, F.\ 2006, \aap, 448, 351 
\bibitem[Sim{\'o}n-D{\'{\i}}az \& Herrero(2007)]{sim07} Sim{\'o}n-D{\'{\i}}az, S., \& Herrero, A.\ 2007, \aap, 468, 1063
\bibitem[Sim{\'o}n-D{\'{\i}}az \& Herrero(2014)]{sim14} Sim{\'o}n-D{\'{\i}}az, S., \& Herrero, A.\ 2014, \aap, 562, A135
\bibitem[Song et al.(2013)]{son13} Song, H.~F., Maeder, A., Meynet, G., et al.\ 2013, \aap, 556, A100
\bibitem[Sota et al.(2011)]{sot11} Sota, A., Ma{\'{\i}}z Apell{\'a}niz, J., Walborn, N.~R., et al.\ 2011, \apjs, 193, 24
\bibitem[Sota et al.(2014)]{sot14} Sota, A., Ma{\'{\i}}z Apell{\'a}niz, J., Morrell, N.~I., et al.\ 2014, \apjs, 211, 10 
\bibitem[Stankov \& Handler(2005)]{sta05} Stankov, A., \& Handler, G.\ 2005, \apjs, 158, 193
\bibitem[Stellingwerf(1978)]{ste78} Stellingwerf, R.~F.\  1978, \apj, 224, 953
\bibitem[Stickland \& Lloyd(2001)]{sti01} Stickland, D.~J., \& Lloyd, C.\ 2001, The Observatory, 121, 1
\bibitem[Strai{\v z}ys \& Laugalys(2007)]{str07} Strai{\v z}ys, V., \& Laugalys, V.\ 2007, Baltic Astronomy, 16, 167
\bibitem[Tetzlaff et al.(2010)]{tez10} Tetzlaff, N., Neuh{\"a}user, R., Hohle, M.~M., \& Maciejewski, G.\ 2010, \mnras, 402, 2369 
\bibitem[Tetzlaff et al.(2011)]{tez11} Tetzlaff, N., Neuh{\"a}user, R., \& Hohle, M.~M.\ 2011, \mnras, 410, 190
\bibitem[Tokovinin et al.(2010)]{tok10} Tokovinin, A., Mason, B.~D., \& Hartkopf, W.~I.\ 2010, \aj, 139, 743 
\bibitem[Turner et al.(2008)]{tur08} Turner, N.~H., ten Brummelaar, T.~A., Roberts, L.~C., et al.\ 2008, \aj, 136, 554 
\bibitem[Tylenda et al.(2011)]{tyl11} Tylenda, R., Hajduk, M., Kami{\'n}ski, T., et al.\ 2011, \aap, 528, A114
\bibitem[Underhill \& Gilroy(1990)]{und90} Underhill, A.~B., \& Gilroy, K.~K.\ 1990, \apj, 364, 626
\bibitem[van Buren \& McCray(1988)]{vanB88} van Buren, D., \& McCray, R.\ 1988, \apjl, 329, L93
\bibitem[van Leeuwen et al.(1997)]{vanl97} van Leeuwen, F., Evans, D.~W., Grenon, M., et al.\ 1997, \aap, 323, L61 
\bibitem[van Rensbergen et al.(1996)]{van96} van Rensbergen, W., Vanbeveren, D., \& De Loore, C.\ 1996, \aap, 305, 825
\bibitem[Villamariz et al.(2002)]{vil02} Villamariz, M.~R., Herrero, A., Becker, S.~R., \& Butler, K.\ 2002, \aap, 388, 940
\bibitem[Villamariz \& Herrero(2005)]{vil05} Villamariz, M.~R., \& Herrero, A.\ 2005, \aap, 442, 263
\bibitem[Walborn(1973)]{wal73} Walborn, N.~R.\ 1973, \aj, 78, 1067
\bibitem[Walborn et al.(2011)]{wal11} Walborn, N.~R., Ma{\'{\i}}z Apell{\'a}niz, J., Sota, A., et al.\ 2011, \aj, 142, 150
\bibitem[Whittaker \& Robinson(1944)]{whi44} Whittaker, E.~T., \& Robinson, G.\ 1944, The calculus of observations; a treatise on numerical mathematics, by Whittaker, E.~T.; Robinson, George.~ London, Blackie [1944],
\bibitem[Williams et al.(2011)]{wil11} Williams, S.~J., Gies, D.~R., Hillwig, T.~C., McSwain, M.~V., \& Huang, W.\ 2011, \aj, 142, 146
\bibitem[Zahn(1975)]{zah75} Zahn, J.-P.\ 1975, \aap, 41, 329
\bibitem[Zechmeister \& K\"urster(2009)]{zec09} Zechmeister, M., K\"urster, M.\ 2009, \aap, 496, 577
\end{thebibliography}
\end{document}